\documentclass[preprints,article,accept,pdftex,moreauthors]{Definitions/mdpi} 
\usepackage{isomath}
\usepackage{dcolumn}
\newcolumntype{.}{D{.}{.}{-1}}
\newcolumntype{0}{D{.}{.}{0}}
\usepackage{siunitx}
\usepackage{threeparttablex}
\usepackage{longtable}
\usepackage{aas_macros}
\usepackage{paracol}

\newcommand{\kms}{km\,s$^{-1}$}

\renewcommand{\vec}[1]{\mathbfit{#1}}

\newcommand{\Var}{\mathrm{Var}}

\newcommand{\los}{\mathrm{los}}

\newcommand{\Vh}{V_\mathrm{h}}

\newcommand{\vperp}{v_\perp}
\newcommand{\vlos}{v_\los} % \parallel

\newcommand{\proj}{\mathrm{p}}

\newcommand{\rproj}{r_\proj}
\newcommand{\Mass}{M}

\newcommand{\Msun}{\mbox{$\Mass_\odot$}}

\newcommand{\Mvir}{\Mass_\mathrm{vir}}
\newcommand{\Mtot}{\Mass_\mathrm{tot}}
\newcommand{\Mproj}{\Mass_\proj}

\newcommand{\Miso}{\Mass_\mathrm{I}}

\newcommand{\Mlos}{\Mass_\mathrm{los}}

\firstpage{1} 
\pubvolume{1}
\issuenum{1}
\articlenumber{0}
\pubyear{2025}
\copyrightyear{2025}
\datereceived{ } 
\daterevised{ } % Comment out if no revised date
\dateaccepted{ } 
\datepublished{ } 
\hreflink{https://doi.org/} % If needed use \linebreak
\Title{Line-of-sight mass estimator and the masses of the Milky Way and Andromeda Galaxy}

\TitleCitation{Line-of-sight mass estimator and the masses of the Milky Way and Andromeda Galaxy}

\Author{Danila Makarov $^{1}$, Dmitry Makarov $^{1}$\orcidA{}, Kirill Kozyrev $^{2, 1}$\orcidB{} and Noam Libeskind $^{3}$\orcidC{}}

\AuthorNames{Danila Makarov, Dmitry Makarov, Kirill Kozyrev and Noam Libeskind}

\address{%
$^{1}$ \quad Special Astrophysical Observatory, Russian Academy of Sciences, Nizhnij Arkhyz, 369167 Russia\\
$^{2}$ \quad Kazan Federal University, Kazan, Russia\\
$^{3}$ \quad Leibniz Institut f\"{u}r Astrophysik Potsdam (AIP), An der Sternwarte 16, D-14482, Potsdam, Germany}

\corres{Correspondence: dim@sao.ru}

\abstract{
The total mass of a galaxy group, such as the Milky Way (MW) and the Andromeda Galaxy (M~31), is typically determined from the kinematics of satellites within their virial zones.
Bahcall and Tremaine (1981) proposed the $v^2r$ estimator as an alternative to the virial theorem.
In this work, we extend their approach by incorporating the three-dimensional spatial distribution of satellites within the system to improve the reliability and accuracy of galaxy mass estimates.
Applying this method to a comprehensive dataset of Local Group satellites based on recent, high-precision distance measurements, we estimate the total mass of the MW to be $(7.9 \pm 2.3) \times 10^{11}$~$\Msun$ and that of M~31 to be $(15.5 \pm 3.4) \times 10^{11}$~$\Msun$.
The effectiveness of the method is constrained by the precision of distance measurements, making it particularly well suited for the Local Group, but challenging to apply to more distant systems.
}


\begin{document} 

\section{Introduction}

Analyzing kinematics within virial zones remains the primary---and sometimes the only---available method for estimating the total mass of gravitating systems.
To this end, various analogues of the virial theorem are used \citep{1981ApJ...244..805B, 1985ApJ...298....8H, 2010MNRAS.406..264W}.
For distant systems, only line-of-sight velocities and projected sky positions of galaxies are typically accessible through observations, leading to substantial uncertainties in mass estimates.
Accounting for the three-dimensional distribution of satellites, derived from distance measurements, eliminates the uncertainty associated with sky projection and, consequently, may improve the reliability of mass estimates for nearby groups.

%Over the past two decades, thanks to the Hubble Space Telescope and advances in high-precision distance measuring techniques, breakthroughs have been made in exploring the structure of the nearest Universe.
%For example, according to the Color-Magnitude Diagram Catalog~\citep{2021AJ....162...80A} of the Extragalactic Distances Database~\citep{2009AJ....138..323T}, at the moment, the tip of the red giant branch (TRGB) distances with a median accuracy of 3.7\% have been measured for about 500 nearby galaxies on scales up to 10~Mpc.
%Thanks to this, we know the real three-dimensional distribution of satellites around giant Local Volume galaxies, and in particular the detailed structure of the groups around our Galaxy and the Andromeda Galaxy.

Over the past two decades, significant breakthroughs have been made in exploring the structure of the nearby Universe, thanks to the Hubble Space Telescope and advances in high-precision distance measurement techniques. 
For instance, according to the Color-Magnitude Diagram Catalog~\citep{2021AJ....162...80A} of the Extragalactic Distances Database~\citep{2009AJ....138..323T}, the tip of the red giant branch (TRGB) distances have been measured for about 500 nearby galaxies with a median accuracy of 3.7\%, covering scales up to 10~Mpc.
As a result, we now have a detailed understanding of the three-dimensional distribution of satellites around giant Local Volume galaxies, particularly those surrounding the Milky Way (MW) and the Andromeda Galaxy (M~31).

In most nearby groups, the central galaxy dominates its system in both luminosity and mass.
The aim of this paper is to develop a simple method, with a minimal number of free parameters, for the mass estimation of a massive galaxy, accounting for the three-dimensional distribution of its satellites.
A simple model of a point mass surrounded by test particles can be used for this purpose.
This is exactly the case considered in the work of \citet{1981ApJ...244..805B}.
Therefore, we extend their projection estimator method by incorporating the three-dimensional satellite distances and averaging the observed parameters over all possible orbits.
Since the only remaining uncertainty is the projection of the satellite velocity onto the line of sight, we refer to this approach as the line-of-sight mass estimator.
In addition, it is also necessary to investigate the applicability of the method to the study of nearby galaxy groups.

The MW and M~31 serve as ideal laboratories for testing this method.
Modern sky surveys offer researchers access to a wealth of high-precision data, which, among other outcomes, has led to a remarkable increase in publications devoted to mass estimates of the two dominant Local Group galaxies, using a variety of methods and tracers.
As an example, we present several recent results that illustrate the diversity of approaches.

Modeling the rotation curve of a galaxy is a classical way to determine the parameters of its gravitational potential.
In this way, \citet{2023ApJ...946...73Z} constructed the circular velocity curve of the MW from 5 to 25~kpc using 54,000 luminous red giant branch stars. 
They determined the parameters of a model that includes the bulge, thin and thick disks, and a Navarro-Frenk-White dark halo~\citep{1995MNRAS.275...56N}, estimating the total mass of the Galaxy as $(8.5 \pm 1.2) \times 10^{11}$~\Msun{} with a corresponding virial radius of $192 \pm 9$~kpc.

Globular clusters allow us to trace the Galactic gravitational potential to much larger distances---typically around 40~kpc and extending up to 100~kpc.
Using Gaia Data Release 2, \citet{2019MNRAS.484.2832V} determined the proper motions of nearly the entire known population of Milky Way globular clusters.
By combining these with the distances and line-of-sight velocities, he analyzed their distribution in six-dimensional phase space.
As a result, he derived the total enclosed mass to be $5.4^{+1.1}_{-0.8} \times 10^{11}$~\Msun{} within 50~kpc, and $8.5^{+3.3}_{-2.0} \times 10^{11}$~\Msun{} within 100~kpc.
The extrapolated virial mass and radius are $12^{+15}_{-5} \times 10^{11}$~\Msun{} and $280^{+80}_{-50}$~kpc, respectively.

Satellites are the only tracers that cover the entire virialized region of a group.
In analyzing satellite kinematics, it is crucial to account for perturbations caused by the passage of a massive galaxy, such as the Large Magellanic Cloud (LMC), which can distort the satellite velocity distribution and bias the virial mass estimate.
\citet{2024OJAp....7E..50K} developed a new robust halo mass estimator, $M\propto r_\mathrm{med}\sigma^3_\mathrm{3D}$, calibrated using simulated MW-size halos.
Based on this, they estimated the virial mass of the Milky Way to be $(9.96 \pm 1.45) \times 10^{11}$~\Msun{}.

The next level involves studying the deceleration of the Hubble flow due to the gravitational influence of the Local Group, allowing for mass estimates on scales of about 1~Mpc. 
For example, \citet{2009MNRAS.393.1265K} measured the radius of the zero-velocity surface to be $R_0 = 0.96 \pm 0.03$~Mpc, which corresponds to the total mass of $\Mass_\mathrm{LG} = (19 \pm 2) \times 10^{11}$~\Msun{}. 
However, this topic is beyond the scope of the present study and will be addressed in future work.

The examples provided above represent only a tiny fraction of the studies devoted to understanding the structure of the Milky Way.
They demonstrate the potential of testing the new method using various approaches. 
In addition, the new method will provide one more independent estimate of the mass to the collection.

The article is organized as follows.
Section~\ref{sec:ProjectionMass} briefly describes the projection mass method.
Section~\ref{sec:LoSMass} is devoted to the line-of-sight mass estimator.
Comparison with numerical models is performed in Sections~\ref{sec:Simulations}.
Its application to the mass estimation of our Galaxy and the Andromeda Galaxy is discussed in Section~\ref{sec:LocalGroup}.
We conclude in Section~\ref{sec:Conclusion}.

\section{Projection mass method}
\label{sec:ProjectionMass}

By considering the motion of test particles around a point mass, \citet{1981ApJ...244..805B} showed that, due to projection effects, the mass of a galaxy group, as given by the virial theorem, $\Mvir$, is statistically:
\begin{enumerate}
\item \emph{biased}, meaning that the average of $\Mvir$ estimates for the same group does not necessarily equal to the true mass for a finite number of particles, $N$;
\item \emph{inefficient} with a large variance;
\item \emph{inconsistent} in some cases, meaning that, $\Mvir$ does not converge to the true mass as $N\to\infty$.
\end{enumerate}
The main reason for the inefficiency of the virial mass estimate is the uncertainty introduced by the projection factor in the distances between galaxies.
However, even without considering projection effects, $\Mvir$ remains ineffective, as nearby particles contribute more than distant ones to both the estimate of kinetic energy (as the sum of squared velocities, $\sum_i v_i^2$) and to the estimate of potential energy (as the harmonic mean of distances, $\sum_i 1/r_i^2$).
Since both distant and nearby particles provide the same information about the central mass, it is evident that the virial estimate, $\Mvir$, does not effectively utilize all available information.

As an alternative to the virial theorem, \citet{1981ApJ...244..805B} propose to use the so-called projection mass estimate, which is based directly on the observed values
\begin{equation}
    G \Mproj = a \vlos^2 \rproj,
\end{equation}
where $\vlos$ is the line-of-sight velocity of the particle with respect to the central body, and $\rproj$ is its projected separation.
This method treats test particles at all distances equally, since on average $\langle \vlos^2 \rproj \rangle = \mathrm{const}$.
The correction coefficient, $a$, is determined by averaging over all possible particle trajectories in the system. 
As a result, for the typical case of observing a distant group with a dominant massive galaxy, \citet{1981ApJ...244..805B} derived
\begin{equation}
    G \Mproj = \frac{32}{\pi (3-2\langle e^2 \rangle) } \langle \vlos^2 \rproj \rangle.
\label{eq:MprojOrig}
\end{equation}

To obtain a realistic estimate of the mass, one must make an assumption about the orbits of the particles.
The most natural assumption is that the velocity distribution is isotropic, with an eccentricity of $\left\langle e^2 \right\rangle = 1/2$.
In this case, the projected mass estimator, $\Miso$, becomes 
\begin{equation}
G \Miso = \frac{16}{\pi} \langle \vlos^2 \rproj \rangle,
\label{eq:MisoOrig}
\end{equation}
with the variance of
\begin{equation}
\Var\Bigl( \langle G\Miso \rangle \Bigr)
  = \frac{1}{N} \left( \frac{128}{5\pi^2} - 1 \right) \langle G\Miso \rangle^2,
\end{equation}
where $N$ is the number of test particles.

%---------------------------------------------------
\section{Line-of-sight mass method}
\label{sec:LoSMass}

Our goal is to derive a correction factor, $a$, for the mass estimator, based on an expression of the form
\begin{equation}
    G \Mlos = a \vlos^2 r,
\label{eq:Mlos}
\end{equation}
where $\vlos$ is the line-of-sight velocity of the test particle, and $r$ is the spatial separation from the central body.
Hereafter, we adopt the notation and approach used in the work of \citet{1981ApJ...244..805B}.
The average of any quantity $\xi(\vec{r},\vec{v})$ can be obtained by integrating over the distribution function $f(\vec{r},\vec{v})$ of a spherically symmetric gravitating system
\begin{equation}
\left\langle \xi \right\rangle = 
   A 
   \int_0^\infty r^2 dr 
   \int_0^\pi \sin\Theta \, d\Theta 
   \int_0^\infty \vperp \, d\vperp
   \int_{-\infty}^\infty \, dv_r 
   \int_0^{2\pi} \xi f
   \, d \phi^\prime.
\label{eq:AverOverDistribution}
\end{equation}
Here, $\Theta$ is the polar angle of the radius vector $\vec{r}$ in a spherical coordinate system, with the polar axis directed toward the observer;
$\phi^\prime$ is the azimuthal angle of the velocity vector $\vec{v}$ in a cylindrical system, where the z-axis is aligned along the radius vector;
$v_r$ and $\vperp$ are the components of the velocity vector directed along and perpendicular to the radius vector, respectively;
and $A$ is the normalization coefficient.

According to Jeans's theorem, any steady-state solution of the collisionless Boltzmann equation depends on the phase-space coordinates only through the integrals of motion. 
For a spherically symmetric system of test particles, the distribution function is determined by the energy, $E = G\Mass/r - v^2/2$, and the square of the angular momentum, $J^2 = r^2\vperp^2$.
For a particle with specific values of energy, $E_0$, and angular momentum, $J_0$, it can be expressed using the Dirac delta function\footnote{The energy, $E_0$, and the angular momentum, $J_0$, define the orbit of a satellite within a spherically symmetric potential, while the delta functions represent the phase-space distribution under the condition that the energy and angular momentum are constrained to these specific values.}:
%The distribution function of a spherically symmetric gravitating system can depend only on the energy, $E=G\Mass/r-v^2/2$, and the square of the angular momentum, $J^2=r^2\vperp^2$.
%Using the exact values of the energy, $E_0$ and the angular momentum $J_0$, the distribution function can be expressed in terms of the Dirac delta function\footnote{Specifically, the Dirac delta function imposes an exact relationship between the parameters. In this case, it binds the distance and velocity of the test particle to its total energy via $E_0 = \frac{GM}{r} - \frac{v^2}{2}$, and to its squared angular momentum via $J_0^2 = r^2 \vperp^2$.}:
\begin{align}
\nonumber
f(\vec{r},\vec{v}) & = f(E,J) = \delta(E-E_0)\,\delta(J^2-J_0^2) \\
  & = \delta(G\Mass/r-v^2/2-E_0)\,\delta(r^2\vperp^2-J_0^2).
\end{align}
Given the property of the delta function, 
\begin{equation*}
    \delta(f(x)) = \sum_i \frac{\delta(x-x_i)}{|f^\prime(x_i)|},
\end{equation*}
where $x_i$ are roots of the function $f(x)$, the equation~(\ref{eq:AverOverDistribution}) is transformed to
\begin{align}
\nonumber
\left\langle \xi \right\rangle
& = A 
   \int_0^\infty r^2 dr 
   \int_0^\pi \sin\Theta \, d\Theta 
   \int_0^\infty \vperp \, d\vperp
   \int_{-\infty}^\infty d v_r \\
\nonumber
&  \quad\times
   \int_0^{2\pi}
   \xi
   \delta(G\Mass/r-v^2/2-E_0)\delta(r^2\vperp^2-J_0^2)
   \, d\phi^\prime \\
\nonumber
& = A 
   \int_{r_{\min{}}}^{r_{\max{}}} r^2 dr 
   \int_0^\pi \sin\Theta \, d\Theta 
   \int_0^{2\pi}
   \xi
   \frac{2}{|v_r|}
   \frac{1}{|2r^2|}
   \, d \phi^\prime \\
& = A 
   \int_{r_{\min{}}}^{r_{\max{}}} dr 
   \int_0^\pi \sin\Theta \, d\Theta 
   \int_0^{2\pi}
   \frac{\xi}{|v_r|}
   \, d \phi^\prime
\label{eq:SimplifiedAverOverDistribution}
\end{align}
where 
\begin{align}
\vperp^2 & =\frac{J_0^2}{r^2} =  \frac{(GM)^2}{2E_0} (1-e^2) \frac{1}{r^2} \label{eq:sqvperp} \\
\nonumber
v_r^2    & = 2\frac{GM}{r}-\frac{J_0^2}{r^2}-2E_0 \\
         & = 2\frac{GM}{r}-\frac{(GM)^2}{2E_0} (1-e^2) \frac{1}{r^2} -2E_0 , \label{eq:sqvr}
\end{align}
and it is taken into account that the maximum angular momentum for a given energy $E_0$ is $J_{\max} = G\Mass/(2E_0)^{1/2}$, and the eccentricity is defined as $e^2 = 1-\left(J/J_{\max}\right)^2$.
The integration limits are defined by the relations $v_r(r_{\min}) = v_r(r_{\max}) = 0$ and are equal to
\begin{align}
r_{\min} = & \frac{GM}{2E_0} (1-e)  \label{eq:rmin} \\
r_{\max} = & \frac{GM}{2E_0} (1+e). \label{eq:rmax}
\end{align}

\citet{1981ApJ...244..805B} determined the normalization coefficient $A$ by considering the condition $\langle 1 \rangle = 1$
\begin{equation}
A = \left( 4\pi \int_{r_{\min}}^{r_{\max}} \frac{d r}{|v_r|} \right)^{-1} = \frac{ \left(2E_0\right)^{3/2} }{ 4\pi^2 G\Mass }.
\end{equation}

Using the dimensionless variable $x = \frac{2E_0}{GM}r$, equations~(\ref{eq:SimplifiedAverOverDistribution})--(\ref{eq:sqvr}) simplify to: 
\begin{align}
r        & = \frac{GM}{2E_0}x \nonumber \\
\vperp^2 & = 2E_0 \frac{1-e^2}{x^2} \nonumber  \\
v_r^2    & = 2E_0 \frac{e^2-(x-1)^2}{x^2} \nonumber  \\
\left\langle \xi \right\rangle & = \frac{1}{4\pi^2} \int_{1-e}^{1+e} \frac{x \, dx}{\sqrt{e^2-(x-1)^2}} \int_0^\pi \sin\Theta \, d\Theta \int_0^{2\pi} \xi \, d \phi^\prime.
\end{align}
It is clear from this that, although we considered the distribution function at the arbitrary fixed energy, $E_0$, of a test particle, the average of observable values of the form $\langle v^2 r \rangle \propto 2E_0\frac{GM}{2E_0} = GM$ depends only on the mass of the central body, not on $E_0$.
This allows us to determine the mass regardless of the particle energy distribution.
However, orbital eccentricity still plays a role.

As noted by \citet{1981ApJ...244..805B}, the variance of the mass estimator of the form given in equation~(\ref{eq:Mlos}) is well-defined and can be determined in a similar way.
Let us define the value
\begin{equation}
    (G \Mlos)^2 = b (\vlos^2 r)^2.
\end{equation}
Then the variance is 
\begin{align}
\nonumber
\Var(G\Mlos) 
    & = \Var( a\vlos^2r ) = a^2 \Var( \vlos^2 r ) \\
\nonumber
    & = a^2 \left( \langle (\vlos^2 r)^2 \rangle - \langle \vlos^2 r \rangle^2 \right)
      = a^2 \left( \frac{(GM)^2}{b} - \left(\frac{GM}{a}\right)^2 \right) \\
    & = \left( \frac{a^2}{b} -1\right) (G\Mlos)^2.
\end{align}
Accordingly, the variance of the mean $\langle GM \rangle$ is equal to
\begin{equation}
\Var\Bigl( \langle G\Mlos \rangle \Bigr) = \frac{1}{N}\left( \frac{a^2}{b} -1\right) \langle G\Mlos \rangle^2,
\end{equation}
where $N$ is the number of the test particles.

Our case of finding the average $\langle \vlos^2 r \rangle$ actually involves two scenarios:
one where the observer sits near the center of the group (as in the case of the Milky Way), and another where the observer views the system from the side (as in the case with all other nearby groups of galaxies).

%---------------------------------------------------
\subsection{Milky Way case}
\label{sec:LoSMassMilkyWay}

In the case of the Milky Way group, we are essentially at the center of the system with respect to the distribution of satellites.
As a consequence, the line of sight nearly coincides with the radial coordinate, allowing us to approximate $\vlos \approx v_r$ with good accuracy.
Thus,
\begin{align}
\langle \vlos^2 r \rangle = \langle v_r^2 r \rangle 
& = 4 \pi A \int_{r_{\min{}}}^{r_{\max{}}} r |v_r| \, dr \nonumber \\
& = \frac{GM}{\pi} \int_{1-e}^{1+e} \sqrt{e^2-(x-1)^2} \, dx
= \frac{ G\Mass }{ 2 } \langle e^2 \rangle.
\label{eq:MlosMilkyWay}
\end{align}
Consequently, the line-of-sight mass estimator for the Milky Way is
\begin{equation}
G\Mlos = \frac{2}{\langle e^2 \rangle} \langle v_r^2 r \rangle.
\end{equation}
To estimate the variance, we obtain the value
\begin{equation}
\langle (\vlos^2 r)^2 \rangle = (G\Mlos)^2 \left( \frac{3}{2}e^2 -1 +(1-e^2)^{3/2}\right),
\end{equation}
which yields the variance of the mass estimate for our Galaxy equal to
\begin{equation}
    \Var\Bigl( \langle G\Mlos \rangle \Bigr) = \frac{1}{N} \frac{4}{\langle e^2 \rangle} \left( \frac{3}{2}\langle e^2\rangle -1 + (1-\langle e^2 \rangle)^{3/2}\right).
\end{equation}

In the case of an isotropic velocity distribution, where $\langle e^2 \rangle=\frac{1}{2}$, the line-of-sight mass estimator and its variance become
\begin{align}
G\Miso & = 4 \langle v_r^2 r \rangle \label{eq:MisoMilkyWay}\\
\Var\Bigl( \langle G\Miso \rangle \Bigr) & = \frac{4(\sqrt{2}-1)}{N} \langle G\Miso \rangle^2 \approx \frac{1.657}{N} \langle G\Miso \rangle^2.
\end{align}

%---------------------------------------------------
\subsection{Nearby group case}
\label{sec:LoSMassNerabyGrp}

Consider the case where the observer is at a sufficient distance from the group.
Assume that the three-dimensional distribution and line-of-sight velocities of the satellites are known from observations.
This case corresponds to nearby galaxy groups in the Local Volume, where high-precision distances have been measured for many galaxies.
The line-of-sight velocity can be represented as
$\vlos=v_r\cos\Theta-\vperp\sin\phi^\prime\sin\Theta$. 
The equation~(\ref{eq:SimplifiedAverOverDistribution}) is transformed into the form
\begin{align}
\langle \vlos^2 r \rangle 
& = \frac{4 \pi A}{3} \int_{r_{\min{}}}^{r_{\max{}}} r v_r \left( 1+\frac{\vperp^2}{v_r^2}\right) \, dr \nonumber \\
& = \frac{GM}{3\pi} \int_{1-e}^{1+e} \frac{1-(x-1)^2}{\sqrt{e^2-(x-1)^2}} \, dx \nonumber \\
& = \frac{ G\Mass }{ 6 } \left( 2 - \langle e^2 \rangle \right).
\label{eq:MlosNearbyGroup}
\end{align}
Similarly, for the mean square, we obtain the following expression:
\begin{equation}
\langle (\vlos^2 r)^2 \rangle  = \frac{ (GM)^2}{10} \left( 2 - \langle e^2 \rangle \right).
\end{equation}

The corresponding mass estimator and its variance are
\begin{align}
G\Mlos & = \frac{6}{2-\langle e^2 \rangle} \langle \vlos^2 r \rangle \\
\Var\Bigl( \langle G\Mlos \rangle \Bigr) & = \frac{1}{N} \left( \frac{36}{10(2-\langle e^2 \rangle)} -1 \right) \langle GM \rangle^2.
\end{align}

For an isotropic distribution over orbits $\langle e^2 \rangle = \frac{1}{2}$ we obtain
\begin{align}
G\Miso & = 4 \langle \vlos^2 r \rangle \label{eq:MisoNearbyGroup}\\
\Var\Bigl( \langle G\Miso \rangle \Bigr) & = \frac{7}{5N} \langle G\Miso \rangle^2 = \frac{1.4}{N} \langle G\Miso \rangle^2.
\end{align}

Note that, despite the fundamental difference in the general expressions for the mass of the Milky Way~(eq.~\ref{eq:MlosMilkyWay}) and the nearby group~(eq.~\ref{eq:MlosNearbyGroup}), the correction factors for the isotropic case, as give in equations~(\ref{eq:MisoMilkyWay}) and (\ref{eq:MisoNearbyGroup}), coincide surprisingly.

%---------------------------------------------------
\subsection{Additional notes}
\label{sec:LoSMassNotes}

\citet{2010MNRAS.406..264W} presented an entire family of estimators of the form $\Mass \propto \langle v^2 r^\alpha \rangle$, similar to the projected mass method initially introduced by \citet{1981ApJ...244..805B} and later expanded by \citet{1985ApJ...298....8H}.
The estimators are derived from solutions of the Jeans's equation, assuming that the tracers follow a power-law density distribution and move within a gravitational potential that also takes a power-law form.

It is worth noting that, despite significant progress in discovering the satellites of the Milky Way and the Andromeda Galaxy, their spatial distribution remains strongly affected by selection effects. 
This effect becomes even more pronounced for more distant systems.
As \citet{2010MNRAS.406..264W} point out, this can alter the observed power-law slope of the tracer number density and, consequently, lead to a bias in the mass estimate.
%An important advantage of our approach is that it averages the observed quantities over all possible orbits, regardless of the model satellite distribution.
%In contrast, our approach averages the observed quantities $\vlos^2 r$ over all possible orbits, independently of any assumed model for the satellite distribution.
Although our approach is simpler, it bypasses this problem by averaging the observables $\vlos^2 r$ over all possible orbits, independent of the radial distribution of satellites, whether measured or model-based.
This may be particularly useful for distant systems, where the number of known satellites is very limited.
%\footnote{Our solution can be considered as a special case of the estimators proposed by \citet{2010MNRAS.406..264W}, corresponding to a specific choice of parameters, namely $\alpha = 1$ and $\gamma = 3$, although it is derived independently.}.
In addition, this approach allowed us to derive explicit expressions for the error in the mass estimation.

%---------------------------------------------------
\section{Comparison with cosmological simulations}
\label{sec:Simulations}

\begin{figure}
    \centering
    \includegraphics[width=0.6\textwidth]{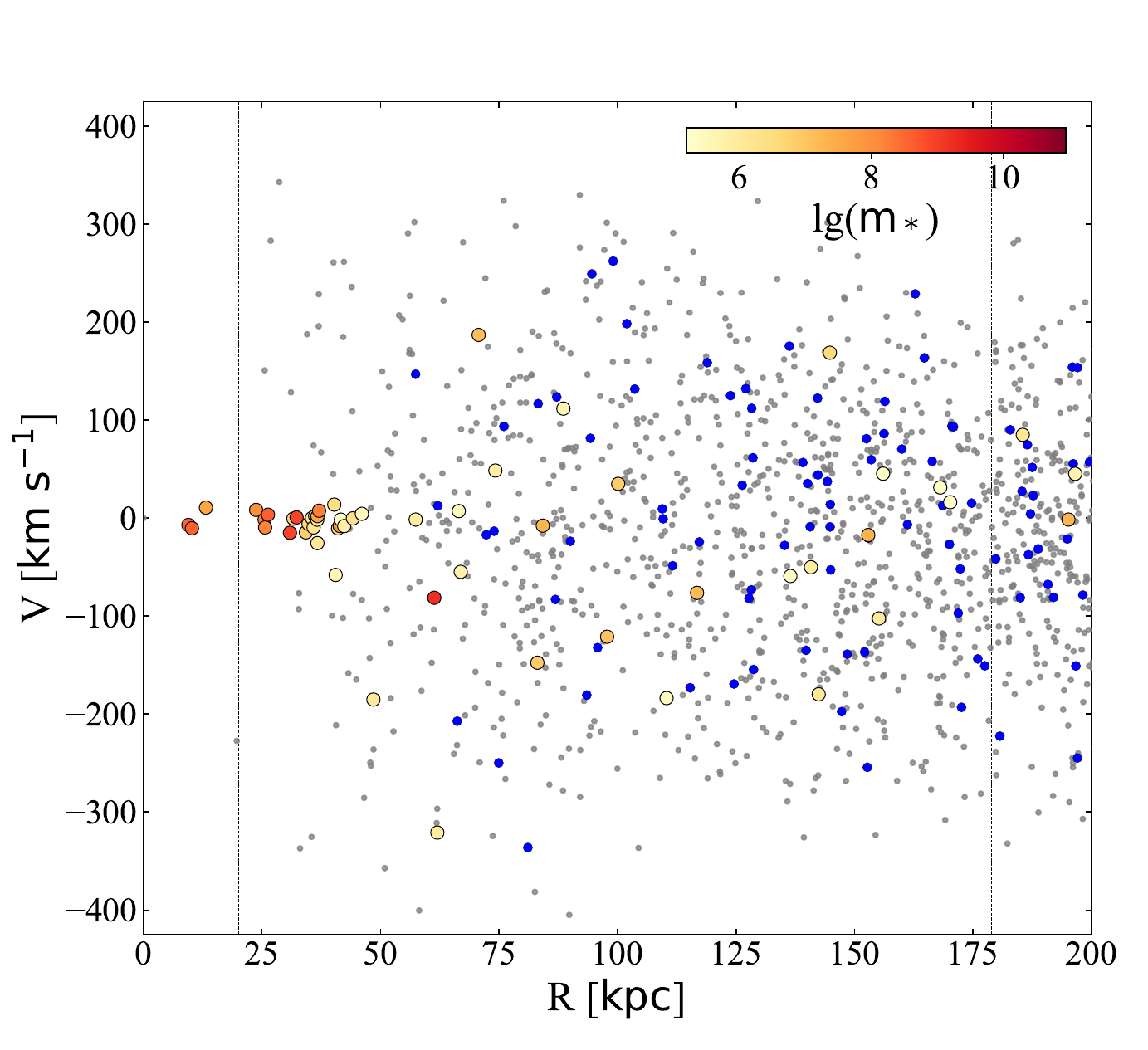}
    
    \centering
    \includegraphics[width=0.49\textwidth]{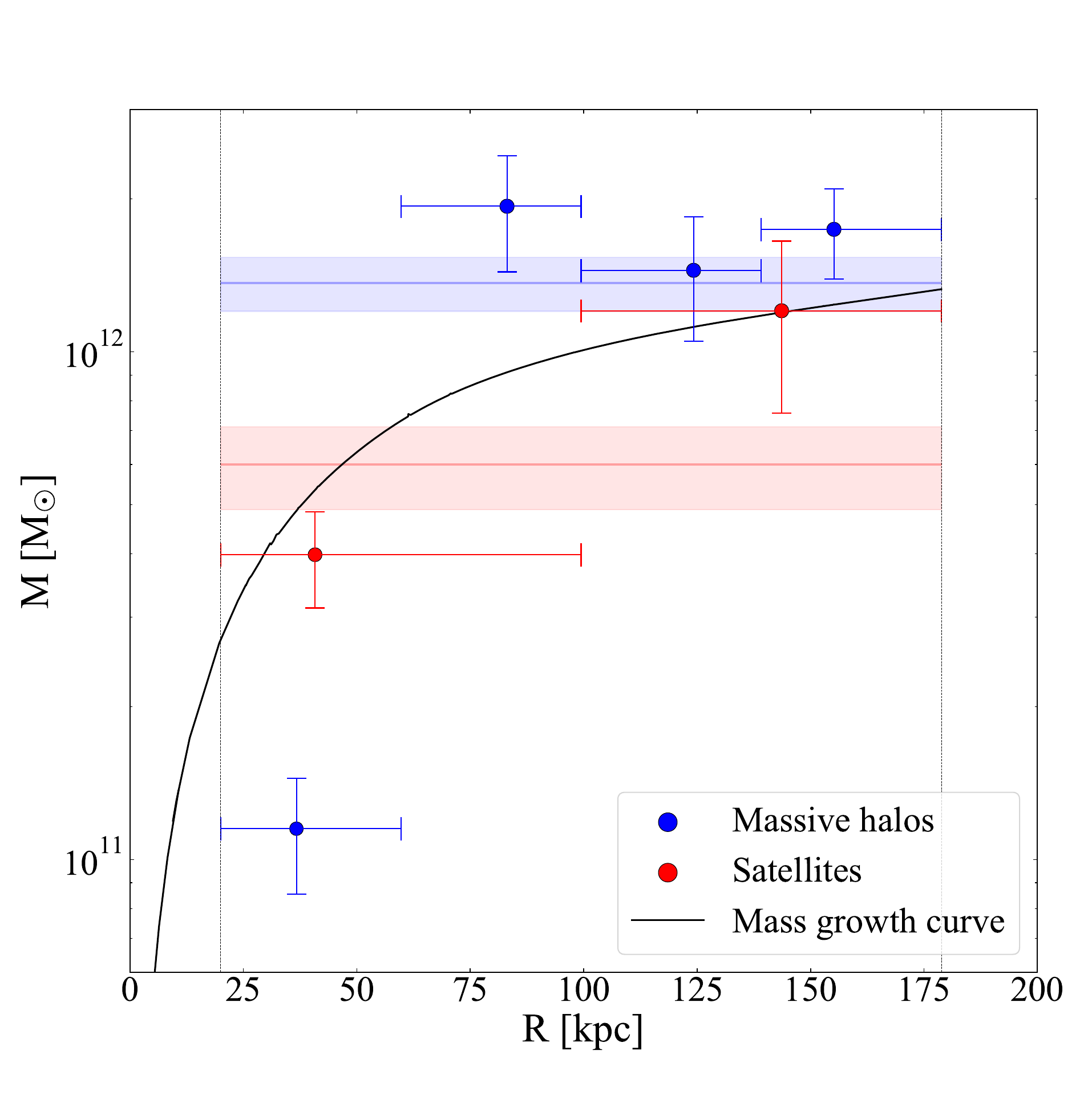}
    \includegraphics[width=0.49\textwidth]{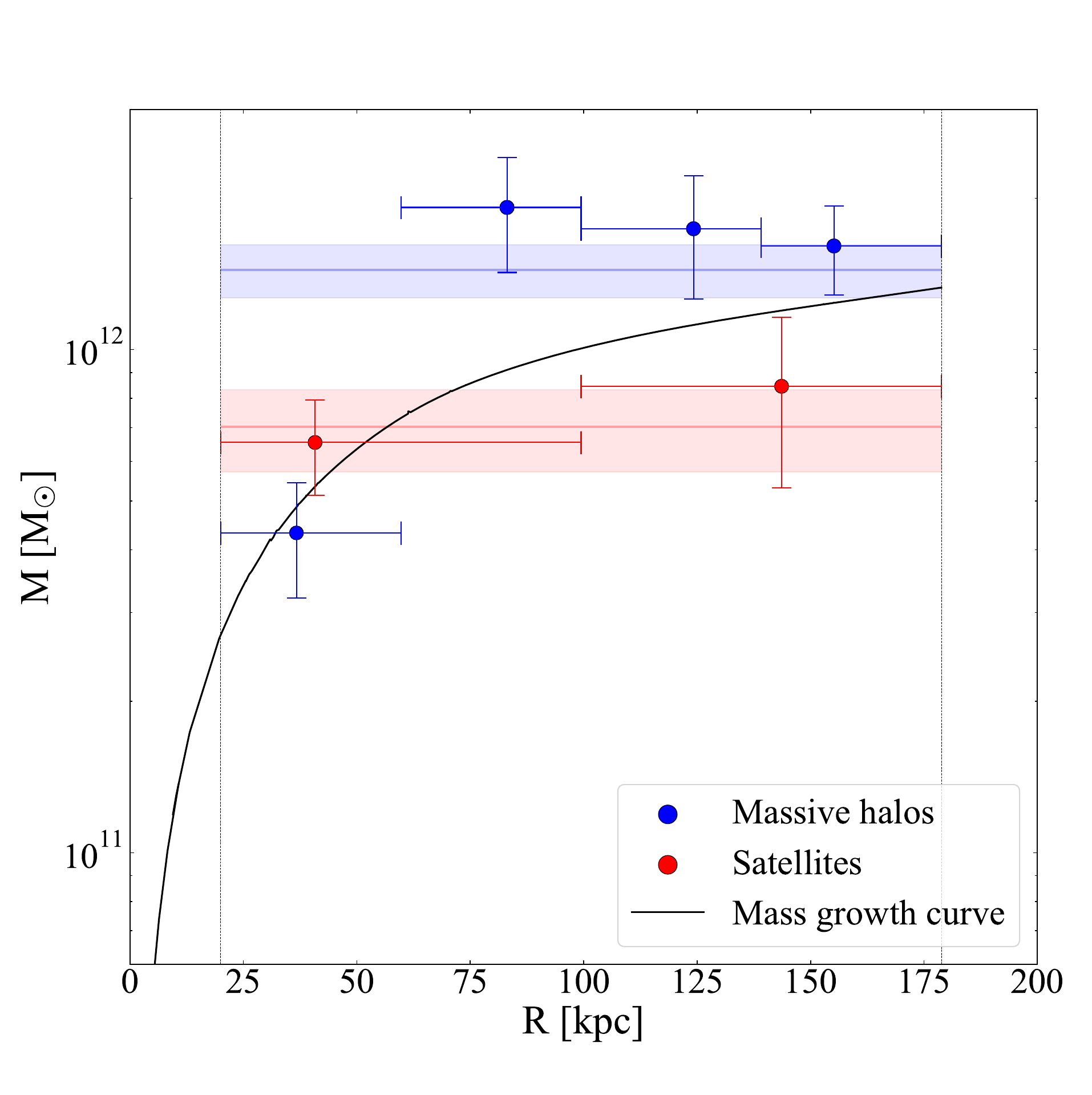}
    
    \caption{
    \textit{Top panel:} 
    Radial velocity–distance diagram for all halos surrounding the MW analog in the HESTIA \texttt{17\_11} simulation.
    Gray dots represent the distribution of low-mass dark matter halos. 
    Blue dots correspond to massive halos with $m_\mathrm{dm} > 4.3 \times 10^7$~\Msun{}.
    Large colored dots indicate satellite-like halos with stellar masses greater than $m_{*} > 1.5\times10^5$~\Msun{}.
    Vertical lines denote the selected region beyond 20~kpc and within the virial radius of 178.9~kpc.
    \textit{Bottom left panel:}
    Mass estimates based on observations from the center of the massive halo, within the radial range of $[20, 178.9]$~kpc.
    Red dots show the masses obtained for two satellite subgroups, while the red line represents the result based on all satellites within the specified range.
    Similarly, four blue dots indicate the masses estimates from 112 massive halos grouped into four equal radial intervals, while the blue line shows the result for the full sample.
    The solid black line depicts the cumulative mass profile $M(<r)$, as extracted directly from the simulation.
    \textit{Bottom right panel:}
    Same as the left panel, but from the viewpoint of the nearest massive halo, imitating observations of M~31.
    }
    \label{fig:HESTIAestimates}
\end{figure}

To test the applicability of the method, we applied it to the High-resolution Environmental Simulations of The Immediate Area~\citep[HESTIA,][]{2020MNRAS.498.2968L}.
It is a set of cosmological models specifically designed to simulate the Local Group of galaxies. 
These constrained dark-matter and Magneto-Hydrodynamic simulations are based on the AREPO code~\citep{2020ApJS..248...32W}.
The initial conditions are derived from the peculiar velocities of the nearby galaxies in the CosmicFlows-2 survey~\citep{2013AJ....146...86T}, using a reverse Zeldovich approximation~\citep{2013MNRAS.430..888D}.
Baryonic physics follows the AURIGA galaxy formation model~\citep{2017MNRAS.467..179G}.

The simulations include a high-resolution region of 3--5~Mpc around the Local Group analog, where galaxies similar to the Milky Way and Andromeda Galaxy are formed.
In this area, a mass resolution of $m_\mathrm{dm} = 1.2 \times 10^6$~\Msun{} and a gas particle resolution of $m_\mathrm{gas} = 1.8 \times 10^5$~\Msun{} are achieved.
These simulated galaxies match observations in terms of various physical and morphological properties, making them useful for studying the role of the surrounding environment in galaxy evolution~\citep{2020MNRAS.498.2968L}.
Thanks to the technique of constrained simulations, the Local Group analog is surrounded by structures that closely resemble real-world environments, such as the Virgo-like cluster, local void, and local sheet.

HESTIA offers three high-resolution models of the Local Group.
We are primarily interested in the positions of the halos in space and their peculiar velocity vectors at $z=0$.
All relevant quantities are expressed in simulation units and can be rescaled to different cosmological parameters if necessary.
The top panel of Fig.~\ref{fig:HESTIAestimates} presents the radial velocity of surrounding halos as a function of distance for the second most massive halo---the MW analog---in the \texttt{17\_11} HESTIA run.
Its total mass is $\Mtot = 13.3 \times 10^{11}$~\Msun{}, while the largest halo, which is the analog of M~31, has a total mass of $\Mtot = 15.6 \times 10^{11}$~\Msun{}.
The first thing that catches the eye is the significant difference in the velocity-distance distribution of satellites compared to that observed around the Milky Way (Fig.~\ref{fig:MWVelDist}).
The HESTIA `satellites' (the halos with stars) within 50~kpc in the \texttt{17\_11} run exhibit a very low line-of-sight velocity dispersion, $\sigma_{V} = 36.3$~\kms{}, compared to $\sigma_{V} = 113.3$~\kms{} for more distant halos.
This reflects the effects of dynamical friction and the reorganization of the orbits of the nearest halos into more circular ones, as well as the impact of the smaller mass enclosed within this radius.
Additionally, below 30--40~kpc, selection effects in halo identification become significant, resulting in a noticeable deficiency of halos at short distances from the central body.
%This leads to fewer statistics and, more importantly, a significant underestimate of the mass inside 50~kpc in the case of observation from the center of the system.
This pattern is typical for all central massive halos in the HESTIA runs.

\begin{table}
\centering
\caption{Statistics on the tests of mass estimation of the MW and M~31 analogs using the HESTIA simulations.}
\label{tab:HestiaTests}
\begin{tabular}{llcccc}
\hline\hline
& Sample  &  \multicolumn{2}{c}{$\langle \frac{\Miso}{\sum \Mass_\mathrm{halo}} \rangle$} & $\sigma$ & Median\\
\hline
\multicolumn{6}{l}{Observations from the outside (the case of M~31)} \\
& Heavy halos  & 1.054 & $\pm 0.042$ & 0.103 & 1.055 \\
& Satellites   & 0.858 & $\pm 0.157$ & 0.384 & 0.848 \\
\hline
\multicolumn{6}{l}{Observations from the inside (the case of MW)} \\
& Heavy halos  & 1.022 & $\pm 0.048$ & 0.116 & 1.047 \\
& Satellites   & 0.704 & $\pm 0.129$ & 0.315 & 0.625 \\
\hline\hline
\end{tabular}
\end{table}

% We focus on region within 200~kpc (which is closely equal to 300 kpc if $H_0 = 67.8$~\kmsMpc{}) around the two most massive halos, the MW and M~31 analogs, and mimic real observations.
To mimic real observations, we placed an observer in the center of one of these massive halos and calculated the line-of-sight velocities of all surrounding halos based on their positions and peculiar velocities.
We focus on virial zones around the two most massive halos: the MW and M~31 analogs.
Their virial radii vary from 143 to 189~kpc, depending on the mass of the central halo.
We also exclude from consideration all halos located within 20 kpc of the main massive halo. 
This decision is motivated by two factors:
first, the difficulty of reliably identifying such objects in cosmological simulations due to the high-density region of the central massive halo; 
and second, observational evidence that satellites at these distances are likely embedded within the central galaxy body, where they are subject to strong tidal forces and eventual accretion.
As a result, we obtain two sets of data: one for observing the satellites of the given halo and another for observing the satellites of the next massive halo.
In total, we obtain six datasets for each scenario.
In our tests, we apply the line-of-sight mass estimator, assuming isotropic orbits (Section~\ref{sec:LoSMass}), to both samples. 
The first consists of `satellites'---halos containing star particles with stellar mass $m_{*} > 1.5 \times 10^{5}$~\Msun{}---and typically includes around 30 objects. 
The second sample of heavy halos includes all halos with $m_\mathrm{dm} > 4.3 \times 10^{7}$~\Msun{}, corresponding to the lower mass threshold for halos capable of hosting star formation, and typically comprises around 100 test particles.
We compared the estimated mass with the total mass of all halos enclosed within the virial radius.
The results are summarized in Table~\ref{tab:HestiaTests}.
The bottom left and right panels of Fig.~\ref{fig:HESTIAestimates} compare the mass estimates with the mass growth curve within the virial radius of the MW analog in the HESTIA simulation \texttt{17\_11}. 
The left panel represents the point of view of the internal observer, while the right panel reflects the view from the center of the neighboring massive halo.

It is important to note that the masses derived from the samples at different distances follow the halo mass growth curve in the simulations.
Because of the limited number of 'satellites', mass estimates based on this sample exhibit significant scatter. 
In contrast, including all massive halos significantly improves the statistics, reduces the scatter to 10--12\%, and enhances the  robustness of the mass estimates.
As shown in Table~\ref{tab:HestiaTests}, the average ratio of the estimated mass to the true value is statistically consistent with unity within the 1-sigma uncertainty.
This supports the conclusion that the proposed approach is suitable for estimating the mass of virialized systems and provides realistic measurements.

%
% satellites: 
% from inside:  <>= 1.26 +-  0.32  Std= 0.78 : Me= 0.95 [ 0.77, 1.80] IQR=1.03 MAD= 0.30
% from outside: <>= 1.29 +-  0.36  Std= 0.89 : Me= 1.06 [ 0.58, 2.00] IQR=1.42 MAD= 0.56
%
% all halos:
% from inside:  <>= 1.35 +-  0.08  Std= 0.20 : Me= 1.32 [ 1.18, 1.52] IQR=0.35 MAD= 0.16
% from outside: <>= 1.27 +-  0.06  Std= 0.15 : Me= 1.28 [ 1.20, 1.35] IQR=0.16 MAD= 0.08

%---------------------------------------------------
\section{Application to the Local Group of galaxies}
\label{sec:LocalGroup}

\begin{figure*}
\centering
\includegraphics[width=0.49\textwidth,bb=87 184 493 580,clip]{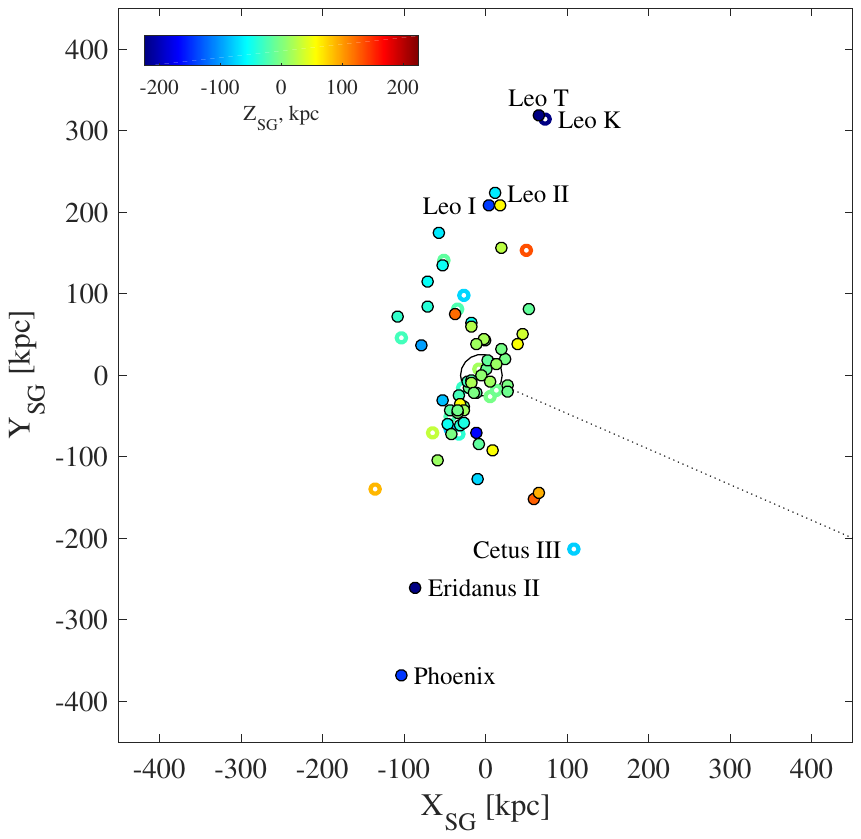}
\includegraphics[width=0.49\textwidth,bb=87 184 493 580,clip]{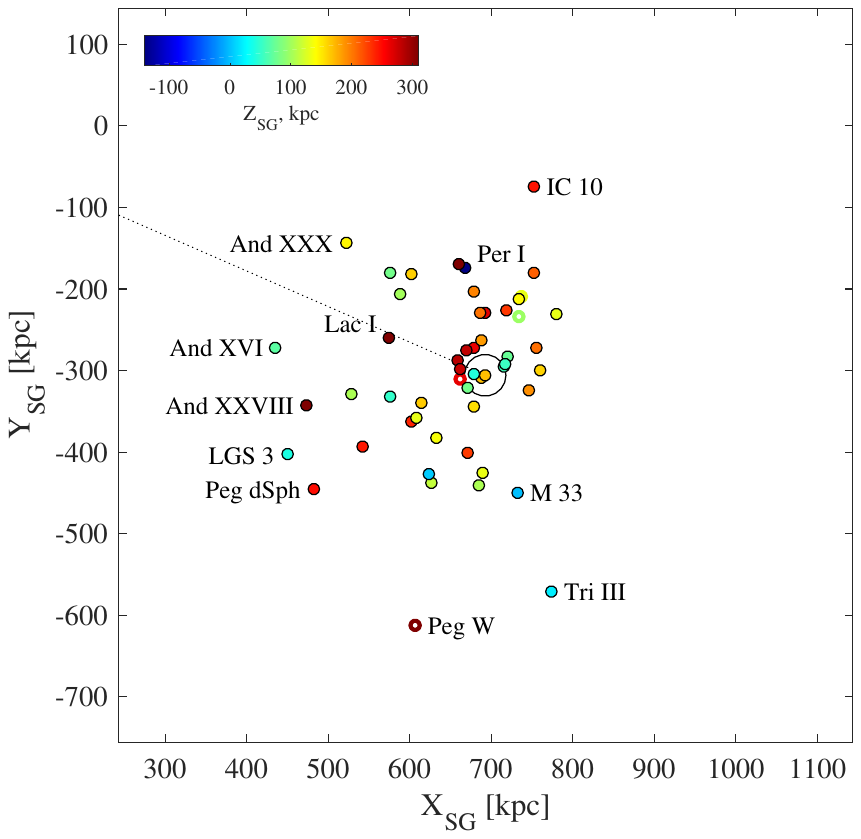}
\caption{
3D-distribution of galaxies within 450~kpc from MW (left panel) and M~31 (right panel) in Cartesian Supergalactic coordinates.
The Z-axis is indicated by color.
Galaxies with unknown line-of-sight velocity are shown by open colored circles.
The dotted line connects MW and M~31.
Satellites located further than 220 kpc from the central galaxy are labeled.
}
\label{fig:LocalGroupXYZ}
\end{figure*}

The Local Group of galaxies requires no special introduction.
It is a fairly isolated system, bound within a radius of approximately 1~Mpc.
The nearest comparable galaxy groups are located at a distance of 3--4~Mpc~\citep{2014MNRAS.440..405M}.
The Local Group is based on two giant spiral galaxies separated by 780~kpc, the Milky Way and the Andromeda Galaxy, which are similar to each other in many ways.
These giants are surrounded by rich suites of satellites that extend up to 300~kpc.
Only about a dozen dwarf galaxies are known in the Local Group outside the virial zones of the MW and M~31.
Thanks to modern deep optical surveys and systematic searches, our knowledge of the Local Group population has expanded dramatically.
The list of satellites is growing with new members all the time.
This allows one to trace their population to extremely low luminosity, which is technically inaccessible for more distant galaxy groups.
The recently discovered galaxy Ursa~Major~III~\citep{2024ApJ...961...92S} has a luminosity of $M_V=+2.2$~mag, corresponding to a stellar mass of only 16~\Msun{}.
The subsystems of the MW and M~31 satellites do not overlap, allowing them to be studied independently, without considering complex interactions.
Various methods and subsystems are used to estimate the mass, from the motion of stars and gas in the galaxies themselves, to the study of the kinematics of globular clusters and satellites, the timing argument for the orbiting of the MW and M~31, and the Hubble flow braking at the Local Group boundary~\citep[see recent reviews by][]{2020SCPMA..6309801W, 2023ARep...67..812B, 2023arXiv230503293B}.
All this allows us to trace the distribution of both baryonic and dark matter in the Local Group, spanning a scale from a few kiloparsecs to approximately one megaparsec.
There is a consensus that the M~31 is a slightly more massive galaxy than MW~\citep{2009MNRAS.393.1265K, 2010MNRAS.406..264W, 2022MNRAS.513.2385C}, although the mass estimates for both galaxies vary widely, sometimes leading to the opposite conclusion~\citep{2014MNRAS.443.2204P}.
New mass estimates for the MW and M~31 remain an important goal in the study of these galaxies and the Local Group as a whole.

The sample of the MW and M~31 satellites was compiled based on the latest version of the Local Volume galaxy database~\citep{2012AstBu..67..115K}, taking into account recent work on studying cosmic flows around nearby massive galaxies~\citep{2018A&A...609A..11K}, measuring proper motions of MW satellites~\citep{2020AJ....160..124M, 2022ApJ...940..136P}, estimating distances from RR Lyrae variables~\citep{2022ApJ...938..101S}, recent galaxy discoveries~\citep[for instance][]{2020ApJ...893...47D}, HyperLeda~\citep{2014A&A...570A..13M} and NED databases, and many other works.
Members of the MW and M~31 groups with compiled distances and heliocentric line-of-sight velocities are presented in Appendices~\ref{sec:MW_datatable} and \ref{sec:M31_datatable}, respectively.

Figure~\ref{fig:LocalGroupXYZ} shows the three-dimensional distribution of satellites of our Galaxy (left panel) and the Andromeda Galaxy (right panel) in Supergalactic coordinates (the Z coordinate is indicated by color).
The famous satellite planes are clearly visible around both galaxies~\citep{2021Galax...9...66P}.
Also, the M~31 satellites demonstrate a well-known skew toward our Galaxy~\citep{2013ApJ...766..120C}.

%---------------------------------------------------
\subsection{Milky Way}

\begin{figure}
\centering
\includegraphics[width=\linewidth, bb=42 189 549 584, clip]{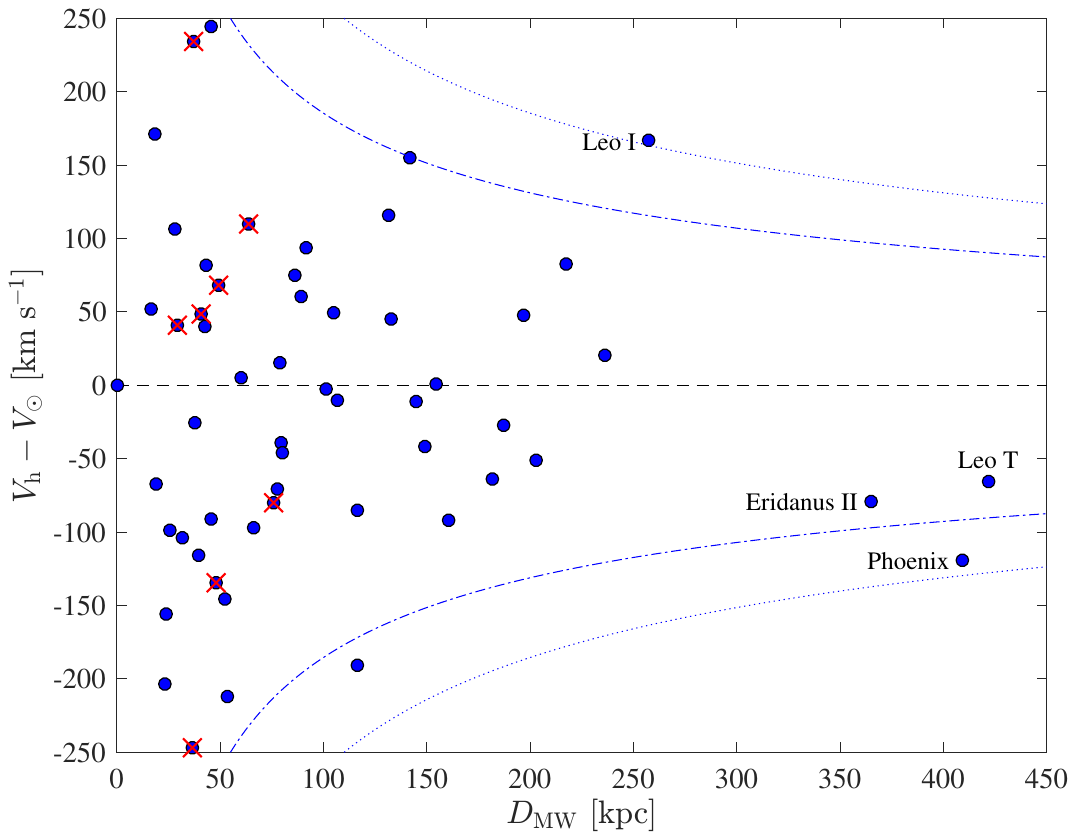}
\caption{
The velocity-distance distribution of the nearest galaxies relative to the center of our Galaxy.
The circular and escape velocities of the point mass $8\times10^{11}$~\Msun{} are shown by the dashed and dotted lines, respectively.
}
\label{fig:MWVelDist}
\end{figure}

Thanks to the Gaia mission~\citep{2018A&A...616A...1G}, the proper motions of most of the satellites of our Galaxy have been measured, which unambiguously allows us to determine all six components of the satellite position in phase space.
This information is actively used to refine the structure of our Galaxy, to estimate its total mass, and to analyze the features of satellite orbits.
It should be emphasized that the line-of-sight velocities of satellites are measured an order of magnitude more accurately than their proper motions.
Therefore, an independent mass estimate based on simple line-of-sight velocities remains extremely important.

Moreover, even a simple analysis of the line-of-sight velocity distribution can reveal unexpected and interesting effects.
Recently, \citet{2023MNRAS.521.3540M} discovered that the dipole component of the line-of-sight velocity field shows an unexpectedly large amplitude of $226\pm50$~\kms{} of the bulk motion of nearby satellites at distances less than 100~kpc.
This anomaly is caused by only eight galaxies (crossed out with red crosses in the left panel of Fig.~\ref{fig:MWVelDist}), which include the LMC and three galaxies from its escort.
Numerical simulations demonstrate that this velocity pattern is consistent with the assumption of the first flyby of the massive LMC around our Galaxy and the perturbation it creates in the motion of the MW satellites.
This example shows the importance of a careful selection of ``the test particles'' for kinematic analysis.
Objects on the first flyby do not have time to virialize and therefore should be excluded from consideration when estimating the total mass of the system.

At present, there are 67 known satellites of our Galaxy within the 260~kpc region.
As can be seen in Fig.~\ref{fig:MWVelDist}, there is a clear separation between the well-randomized MW satellites up to distances of 260~kpc, indicating the extension of the virial zone, and three galaxies with systematic negative velocity at a distance of 400~kpc, which are probably just entering the MW halo for the first time.
Thus, we limited the analysis area to a distance of 260~kpc.

As shown by \citet{2023MNRAS.521.3540M}, after excluding eight objects that most strongly distort the behavior of the solar apex, the observed collective motion of the remaining satellites is consistent within errors with the motion of the Sun in our Galaxy.
Thus, in further analysis, the observed line-of-sight velocities were simply corrected for the solar velocity vector of $(9.5, 250.7, 8.56)$~\kms{}~\citep{2024MNRAS.530..710A} with respect to the Galactic center, determined from the proper motion of Sgr~A* of $(-6.411\pm0.008,-0.219 \pm0.007)$~mas\,yr$^{-1}$ in Galactic coordinates~\citep{2020ApJ...892...39R} and the distance of $8249 \pm9 \pm45$~pc to the central supermassive black hole~\citep{2021A&A...647A..59G}.

The case of Leo~I deserves special attention.
As can be seen in Fig.~\ref{fig:MWVelDist}, this most distant satellite near the border of the virial zone has an extremely high line-of-sight velocity.
The inclusion of Leo~I (just one galaxy) in the analysis leads to a dramatic 33\% increase of the total MW mass~\citep{2010MNRAS.406..264W}.
Currently, it is not entirely clear whether the orbit of Leo~I is elliptic or hyperbolic~\citep{2017ARep...61..727B}.
Nevertheless, proper motion measurements indicate that Leo~I is likely bound to the Milky Way~\citep{2013ApJ...768..140B}, but it is on an extremely elongated orbit with an eccentricity of $0.79^{+0.10}_{-0.09}$~\citep{2022ApJ...940..136P}.
Nevertheless, we prefer to exclude it from the analysis, since it is drastically outliers from the general behavior of the satellites.
Its orbit also obviously does not satisfy the isotropy condition that we use in the following analysis.
Moreover, there is evidence of tidal stripping in this galaxy~\citep{2023ApJ...956L..37P}, indicating its violent prehistory, and as a consequence, it cannot be considered as a simple test particle.

\begin{figure}
\centering
\includegraphics[width=\linewidth, bb=42 189 549 597, clip]{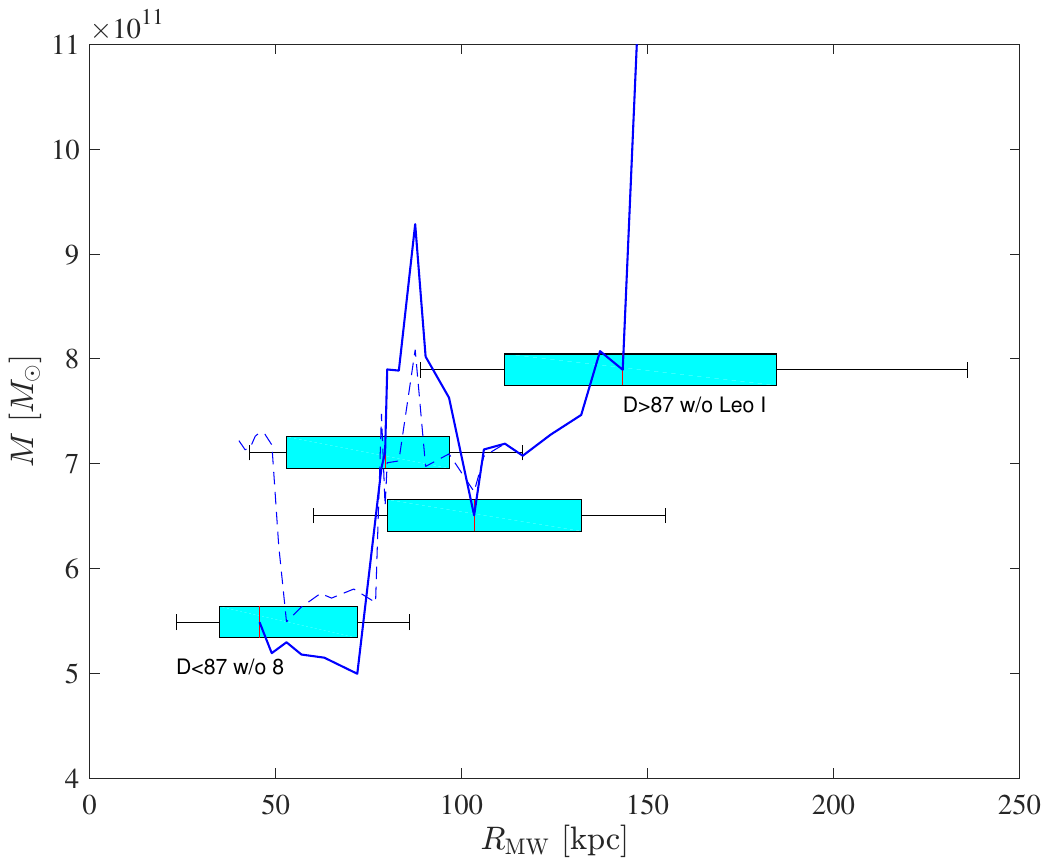}
\caption{
Mass estimates of the Milky Way based on subsamples of 20 satellites. 
Cyan boxplots represent the measured values and illustrate the nonparametric statistics of the distance distribution for the satellites used. 
Specifically, the whiskers indicate the minimum and maximum distances within the sample, the box denotes the first and third quartiles, and the median is shown by a vertical line segment within the box.
The first box corresponds to the `$D<86$~kpc w/o 8' sample, while the fourth represents the `$D<86$~kpc w/o Leo~I' sample (see Table~\ref{tab:MWmass}).
The solid blue line indicates the running mass based on subsamples of 20 satellites, excluding eight galaxies.
The dashed line shows the running mass for the full sample.
}
\label{fig:MWmass}
\end{figure}

\begin{table}
\centering
\caption{
Summary of the MW mass estimates.
}
\label{tab:MWmass}
\begin{tabular}{lr@{--}lcr}
\hline\hline
Sample                  & \multicolumn{2}{c}{range}  & \# & \multicolumn{1}{c}{$M_\mathrm{I}$} \\
                        & \multicolumn{2}{c}{[kpc]}  &    & \multicolumn{1}{c}{[$\times10^{11}$~\Msun{}]} \\
\hline
all                     & 23 & 257 & 49& $8.1\pm1.5$ \\
\enspace w/o Leo~I      & 23 & 236 & 48& $6.9\pm1.3$ \\
\enspace w/o 8          & 23 & 257 & 41& $8.1\pm1.6$ \\
\enspace w/o 8 \& Leo~I & 23 & 236 & 40& $6.7\pm1.4$ \\
\hline
$D<86$~kpc              & 23 & 86  & 28& $6.2\pm1.5$ \\
\enspace w/o 8          & 23 & 86  & 20& $5.5\pm1.6$ \\
\hline
$D>86$~kpc              & 89 & 257  & 21& $10.7\pm3.0$ \\
\enspace w/o Leo~I      & 89 & 236  & 20& $7.9\pm2.3$ \\
%\hline
%$D>100$~kpc     & 102 & 257  & 19& $11.3\pm3.3$ \\
%w/o Leo~I       & 102 & 236  & 18& $8.2\pm2.5$ \\
\hline\hline
\end{tabular}

\end{table}

Based on the results obtained in Section~\ref{sec:LoSMassMilkyWay} and assuming an isotropic distribution of satellite orbits, we estimate the MW mass using different subsamples of satellites.
Figure~\ref{fig:MWmass} presents the behavior of the running line-of-sight mass estimated from 20 satellite sequences for the case of all galaxies (dashed line) and after excluding the eight members of the group with a large collective motion (solid blue line).
As you can see, the exclusion of these eight galaxies does not radically change the mass estimates, but the trend of mass growth with distance appears more clearly.
This behavior is a natural consequence of the density distribution in the dark matter halos surrounding galaxies.
However, because of the small statistics, errors and random fluctuations turn out to be quite large.

The results are summarized in Table~\ref{tab:MWmass}.
It gives the range of galactocentric distances of satellites, the sample size, and the corresponding mass estimate, $M_\mathrm{I}$, assuming isotropic orbits.
The distance of 86~kpc is chosen on the basis that at large distances the motion of the satellites is well randomized and does not show significant collective motions relative to the Galactic center~\citep{2023MNRAS.521.3540M}, and this boundary also divides the sample of satellites into equal parts.
It is evident that Leo~I gives a significant increase in the total mass of the Galaxy, which is especially noticeable in the case of external satellites, where its contribution increases the mass estimate by 35\%.
Based on this analysis, we estimate the mass of the Galaxy within 86~kpc to be in the range [5.5--$6.2] \times 10^{11}$~\Msun{} with uncertainty of $\pm 1.5 \times 10^{11}$~\Msun{}, and the total mass within 240~kpc to be $(7.9 \pm 2.3) \times 10^{11}$~\Msun{}.
%The latter value is in good agreement with numerous estimates by other authors~\citep[see Fig.~\ref{fig:MWmassLiterature} based on recent reviews by][]{2020SCPMA..6309801W,2023ARep...67..812B}.

\begin{figure}
\centering
\includegraphics[width=0.97\linewidth]{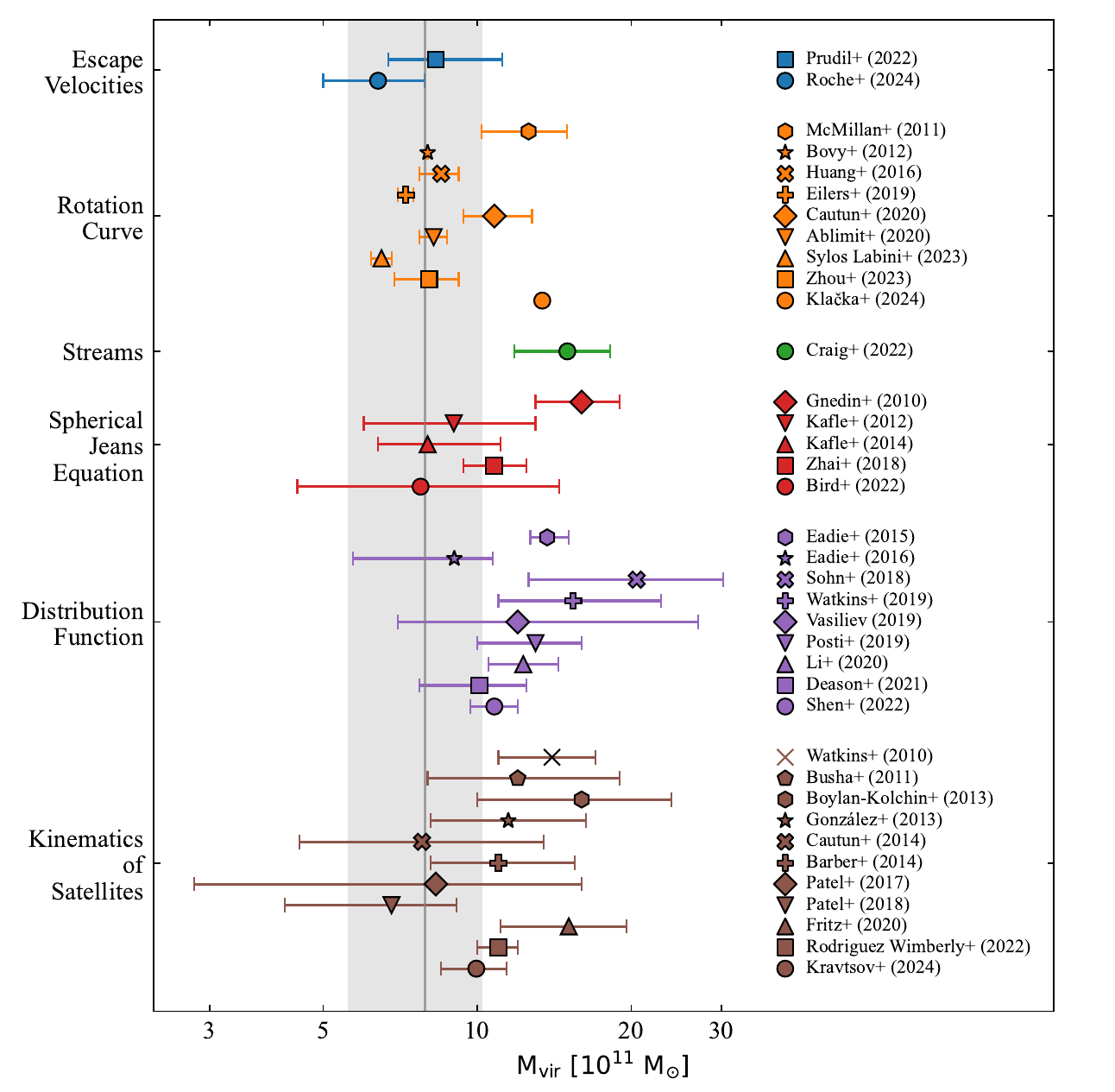}
\caption{
%Comparison of our estimate of the MW mass with recent measurements obtained by different authors.
Comparison of our estimate of the MW mass with recent measurements obtained by different authors:
\citet{2022AandA...664A.148P}, 
\citet{2024ApJ...972...70R}, 
\citet{2011MNRAS.414.2446M},
\citet{2012ApJ...759..131B},
\citet{2016MNRAS.463.2623H},
\citet{2019ApJ...871..120E},
\citet{2020MNRAS.494.4291C},
\citet{2020ApJ...895L..12A},
\citet{2023ApJ...945....3S},
\citet{2023ApJ...946...73Z},
\citet{2024arXiv240712551K},
\citet{2022MNRAS.517.1737C},
\citet{2010ApJ...720L.108G},
\citet{2012ApJ...761...98K},
\citet{2014ApJ...794...59K},
\citet{2018RAA....18..113Z},
\citet{2022MNRAS.516..731B},
\citet{2015ApJ...806...54E},
\citet{2016ApJ...829..108E},
\citet{2018ApJ...862...52S},
\citet{2019ApJ...873..118W},
\citet{2019MNRAS.484.2832V},
\citet{2019AandA...621A..56P},
\citet{2020ApJ...894...10L},
\citet{2021MNRAS.501.5964D},
\citet{2022ApJ...925....1S},
\citet{2010MNRAS.406..264W},
\citet{2011ApJ...743...40B},
\citet{2013ApJ...768..140B},
\citet{2013ApJ...770...96G},
\citet{2014MNRAS.445.2049C},
\citet{2014MNRAS.437..959B},
\citet{2017MNRAS.464.3825P},
\citet{2018ApJ...857...78P},
\citet{2020MNRAS.494.5178F},
\citet{2022MNRAS.513.4968R},
\citet{2024OJAp....7E..50K}.
}
\label{fig:MWmassLiterature}
\end{figure}

The diversity and abundance of new observational data has triggered a surge in estimates of the mass of our Galaxy.
Over the past decade, four reviews on the dynamics and mass of MW have been published.
\citet{2016ARA&A..54..529B}, in their review ``The Galaxy in Context'', obtained an average of individual mass estimates equal to $(11 \pm 3) \times 10^{11}$~\Msun{}.
Based on a compilation of 47 individual mass measurements obtained using seven different methodologies, \citet{2020SCPMA..6309801W} concluded that the virial mass of our Galaxy is still determined with a scatter of a factor of two and is likely to be confined between $5\times10^{11}$ and $20\times10^{11}$~\Msun{}.
In a recent review, \citet{2023ARep...67..812B} used 20 published measurements of the total mass within a radius of 200 kpc. 
They determined a mean value of $(8.8 \pm 0.6) \times 10^{11}$~\Msun{} with a standard deviation of $2.4\times 10^{11}$~\Msun{}.
\citet{2025NewAR.10001721H} claim that the MW mass estimates converge around $10^{12}$~\Msun{} with an uncertainty of about 20-–30\%.

Building upon the compilations by \citet{2020SCPMA..6309801W} and \citet{2023ARep...67..812B}, we supplemented their data with 12 new measurements published after 2020.
Table~\ref{tab:mw_mass_compilation} presents a list of 37 mass estimates for our Galaxy within at least 200~kpc, conducted over the past 15 years.
A comparison of our measurements with data from the literature is shown in Fig.~\ref{fig:MWmassLiterature}.
The measurements are grouped by the classes of methods used.
Our mass estimate is indicated by a vertical line, with its uncertainty represented by the shaded gray region.
It can be seen that our value of $(7.9 \pm 2.3) \times 10^{11}$~\Msun{} within 240~kpc is in good agreement with numerous estimates reported by other authors.
Only measurements based on the phase-space distribution function systematically yield higher masses.
As noted in the review by \citet{2020SCPMA..6309801W}, this may be due to a mismatch between the model and real distribution functions, as well as a violation of the steady-state assumption caused by phase-space correlations and the influence of LMC.
After excluding the phase-space distribution function methods and considering 28 individual measurements, we derive a mean value of $(10.3\pm0.6) \times 10^{11}$~\Msun{} with a standard deviation of $3.0 \times 10^{11}$~\Msun{}, and a median value of $9.5 \times 10^{11}$~\Msun{}, which lies within 1 sigma of our estimate and closely matches the case where the Leo~I is included into consideration.

%---------------------------------------------------
\subsection{Andromeda Galaxy}

\begin{figure}
\centering
\includegraphics[width=\linewidth]{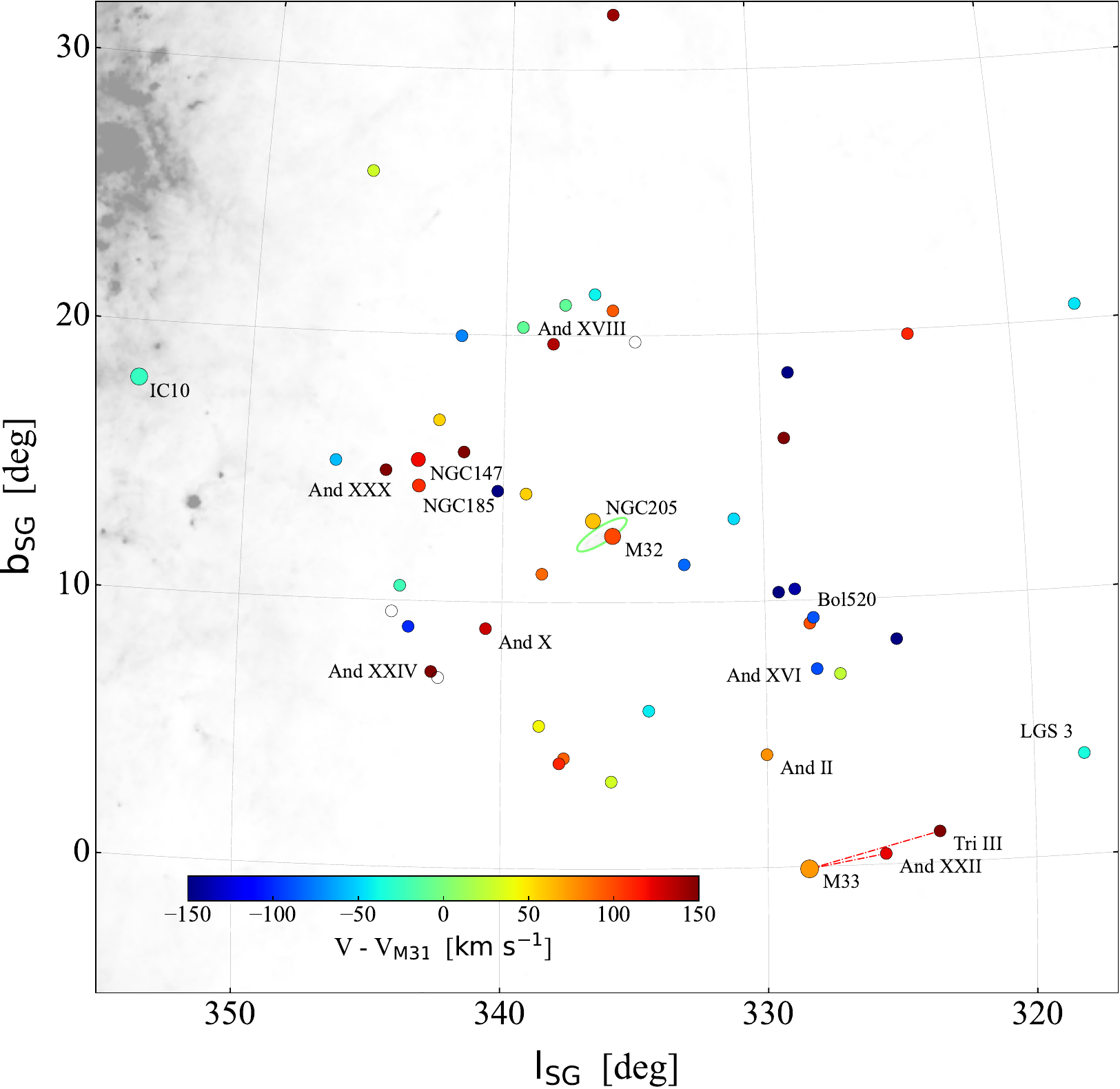}
\caption{
Distribution of the M~31 satellites in the sky in the Aitoff projection.
M~31 is shown by a central ellipse.
The color reflects the velocity difference relative to the central galaxy.
The open dots mark galaxies with unknown line-of-sight velocities.
The IRAS interstellar extinction map is given by gray clouds.
}
\label{fig:M31GroupMap}
\end{figure}

The sky distribution of the 50 known neighbors of M~31 within 500~kpc of it is shown in Fig.~\ref{fig:M31GroupMap}.
%Section~\ref{sec:M31_datatable} compiles recent distance and velocity measurements for them.
Unlike the MW satellites, the proper motions in the Andromeda group are known only for a few galaxies: M~31~\citep{2019ApJ...872...24V, 2021MNRAS.507.2592S}, M~33~\citep{2019ApJ...872...24V, 2023ApJ...948..104P}, NGC~185, NGC~147, IC~10~\citep{2023ApJ...948..104P} and And~III~\citep{2024ApJ...975..138C}.
However, the tangential velocity is determined with large errors and the significance of such measurements rarely exceeds 4 sigma.
Therefore, to estimate the M~31 mass from the satellite motions, we must still rely only on their line-of-sight velocities.

The transverse velocity of the Andromeda Galaxy relative to our Galaxy also remains highly controversial~\citep[see fig.~6 in][]{2021MNRAS.507.2592S}. 
\citet{2021MNRAS.507.2592S} conclude that the approach of these two giant galaxies is radial.
Correction of the observed velocity of M~31 for the solar motion in the Galaxy gives the approach velocity equal to $-109$~\kms{}.
Unfortunately, the fairly compact location of the M~31 satellites in the sky leaves no chance to estimate the tangential velocity of the system using only line-of-sight velocities.
We can only test the velocity component toward M~31.
The average line-of-sight velocity of the M~31 satellite system within 200~kpc is $-105\pm20$~\kms{}, which is almost identical to the velocity of the central galaxy.
Beyond 200~kpc, the average velocity of the satellites diverges from that of M~31, reaching $\Delta V \approx 35$~\kms{}, under the influence of M~33, its satellites And~XXII and Tri~III, and NGC~185.
Therefore, for further analysis, we assume a head-on approach of our Galaxy and the Andromeda Galaxy with $V=-109$~\kms{}.

\begin{table}
\centering
\caption{
Summary of the M~31 mass estimates.
}
\begin{tabular}{lccccc}
\hline\hline
Sample  & \# & $\sigma_{V}$ & 
    \multicolumn{1}{c}{$M_\mathrm{I}$} & 
    \multicolumn{1}{c}{$M^c_\mathrm{I}$} & 
    \multicolumn{1}{c}{$M^p_\mathrm{I}$} \\
\cline{4-6}
        &    
        & \kms{} 
        & \multicolumn{3}{c}{$\times10^{11}$~\Msun{}} \\
\hline
$cz$                        & 47 & 121.1 &              &              & $22.7\pm4.2$ \\ 
$D$ \& $cz$                 & 33 & 114.1 & $18.5\pm3.8$ & $16.9\pm3.5$             & $21.7\pm4.8$ \\
\enspace w/o 2$^\dagger$    & 31 & 113.6 & $18.5\pm3.9$ & $16.9\pm3.6$             & $19.5\pm4.4$ \\
\enspace w/o 3$^\ddagger$   & 30 & 112.7 & $17.0\pm3.7$ & $15.5\pm3.4$ & $17.4\pm4.0$ \\
\hline\hline
\multicolumn{6}{l}{$^\dagger$ excluding M~33 subgroup} \\
\multicolumn{6}{l}{$^\ddagger$ excluding M~33 subgroup and And~XXX} \\
\end{tabular}
\label{tab:M31mass}
\end{table}

We estimate the total mass of the Andromeda Galaxy using the line-of-sight mass estimator assuming isotropy of the orbits for the case of observing the system from the outside (Section~\ref{sec:LoSMassNerabyGrp}).
The results for different satellite subsamples are summarized in Table~\ref{tab:M31mass}.
In addition to the sample size (\#), the velocity dispersion ($\sigma_{V}$), and the line-of-sight mass ($M_\mathrm{I}$), it also provides the line-of-sight mass ($M^c_\mathrm{I}$) corrected for the distance errors (see Section~\ref{sec:ErrorInfluence}), and the mass obtained by the classical projected mass method ($M^p_\mathrm{I}$) proposed by~\citet{1981ApJ...244..805B}.
The description of the subsamples is given below.

As in the case of our Galaxy, the sample of M~31 satellites for mass determination requires some careful consideration.
From the total list of 45 satellites around the Andromeda Galaxy within 300~kpc, we excluded all objects with unknown, questionable, or imprecise distance measurements because they could be a source of large systematic errors.
Among them are 8 objects from the Pan-Andromeda Archaeological Survey (PAndAS) without distance measurements.
We also excluded all objects, whose distance errors are greater than 5\%, namely: 
Tri~III (6.4\%), probably satellite of M~33~\citep{2024MNRAS.528.2614C};
a globular cluster Bol~520 (9.6\%);
IC~10 (5.7\%) and And~XXVII (5.7\%), which reside in the region of high MW extinction.
This sample is designated as `$D$ \& $cz$' in Table~\ref{tab:M31mass}.

The Andromeda group has its own LMC analog.
As shown by \citet{2018MNRAS.480.1883P}, M~33 is located near the apocenter of its first flyby around the Andromeda Galaxy.
A plume of neutral hydrogen stretches along its trajectory.
Therefore, we excluded from consideration M~33 with its suit of And~XXII and Tri~III\footnote{Tri~III has already been excluded in the previous step due to insufficient distance measurement accuracy.} as unsatisfying the requirement of isotropy of their orbits and virialization in the system.
This sample is labeled as `w/o 2' because it excludes 2 galaxies.
For a similar reason, we do not include in consideration And~XVIII at a distance of $1.33_{-0.09}^{+0.06}$~Mpc~\citep{2017MNRAS.464.2281M}, which is located beyond the Andromeda Galaxy at a distance of 579~kpc and has not yet reached its virial zone.
And~XXX has an extremely high line-of-sight velocity $V = 159.6$~\kms{} relative to Andromeda, so in addition to the M~33 subgroup, we also excluded it from the analysis, as it could significantly distort the data.
We call the sample `w/o 3'.

\begin{figure}
\centering
\includegraphics[width=0.8\linewidth]{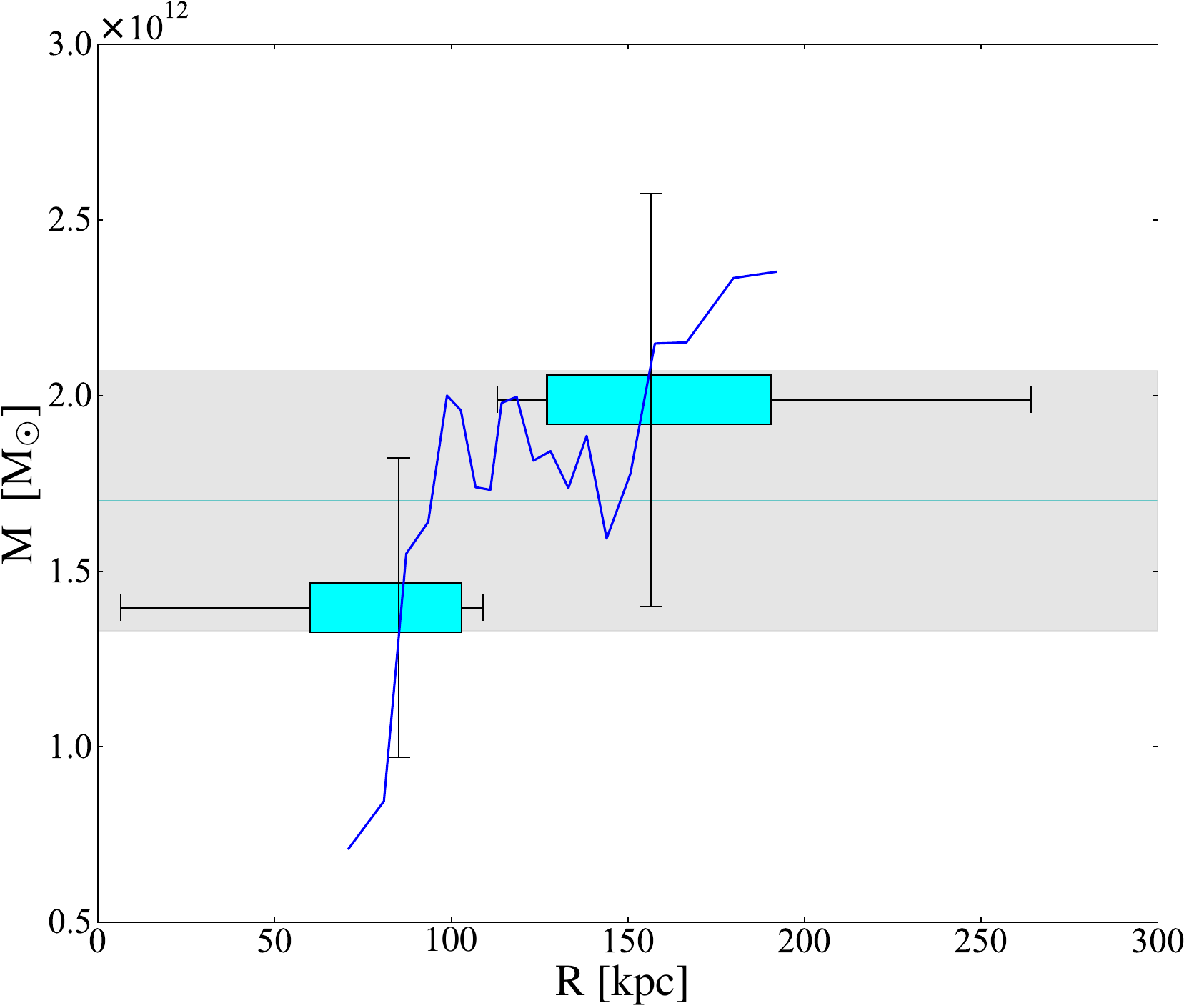}
\caption{
Resulted mass estimate of the Andromeda Galaxy as a function of distance. 
Two boxes correspond with both near and distant subsamples, each containing 15 satellites. 
The meaning of the boxplot is the same as in Fig.~\ref{fig:MWmass}.
The total mass estimate $M = (17.0 \pm 3.7) \times 10^{11}$~\Msun{} (cyan line) with its error (gray filling) is provided. 
The solid blue line shows the running mass for the entire sample.
}
\label{fig:M31massEstimate}    
\end{figure}

We believe that the latter sample provides the most realistic estimate of the mass.
Thus, our final sample contains 30 satellites within 300~kpc of M~31.
Their distribution is characterized by the standard deviation of the line-of-sight velocity $\sigma_{V} = 112.7$~\kms{} and the average distance of $\langle r \rangle = 138$~kpc from the Andromeda Galaxy.
Using the line-of-sight velocity method for nearby groups (Section~\ref{sec:LoSMassNerabyGrp}), we estimate the total mass of M~31 equal to $M_\mathrm{I} = (17.0 \pm 3.7) \times 10^{11}$~\Msun{}.
As in the case of our Galaxy, the running mass and subsamples of nearby and more distant satellites show a trend in mass on the scale from 100 to 200--300~kpc~(Fig.~\ref{fig:M31massEstimate}).

It would be useful to compare this result with the classical approach based on the use of projected distances and line-of-sight velocities only.
Using the same list of satellites, the projected mass method~\citep{1981ApJ...244..805B} gives a very close mass estimate of $M^p_\mathrm{I} = (17.4 \pm 4.0) \times 10^{11}$~\Msun{}, but with worse accuracy.
However, it also allows us to include objects in the analysis without precise distance measurements and improve accuracy through statistics.
Within a \textit{projected} radius of 300~kpc from M~31, there are 47 galaxies with known velocities. 
This sample is labeled `$cz$' in Table~\ref{tab:M31mass}. 
Based on it, we obtain a projected mass of $M^p_\mathrm{I} = (22.7 \pm 4.2) \times 10^{11}$~\Msun{}.
It can be seen that the accuracy is still lower than in our approach.
This indicates the importance of knowing the distribution of satellites in space for a more accurate estimate of the galaxy mass.

%---------------------------------------------------
\subsection{Influence of the distance measurement errors}
\label{sec:ErrorInfluence}

The previous result was obtained under the assumption that we know the galaxy distances absolutely precisely.
Unfortunately, random errors in distance measurements lead to a systematic bias in the measured average separation of the satellites from the central galaxy and to a fictitious stretching of the system along the line of sight.
As a result, we will overestimate the value of $v^2r$ and as a consequence the mass of the galaxy group.
This effect can be especially significant when distance errors become comparable to the characteristic size of the system.

The typical accuracy of distance measurements in the near Universe is 3--4\%.
For the satellites of our Galaxy, this corresponds to a typical distance error of the order of a few kpc, which is negligible compared to the virial radius of the Milky Way.
In the case of the Andromeda Galaxy, the situation is worse.
At the distance of M~31, the typical error is 20--30~kpc, which is on the order of 10\% of its virial radius.

\begin{figure}
\centering
\includegraphics[width=0.90\linewidth]{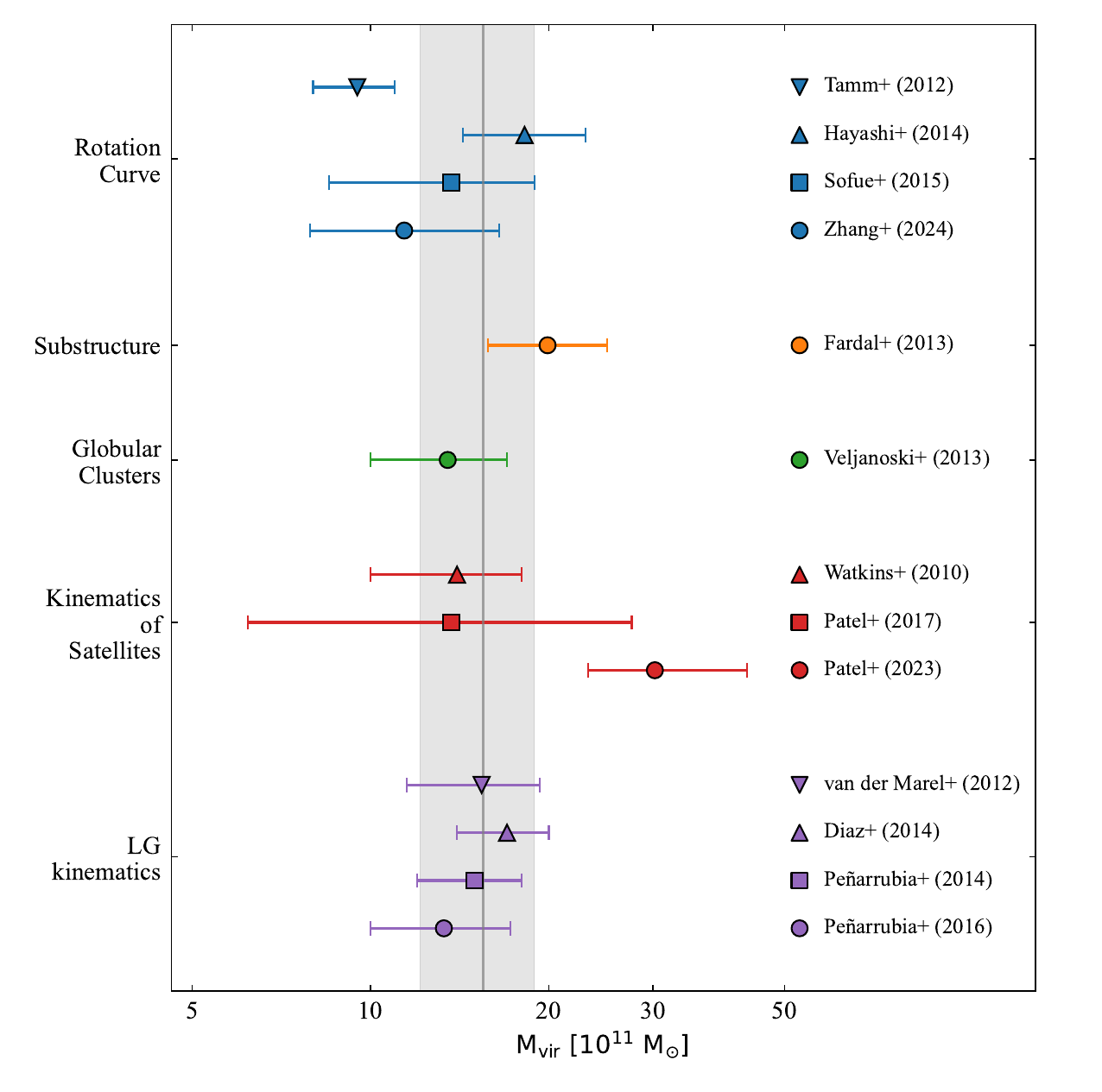}
\caption{
Comparison of our M~31 mass estimate (grey line) with recently published mass measurements on the scale of 200~kpc using different methods.
The compilation contains the following works:
\citet{2012AandA...546A...4T},
\citet{2014ApJ...789...62H},
\citet{2015PASJ...67...75S},
\citet{2024MNRAS.528.2653Z},
\citet{2013MNRAS.434.2779F},
\citet{2013ApJ...768L..33V},
\citet{2010MNRAS.406..264W},
\citet{2017MNRAS.468.3428P},
\citet{2023ApJ...948..104P},
\citet{2012ApJ...753....8V},
\citet{2014MNRAS.443.1688D},
\citet{2014MNRAS.443.2204P},
\citet{2016MNRAS.456L..54P}.
}
\label{fig:M31massLiterature}    
\end{figure}

To estimate the contribution of this systematic uncertainty to the mass of M~31, we apply the Monte Carlo method, which allows us to simulate distance measurements and the propagation of errors in a controlled manner.
Assuming a Gaussian distribution of distance modulus errors, we randomly vary the measured distance modulus according to its measurement accuracy for all galaxies in the group including M~31 itself.
The resulting modulii are then converted into linear distances in the standard way.
All this leads to an asymmetric shift of galaxies along the line of sight relative to their original position.
The mass estimate is made using exactly the same procedure as for the real galaxy group.
Based on 100,000 random generations, we find that the observed distance errors lead to an effective overestimation of the mass by 9.7\% with a standard deviation of 3\%.
As a result, for our final sample of 30 satellites, we obtain a corrected M~31 mass of $M^c_\mathrm{I} = (15.5 \pm 3.4) \times 10^{11}$~\Msun{}, where the error takes into account the small spread introduced by distance measurement errors.

On the one hand, the Andromeda Galaxy is easier to observe than the Milky Way because we can see it from the outside. 
However, its significantly greater distance imposes certain observational limitations.
Among other factors, this is reflected in the smaller number of studies that focus on the M~31 mass estimate.
Over the same period since 2010, we have compiled only 13 measurements of the total mass within 200~kpc, which is three times less than the number of estimates available for the Milky Way.
We added only three new measurements to the review by \citet{2023arXiv230503293B}.
The list is presented in Table~\ref{tab:m31_mass_compilation}.
As illustrated in Fig.~\ref{fig:M31massLiterature}, our value $M^c_\mathrm{I} = (15.5 \pm 3.4) \times 10^{11}$~\Msun{}, indicated by a vertical line with a shaded region representing its uncertainty, is in excellent agreement with 13 recent measurements of the M~31 mass within 200~kpc. 
The mean of the data from the literature is $(15.8 \pm 1.4) \times 10^{11}$~\Msun{} with a standard deviation of $5.1 \times 10^{11}$~\Msun{}, and a median of $14.0 \times 10^{11}$~\Msun{}.

%---------------------------------------------------
\section{Discussion and conclusion}
\label{sec:Conclusion}

We have developed a method for the mass estimation of groups with a dominant central galaxy, based on the $v^2r$ estimator proposed by~\citet{1981ApJ...244..805B}, for the case of a known three-dimensional distribution of satellites in the system.
This situation is realized in the nearby Universe, where high-precision distances for a large number of galaxies are available.
Since the only unaccounted uncertainty in this case is related to the projection of the three-dimensional velocity onto the line of sight, we call this approach the line-of-sight mass estimator.
As with the original projected-mass methodology, this approach requires an assumption about the nature of the satellite orbits in the system, which is expressed in terms of the mean square of the eccentricity $\langle e^2 \rangle$.
We considered two cases: observations from the central galaxy (the case of the Milky Way and its satellites) and observations from outside the system.
Despite the significant difference in the observation conditions, which is reflected in equations~(\ref{eq:MlosMilkyWay}) and (\ref{eq:MlosNearbyGroup}), in the case of isotropic orbits, $\langle e^2 \rangle = 1/2$, the correction factor for both cases turned out to be the same and equal to 4 (equations~(\ref{eq:MisoMilkyWay}) and (\ref{eq:MisoNearbyGroup})).
This coefficient differs from the naively expected correction of 3 for the case of a random projection of the velocity vector.

Comparison with numerical cosmological simulations confirmed the applicability of the method and showed good agreement with the true mass of the systems under the assumption of the isotropic distribution of satellite orbits.
In addition, this methodology allows us to trace the distribution of the mass of the system with distance.

Unfortunately, measurement errors in distance can lead to significant bias in the mass estimate.
Since the distance errors increase with distance, the applicability of the method outside the Local Group is severely limited and requires a more sophisticated Bayesian approach and a joint consideration of the assumed distribution of satellites and their velocities to obtain more reliable determination.

Application of this methodology to the Local Group allowed us to estimate the masses of the Milky Way and the Andromeda Galaxy assuming an isotropic distribution of satellite orbits.

We estimate the total mass of our Galaxy on a scale of 240~kpc of $M_\mathrm{MW} = (7.9 \pm 2.3) \times 10^{11}$~\Msun{}.
This value is in good agreement with recent measurements by other authors.
The mean of 28 individual measurements performed since 2010 is $(10.3 \pm 0.6) \times 10^{11}$~\Msun{} with a standard deviation of $3.0 \times 10^{11}$~\Msun{}, and a median value of $9.5 \times 10^{11}$~\Msun{}.
Similar values have been reported in recent reviews.
\citet{2016ARA&A..54..529B} obtained an average of individual mass estimates equal to $(11 \pm 3) \times 10^{11}$~\Msun{}.
\citet{2020SCPMA..6309801W} concluded that the mass of our Galaxy is known with an uncertainty of a factor of two and is constrained within the range of $5\times10^{11}$ and $20\times10^{11}$~\Msun{}.
\citet{2023ARep...67..812B} derived a mean of $(8.8 \pm 0.6) \times 10^{11}$~\Msun{}, and a standard deviation of $2.4\times 10^{11}$~\Msun{}, using a compilation of 20 measurements from the literature.

A significantly larger mass of our Galaxy of $(10.7\pm3.0) \times 10^{11}$~\Msun{} is obtained when Leo~I is included in the consideration.
Its contribution outweighs all other satellites and increases the total mass by 35\%.
Measurements of its proper motion~\citep{2013ApJ...768..140B, 2022ApJ...940..136P} indicate that Leo~I is in an extremely elongated orbit with an eccentricity of $\sim0.8$.
Taking this effect into account should slightly reduce the contribution of Leo~I compared to the other satellites, but in any case the impact remains significant.
Thus, improving the accuracy of 3D velocity measurements and clarifying the orbital parameters of Leo~I is extremely important for understanding the halo structure of our Galaxy and estimating its total mass.

We estimate the total mass of M~31 of $M = (15.5 \pm 3.4) \times 10^{11}$~\Msun{} within a radius of 300~kpc.
This value is also in excellent agreement with recent measurements.
We supplemented the compilation by \citet{2023arXiv230503293B} with three of the most recent measurements.
The average of the Andromeda Galaxy mass derived from publications over the past 15 years is $(15.8 \pm 1.4) \times 10^{11}$~\Msun{} with a scatter of $5.1 \times 10^{11}$~\Msun{}.
%\citet{2023arXiv230503293B} traced the evolution of the Andromeda Galaxy mass measurements by various methods over the last 80 years.
%He concluded that the average of the most accurate and reliable measurements of the last 20 years gives the M~31 mass of $\sim15.6 \times 10^{11}$~\Msun{}.

Thus, measurement of the total mass of the two main galaxies of the Local Group by their satellite kinematics indicates that the M~31 is twice as massive as MW, $M_\mathrm{M31}/M_\mathrm{MW}=2.0\pm0.7$.
The similar ratio of $M_\mathrm{M31}/M_\mathrm{MW}=1.75^{+0.54}_{-0.28}$ is obtained by \citet{2022MNRAS.513.2385C} using a supervised machine learning algorithm on cosmological simulation within the standard $\Lambda$CDM model and applying it to the distance and velocity distribution of the MW and M~31 satellites.
However, it should be noted that the Hubble flow outside the virial radii of MW and M~31 indicates that the Local Group barycenter is located approximately midway between these galaxies.
\citet{2009MNRAS.393.1265K} estimated the MW to M~31 mass ratio of $M_\mathrm{M31}/M_\mathrm{MW}\simeq4/5$.
Unfortunately, the errors in determining the masses remain large enough to speak of a statistically significant difference between these values.

%%%%%%%%%%%%%%%%%%%%%%%%%%%%%%%%%%%%%%%%%%
\vspace{6pt}

\authorcontributions{
Conceptualization, Dmitry Makarov; 
Methodology, Danila Makarov and Dmitry Makarov; 
Validation, Noam Libeskind; 
Formal analysis, Danila Makarov, Kirill Kozyrev and Noam Libeskind; 
Investigation, Danila Makarov, Dmitry Makarov and Kirill Kozyrev; 
Resources, Noam Libeskind; 
Writing –-- original draft, Danila Makarov, Dmitry Makarov, Kirill Kozyrev and Noam Libeskind; 
Supervision, Dmitry Makarov.
}

\funding{This work was supported by the Russian Science Foundation grant \textnumero~24--12--00277}

\dataavailability{
The original contributions presented in this study are included in the article material. 
Further inquiries can be directed to the corresponding author.
}

\acknowledgments{
The authors thank Lidia Makarova for her valuable comments.
We acknowledge the usage of the HyperLeda database\footnote{\url{http://leda.univ-lyon1.fr}}~\citep{2014A&A...570A..13M}.
This research has made use of the NASA/IPAC Extragalactic Database (NED), which is funded by the National Aeronautics and Space Administration and operated by the California Institute of Technology.
This research has made use of the Astrophysics Data System, funded by NASA under Cooperative Agreement 80NSSC21M00561.
}

\conflictsofinterest{The authors declare no conflict of interest.}

%\begin{adjustwidth}{-\extralength}{0cm}
\reftitle{References}
\bibliography{ref}

\begin{thebibliography}{999}

\bibitem[{Bahcall} and {Tremaine}(1981)]{1981ApJ...244..805B}
{Bahcall}, J.N.; {Tremaine}, S.
\newblock {Methods for determining the masses of spherical systems. I. Test
  particles around a point mass.}
\newblock {\em \apj} {\bf 1981}, {\em 244},~805--819.
\newblock {\url{https://doi.org/10.1086/158756}}.

\bibitem[{Heisler} et~al.(1985){Heisler}, {Tremaine}, and
  {Bahcall}]{1985ApJ...298....8H}
{Heisler}, J.; {Tremaine}, S.; {Bahcall}, J.N.
\newblock {Estimating the masses of galaxy groups: alternatives to the virial
  theorem.}
\newblock {\em \apj} {\bf 1985}, {\em 298},~8--17.
\newblock {\url{https://doi.org/10.1086/163584}}.

\bibitem[{Watkins} et~al.(2010){Watkins}, {Evans}, and
  {An}]{2010MNRAS.406..264W}
{Watkins}, L.L.; {Evans}, N.W.; {An}, J.H.
\newblock {The masses of the Milky Way and Andromeda galaxies}.
\newblock {\em \mnras} {\bf 2010}, {\em 406},~264--278,
  \href{http://arxiv.org/abs/1002.4565}{{\normalfont
  [arXiv:astro-ph.GA/1002.4565]}}.
\newblock {\url{https://doi.org/10.1111/j.1365-2966.2010.16708.x}}.

\bibitem[{Anand} et~al.(2021){Anand}, {Rizzi}, {Tully}, {Shaya},
  {Karachentsev}, {Makarov}, {Makarova}, {Wu}, {Dolphin}, and
  {Kourkchi}]{2021AJ....162...80A}
{Anand}, G.S.; {Rizzi}, L.; {Tully}, R.B.; {Shaya}, E.J.; {Karachentsev}, I.D.;
  {Makarov}, D.I.; {Makarova}, L.; {Wu}, P.F.; {Dolphin}, A.E.; {Kourkchi}, E.
\newblock {The Extragalactic Distance Database: The Color-Magnitude
  Diagrams/Tip of the Red Giant Branch Distance Catalog}.
\newblock {\em \aj} {\bf 2021}, {\em 162},~80,
  \href{http://arxiv.org/abs/2104.02649}{{\normalfont
  [arXiv:astro-ph.GA/2104.02649]}}.
\newblock {\url{https://doi.org/10.3847/1538-3881/ac0440}}.

\bibitem[{Tully} et~al.(2009){Tully}, {Rizzi}, {Shaya}, {Courtois}, {Makarov},
  and {Jacobs}]{2009AJ....138..323T}
{Tully}, R.B.; {Rizzi}, L.; {Shaya}, E.J.; {Courtois}, H.M.; {Makarov}, D.I.;
  {Jacobs}, B.A.
\newblock {The Extragalactic Distance Database}.
\newblock {\em \aj} {\bf 2009}, {\em 138},~323--331.
\newblock {\url{https://doi.org/10.1088/0004-6256/138/2/323}}.

\bibitem[{Zhou} et~al.(2023){Zhou}, {Li}, {Huang}, and
  {Zhang}]{2023ApJ...946...73Z}
{Zhou}, Y.; {Li}, X.; {Huang}, Y.; {Zhang}, H.
\newblock {The Circular Velocity Curve of the Milky Way from 5-25 kpc Using
  Luminous Red Giant Branch Stars}.
\newblock {\em \apj} {\bf 2023}, {\em 946},~73,
  \href{http://arxiv.org/abs/2212.10393}{{\normalfont
  [arXiv:astro-ph.GA/2212.10393]}}.
\newblock {\url{https://doi.org/10.3847/1538-4357/acadd9}}.

\bibitem[{Navarro} et~al.(1995){Navarro}, {Frenk}, and
  {White}]{1995MNRAS.275...56N}
{Navarro}, J.F.; {Frenk}, C.S.; {White}, S.D.M.
\newblock {The assembly of galaxies in a hierarchically clustering universe}.
\newblock {\em \mnras} {\bf 1995}, {\em 275},~56--66,
  \href{http://arxiv.org/abs/astro-ph/9408067}{{\normalfont
  [arXiv:astro-ph/astro-ph/9408067]}}.
\newblock {\url{https://doi.org/10.1093/mnras/275.1.56}}.

\bibitem[{Vasiliev}(2019)]{2019MNRAS.484.2832V}
{Vasiliev}, E.
\newblock {Proper motions and dynamics of the Milky Way globular cluster system
  from Gaia DR2}.
\newblock {\em \mnras} {\bf 2019}, {\em 484},~2832--2850,
  \href{http://arxiv.org/abs/1807.09775}{{\normalfont
  [arXiv:astro-ph.GA/1807.09775]}}.
\newblock {\url{https://doi.org/10.1093/mnras/stz171}}.

\bibitem[{Kravtsov} and {Winney}(2024)]{2024OJAp....7E..50K}
{Kravtsov}, A.; {Winney}, S.
\newblock {Effect of the Large Magellanic Cloud on the kinematics of Milky Way
  satellites and virial mass estimate}.
\newblock {\em The Open Journal of Astrophysics} {\bf 2024}, {\em 7},~50,
  \href{http://arxiv.org/abs/2405.06017}{{\normalfont
  [arXiv:astro-ph.GA/2405.06017]}}.
\newblock {\url{https://doi.org/10.33232/001c.120316}}.

\bibitem[{Karachentsev} et~al.(2009){Karachentsev}, {Kashibadze}, {Makarov},
  and {Tully}]{2009MNRAS.393.1265K}
{Karachentsev}, I.D.; {Kashibadze}, O.G.; {Makarov}, D.I.; {Tully}, R.B.
\newblock {The Hubble flow around the Local Group}.
\newblock {\em \mnras} {\bf 2009}, {\em 393},~1265--1274,
  \href{http://arxiv.org/abs/0811.4610}{{\normalfont
  [arXiv:astro-ph/0811.4610]}}.
\newblock {\url{https://doi.org/10.1111/j.1365-2966.2008.14300.x}}.

\bibitem[{Libeskind} et~al.(2020){Libeskind}, {Carlesi}, {Grand}, {Khalatyan},
  {Knebe}, {Pakmor}, {Pilipenko}, {Pawlowski}, {Sparre}, {Tempel}, {Wang},
  {Courtois}, {Gottl{\"o}ber}, {Hoffman}, {Minchev}, {Pfrommer}, {Sorce},
  {Springel}, {Steinmetz}, {Tully}, {Vogelsberger}, and
  {Yepes}]{2020MNRAS.498.2968L}
{Libeskind}, N.I.; {Carlesi}, E.; {Grand}, R.J.J.; {Khalatyan}, A.; {Knebe},
  A.; {Pakmor}, R.; {Pilipenko}, S.; {Pawlowski}, M.S.; {Sparre}, M.; {Tempel},
  E.;  et~al.
\newblock {The HESTIA project: simulations of the Local Group}.
\newblock {\em \mnras} {\bf 2020}, {\em 498},~2968--2983,
  \href{http://arxiv.org/abs/2008.04926}{{\normalfont
  [arXiv:astro-ph.GA/2008.04926]}}.
\newblock {\url{https://doi.org/10.1093/mnras/staa2541}}.

\bibitem[{Weinberger} et~al.(2020){Weinberger}, {Springel}, and
  {Pakmor}]{2020ApJS..248...32W}
{Weinberger}, R.; {Springel}, V.; {Pakmor}, R.
\newblock {The AREPO Public Code Release}.
\newblock {\em \apjs} {\bf 2020}, {\em 248},~32,
  \href{http://arxiv.org/abs/1909.04667}{{\normalfont
  [arXiv:astro-ph.IM/1909.04667]}}.
\newblock {\url{https://doi.org/10.3847/1538-4365/ab908c}}.

\bibitem[{Tully} et~al.(2013){Tully}, {Courtois}, {Dolphin}, {Fisher},
  {H{\'e}raudeau}, {Jacobs}, {Karachentsev}, {Makarov}, {Makarova},
  {Mitronova}, {Rizzi}, {Shaya}, {Sorce}, and {Wu}]{2013AJ....146...86T}
{Tully}, R.B.; {Courtois}, H.M.; {Dolphin}, A.E.; {Fisher}, J.R.;
  {H{\'e}raudeau}, P.; {Jacobs}, B.A.; {Karachentsev}, I.D.; {Makarov}, D.;
  {Makarova}, L.; {Mitronova}, S.;  et~al.
\newblock {Cosmicflows-2: The Data}.
\newblock {\em \aj} {\bf 2013}, {\em 146},~86,
  \href{http://arxiv.org/abs/1307.7213}{{\normalfont
  [arXiv:astro-ph.CO/1307.7213]}}.
\newblock {\url{https://doi.org/10.1088/0004-6256/146/4/86}}.

\bibitem[{Doumler} et~al.(2013){Doumler}, {Hoffman}, {Courtois}, and
  {Gottl{\"o}ber}]{2013MNRAS.430..888D}
{Doumler}, T.; {Hoffman}, Y.; {Courtois}, H.; {Gottl{\"o}ber}, S.
\newblock {Reconstructing cosmological initial conditions from galaxy peculiar
  velocities - I. Reverse Zeldovich Approximation}.
\newblock {\em \mnras} {\bf 2013}, {\em 430},~888--901,
  \href{http://arxiv.org/abs/1212.2806}{{\normalfont
  [arXiv:astro-ph.CO/1212.2806]}}.
\newblock {\url{https://doi.org/10.1093/mnras/sts613}}.

\bibitem[{Grand} et~al.(2017){Grand}, {G{\'o}mez}, {Marinacci}, {Pakmor},
  {Springel}, {Campbell}, {Frenk}, {Jenkins}, and {White}]{2017MNRAS.467..179G}
{Grand}, R.J.J.; {G{\'o}mez}, F.A.; {Marinacci}, F.; {Pakmor}, R.; {Springel},
  V.; {Campbell}, D.J.R.; {Frenk}, C.S.; {Jenkins}, A.; {White}, S.D.M.
\newblock {The Auriga Project: the properties and formation mechanisms of disc
  galaxies across cosmic time}.
\newblock {\em \mnras} {\bf 2017}, {\em 467},~179--207,
  \href{http://arxiv.org/abs/1610.01159}{{\normalfont
  [arXiv:astro-ph.GA/1610.01159]}}.
\newblock {\url{https://doi.org/10.1093/mnras/stx071}}.

\bibitem[{McCall}(2014)]{2014MNRAS.440..405M}
{McCall}, M.L.
\newblock {A Council of Giants}.
\newblock {\em \mnras} {\bf 2014}, {\em 440},~405--426,
  \href{http://arxiv.org/abs/1403.3667}{{\normalfont
  [arXiv:astro-ph.GA/1403.3667]}}.
\newblock {\url{https://doi.org/10.1093/mnras/stu199}}.

\bibitem[{Smith} et~al.(2024){Smith}, {Cerny}, {Hayes}, {Sestito}, {Jensen},
  {McConnachie}, {Geha}, {Navarro}, {Li}, {Cuillandre}, {Errani}, {Chambers},
  {Gwyn}, {Hammer}, {Hudson}, {Magnier}, and {Martin}]{2024ApJ...961...92S}
{Smith}, S.E.T.; {Cerny}, W.; {Hayes}, C.R.; {Sestito}, F.; {Jensen}, J.;
  {McConnachie}, A.W.; {Geha}, M.; {Navarro}, J.F.; {Li}, T.S.; {Cuillandre},
  J.C.;  et~al.
\newblock {The Discovery of the Faintest Known Milky Way Satellite Using
  UNIONS}.
\newblock {\em \apj} {\bf 2024}, {\em 961},~92,
  \href{http://arxiv.org/abs/2311.10147}{{\normalfont
  [arXiv:astro-ph.GA/2311.10147]}}.
\newblock {\url{https://doi.org/10.3847/1538-4357/ad0d9f}}.

\bibitem[{Wang} et~al.(2020){Wang}, {Han}, {Cautun}, {Li}, and
  {Ishigaki}]{2020SCPMA..6309801W}
{Wang}, W.; {Han}, J.; {Cautun}, M.; {Li}, Z.; {Ishigaki}, M.N.
\newblock {The mass of our Milky Way}.
\newblock {\em Science China Physics, Mechanics, and Astronomy} {\bf 2020},
  {\em 63},~109801,  \href{http://arxiv.org/abs/1912.02599}{{\normalfont
  [arXiv:astro-ph.GA/1912.02599]}}.
\newblock {\url{https://doi.org/10.1007/s11433-019-1541-6}}.

\bibitem[{Bobylev} and {Baykova}(2023)]{2023ARep...67..812B}
{Bobylev}, V.V.; {Baykova}, A.T.
\newblock {Modern Estimates of the Mass of the Milky Way}.
\newblock {\em Astronomy Reports} {\bf 2023}, {\em 67},~812--823.
\newblock {\url{https://doi.org/10.1134/S1063772923080024}}.

\bibitem[{Bhattacharya}(2023)]{2023arXiv230503293B}
{Bhattacharya}, S.
\newblock {Weighing Andromeda: Mass estimates of the
  M\raisebox{-0.5ex}\textasciitilde31 galaxy}.
\newblock {\em arXiv e-prints} {\bf 2023}, p. arXiv:2305.03293,
  \href{http://arxiv.org/abs/2305.03293}{{\normalfont
  [arXiv:astro-ph.GA/2305.03293]}}.
\newblock {\url{https://doi.org/10.48550/arXiv.2305.03293}}.

\bibitem[{Carlesi} et~al.(2022){Carlesi}, {Hoffman}, and
  {Libeskind}]{2022MNRAS.513.2385C}
{Carlesi}, E.; {Hoffman}, Y.; {Libeskind}, N.I.
\newblock {Estimation of the masses in the local group by gradient boosted
  decision trees}.
\newblock {\em \mnras} {\bf 2022}, {\em 513},~2385--2393,
  \href{http://arxiv.org/abs/2204.03334}{{\normalfont
  [arXiv:astro-ph.CO/2204.03334]}}.
\newblock {\url{https://doi.org/10.1093/mnras/stac897}}.

\bibitem[{Pe{\~n}arrubia} et~al.(2014){Pe{\~n}arrubia}, {Ma}, {Walker}, and
  {McConnachie}]{2014MNRAS.443.2204P}
{Pe{\~n}arrubia}, J.; {Ma}, Y.Z.; {Walker}, M.G.; {McConnachie}, A.
\newblock {A dynamical model of the local cosmic expansion}.
\newblock {\em \mnras} {\bf 2014}, {\em 443},~2204--2222,
  \href{http://arxiv.org/abs/1405.0306}{{\normalfont
  [arXiv:astro-ph.GA/1405.0306]}}.
\newblock {\url{https://doi.org/10.1093/mnras/stu879}}.

\bibitem[{Kaisina} et~al.(2012){Kaisina}, {Makarov}, {Karachentsev}, and
  {Kaisin}]{2012AstBu..67..115K}
{Kaisina}, E.I.; {Makarov}, D.I.; {Karachentsev}, I.D.; {Kaisin}, S.S.
\newblock {Observational database for studies of nearby universe}.
\newblock {\em Astrophysical Bulletin} {\bf 2012}, {\em 67},~115--122.
\newblock {\url{https://doi.org/10.1134/S1990341312010105}}.

\bibitem[{Kashibadze} and {Karachentsev}(2018)]{2018A&A...609A..11K}
{Kashibadze}, O.G.; {Karachentsev}, I.D.
\newblock {Cosmic flow around local massive galaxies}.
\newblock {\em \aap} {\bf 2018}, {\em 609},~A11,
  \href{http://arxiv.org/abs/1709.09420}{{\normalfont
  [arXiv:astro-ph.CO/1709.09420]}}.
\newblock {\url{https://doi.org/10.1051/0004-6361/201731645}}.

\bibitem[{McConnachie} and {Venn}(2020)]{2020AJ....160..124M}
{McConnachie}, A.W.; {Venn}, K.A.
\newblock {Revised and New Proper Motions for Confirmed and Candidate Milky Way
  Dwarf Galaxies}.
\newblock {\em \aj} {\bf 2020}, {\em 160},~124,
  \href{http://arxiv.org/abs/2007.05011}{{\normalfont
  [arXiv:astro-ph.GA/2007.05011]}}.
\newblock {\url{https://doi.org/10.3847/1538-3881/aba4ab}}.

\bibitem[{Pace} et~al.(2022){Pace}, {Erkal}, and {Li}]{2022ApJ...940..136P}
{Pace}, A.B.; {Erkal}, D.; {Li}, T.S.
\newblock {Proper Motions, Orbits, and Tidal Influences of Milky Way Dwarf
  Spheroidal Galaxies}.
\newblock {\em \apj} {\bf 2022}, {\em 940},~136,
  \href{http://arxiv.org/abs/2205.05699}{{\normalfont
  [arXiv:astro-ph.GA/2205.05699]}}.
\newblock {\url{https://doi.org/10.3847/1538-4357/ac997b}}.

\bibitem[{Savino} et~al.(2022){Savino}, {Weisz}, {Skillman}, {Dolphin},
  {Kallivayalil}, {Wetzel}, {Anderson}, {Besla}, {Boylan-Kolchin}, {Bullock},
  {Cole}, {Collins}, {Cooper}, {Deason}, {Dotter}, {Fardal}, {Ferguson},
  {Fritz}, {Geha}, {Gilbert}, {Guhathakurta}, {Ibata}, {Irwin}, {Jeon},
  {Kirby}, {Lewis}, {Mackey}, {Majewski}, {Martin}, {McConnachie}, {Patel},
  {Rich}, {Simon}, {Sohn}, {Tollerud}, and {van der
  Marel}]{2022ApJ...938..101S}
{Savino}, A.; {Weisz}, D.R.; {Skillman}, E.D.; {Dolphin}, A.; {Kallivayalil},
  N.; {Wetzel}, A.; {Anderson}, J.; {Besla}, G.; {Boylan-Kolchin}, M.;
  {Bullock}, J.S.;  et~al.
\newblock {The Hubble Space Telescope Survey of M31 Satellite Galaxies. I. RR
  Lyrae-based Distances and Refined 3D Geometric Structure}.
\newblock {\em \apj} {\bf 2022}, {\em 938},~101,
  \href{http://arxiv.org/abs/2206.02801}{{\normalfont
  [arXiv:astro-ph.GA/2206.02801]}}.
\newblock {\url{https://doi.org/10.3847/1538-4357/ac91cb}}.

\bibitem[{Drlica-Wagner} et~al.(2020){Drlica-Wagner}, {Bechtol}, {Mau},
  {McNanna}, {Nadler}, {Pace}, {Li}, {Pieres}, {Rozo}, {Simon}, {Walker},
  {Wechsler}, {Abbott}, {Allam}, {Annis}, {Bertin}, {Brooks}, {Burke},
  {Rosell}, {Carrasco Kind}, {Carretero}, {Costanzi}, {da Costa}, {De Vicente},
  {Desai}, {Diehl}, {Doel}, {Eifler}, {Everett}, {Flaugher}, {Frieman},
  {Garc{\'\i}a-Bellido}, {Gaztanaga}, {Gruen}, {Gruendl}, {Gschwend},
  {Gutierrez}, {Honscheid}, {James}, {Krause}, {Kuehn}, {Kuropatkin}, {Lahav},
  {Maia}, {Marshall}, {Melchior}, {Menanteau}, {Miquel}, {Palmese}, {Plazas},
  {Sanchez}, {Scarpine}, {Schubnell}, {Serrano}, {Sevilla-Noarbe}, {Smith},
  {Suchyta}, {Tarle}, and {DES Collaboration}]{2020ApJ...893...47D}
{Drlica-Wagner}, A.; {Bechtol}, K.; {Mau}, S.; {McNanna}, M.; {Nadler}, E.O.;
  {Pace}, A.B.; {Li}, T.S.; {Pieres}, A.; {Rozo}, E.; {Simon}, J.D.;  et~al.
\newblock {Milky Way Satellite Census. I. The Observational Selection Function
  for Milky Way Satellites in DES Y3 and Pan-STARRS DR1}.
\newblock {\em \apj} {\bf 2020}, {\em 893},~47,
  \href{http://arxiv.org/abs/1912.03302}{{\normalfont
  [arXiv:astro-ph.GA/1912.03302]}}.
\newblock {\url{https://doi.org/10.3847/1538-4357/ab7eb9}}.

\bibitem[{Makarov} et~al.(2014){Makarov}, {Prugniel}, {Terekhova}, {Courtois},
  and {Vauglin}]{2014A&A...570A..13M}
{Makarov}, D.; {Prugniel}, P.; {Terekhova}, N.; {Courtois}, H.; {Vauglin}, I.
\newblock {HyperLEDA. III. The catalogue of extragalactic distances}.
\newblock {\em \aap} {\bf 2014}, {\em 570},~A13,
  \href{http://arxiv.org/abs/1408.3476}{{\normalfont
  [arXiv:astro-ph.GA/1408.3476]}}.
\newblock {\url{https://doi.org/10.1051/0004-6361/201423496}}.

\bibitem[{Pawlowski}(2021)]{2021Galax...9...66P}
{Pawlowski}, M.S.
\newblock {Phase-Space Correlations among Systems of Satellite Galaxies}.
\newblock {\em Galaxies} {\bf 2021}, {\em 9},~66,
  \href{http://arxiv.org/abs/2109.02654}{{\normalfont
  [arXiv:astro-ph.GA/2109.02654]}}.
\newblock {\url{https://doi.org/10.3390/galaxies9030066}}.

\bibitem[{Conn} et~al.(2013){Conn}, {Lewis}, {Ibata}, {Parker}, {Zucker},
  {McConnachie}, {Martin}, {Valls-Gabaud}, {Tanvir}, {Irwin}, {Ferguson}, and
  {Chapman}]{2013ApJ...766..120C}
{Conn}, A.R.; {Lewis}, G.F.; {Ibata}, R.A.; {Parker}, Q.A.; {Zucker}, D.B.;
  {McConnachie}, A.W.; {Martin}, N.F.; {Valls-Gabaud}, D.; {Tanvir}, N.;
  {Irwin}, M.J.;  et~al.
\newblock {The Three-dimensional Structure of the M31 Satellite System; Strong
  Evidence for an Inhomogeneous Distribution of Satellites}.
\newblock {\em \apj} {\bf 2013}, {\em 766},~120,
  \href{http://arxiv.org/abs/1301.7131}{{\normalfont
  [arXiv:astro-ph.CO/1301.7131]}}.
\newblock {\url{https://doi.org/10.1088/0004-637X/766/2/120}}.

\bibitem[{Gaia Collaboration} et~al.(2018){Gaia Collaboration}, {Brown},
  {Vallenari}, {Prusti}, {de Bruijne}, {Babusiaux}, {Bailer-Jones}, {Biermann},
  {Evans}, {Eyer}, {Jansen}, {Jordi}, {Klioner}, {Lammers}, {Lindegren},
  {Luri}, {Mignard}, {Panem}, {Pourbaix}, {Randich}, {Sartoretti}, {Siddiqui},
  {Soubiran}, {van Leeuwen}, {Walton}, {Arenou}, {Bastian}, {Cropper},
  {Drimmel}, {Katz}, {Lattanzi}, {Bakker}, {Cacciari}, {Casta{\~n}eda},
  {Chaoul}, {Cheek}, {De Angeli}, {Fabricius}, {Guerra}, {Holl}, {Masana},
  {Messineo}, {Mowlavi}, {Nienartowicz}, {Panuzzo}, {Portell}, {Riello},
  {Seabroke}, {Tanga}, {Th{\'e}venin}, {Gracia-Abril}, {Comoretto},
  {Garcia-Reinaldos}, {Teyssier}, {Altmann}, {Andrae}, {Audard},
  {Bellas-Velidis}, {Benson}, {Berthier}, {Blomme}, {Burgess}, {Busso},
  {Carry}, {Cellino}, {Clementini}, {Clotet}, {Creevey}, {Davidson}, {De
  Ridder}, {Delchambre}, {Dell'Oro}, {Ducourant},
  {Fern{\'a}ndez-Hern{\'a}ndez}, {Fouesneau}, {Fr{\'e}mat}, {Galluccio},
  {Garc{\'\i}a-Torres}, {Gonz{\'a}lez-N{\'u}{\~n}ez}, {Gonz{\'a}lez-Vidal},
  {Gosset}, {Guy}, {Halbwachs}, {Hambly}, {Harrison}, {Hern{\'a}ndez},
  {Hestroffer}, {Hodgkin}, {Hutton}, {Jasniewicz}, {Jean-Antoine-Piccolo},
  {Jordan}, {Korn}, {Krone-Martins}, {Lanzafame}, {Lebzelter}, {L{\"o}ffler},
  {Manteiga}, {Marrese}, {Mart{\'\i}n-Fleitas}, {Moitinho}, {Mora}, {Muinonen},
  {Osinde}, {Pancino}, {Pauwels}, {Petit}, {Recio-Blanco}, {Richards},
  {Rimoldini}, {Robin}, {Sarro}, {Siopis}, {Smith}, {Sozzetti}, {S{\"u}veges},
  {Torra}, {van Reeven}, {Abbas}, {Abreu Aramburu}, {Accart}, {Aerts},
  {Altavilla}, {{\'A}lvarez}, {Alvarez}, {Alves}, {Anderson}, {Andrei},
  {Anglada Varela}, {Antiche}, {Antoja}, {Arcay}, {Astraatmadja}, {Bach},
  {Baker}, {Balaguer-N{\'u}{\~n}ez}, {Balm}, {Barache}, {Barata}, {Barbato},
  {Barblan}, {Barklem}, {Barrado}, {Barros}, {Barstow}, {Bartholom{\'e}
  Mu{\~n}oz}, {Bassilana}, {Becciani}, {Bellazzini}, {Berihuete}, {Bertone},
  {Bianchi}, {Bienaym{\'e}}, {Blanco-Cuaresma}, {Boch}, {Boeche}, {Bombrun},
  {Borrachero}, {Bossini}, {Bouquillon}, {Bourda}, {Bragaglia}, {Bramante},
  {Breddels}, {Bressan}, {Brouillet}, {Br{\"u}semeister}, {Brugaletta},
  {Bucciarelli}, {Burlacu}, {Busonero}, {Butkevich}, {Buzzi}, {Caffau},
  {Cancelliere}, {Cannizzaro}, {Cantat-Gaudin}, {Carballo}, {Carlucci},
  {Carrasco}, {Casamiquela}, {Castellani}, {Castro-Ginard}, {Charlot},
  {Chemin}, {Chiavassa}, {Cocozza}, {Costigan}, {Cowell}, {Crifo}, {Crosta},
  {Crowley}, {Cuypers}, {Dafonte}, {Damerdji}, {Dapergolas}, {David}, {David},
  {de Laverny}, {De Luise}, {De March}, {de Martino}, {de Souza}, {de Torres},
  {Debosscher}, {del Pozo}, {Delbo}, {Delgado}, {Delgado}, {Di Matteo},
  {Diakite}, {Diener}, {Distefano}, {Dolding}, {Drazinos}, {Dur{\'a}n},
  {Edvardsson}, {Enke}, {Eriksson}, {Esquej}, {Eynard Bontemps}, {Fabre},
  {Fabrizio}, {Faigler}, {Falc{\~a}o}, {Farr{\`a}s Casas}, {Federici},
  {Fedorets}, {Fernique}, {Figueras}, {Filippi}, {Findeisen}, {Fonti},
  {Fraile}, {Fraser}, {Fr{\'e}zouls}, {Gai}, {Galleti}, {Garabato},
  {Garc{\'\i}a-Sedano}, {Garofalo}, {Garralda}, {Gavel}, {Gavras}, {Gerssen},
  {Geyer}, {Giacobbe}, {Gilmore}, {Girona}, {Giuffrida}, {Glass}, {Gomes},
  {Granvik}, {Gueguen}, {Guerrier}, {Guiraud}, {Guti{\'e}rrez-S{\'a}nchez},
  {Haigron}, {Hatzidimitriou}, {Hauser}, {Haywood}, {Heiter}, {Helmi}, {Heu},
  {Hilger}, {Hobbs}, {Hofmann}, {Holland}, {Huckle}, {Hypki}, {Icardi},
  {Jan{\ss}en}, {Jevardat de Fombelle}, {Jonker}, {Juh{\'a}sz}, {Julbe},
  {Karampelas}, {Kewley}, {Klar}, {Kochoska}, {Kohley}, {Kolenberg},
  {Kontizas}, {Kontizas}, {Koposov}, {Kordopatis}, {Kostrzewa-Rutkowska},
  {Koubsky}, {Lambert}, {Lanza}, {Lasne}, {Lavigne}, {Le Fustec}, {Le
  Poncin-Lafitte}, {Lebreton}, {Leccia}, {Leclerc}, {Lecoeur-Taibi},
  {Lenhardt}, {Leroux}, {Liao}, {Licata}, {Lindstr{\o}m}, {Lister}, {Livanou},
  {Lobel}, {L{\'o}pez}, {Managau}, {Mann}, {Mantelet}, {Marchal}, {Marchant},
  {Marconi}, {Marinoni}, {Marschalk{\'o}}, {Marshall}, {Martino}, {Marton},
  {Mary}, {Massari}, {Matijevi{\v{c}}}, {Mazeh}, {McMillan}, {Messina},
  {Michalik}, {Millar}, {Molina}, {Molinaro}, {Moln{\'a}r}, {Montegriffo},
  {Mor}, {Morbidelli}, {Morel}, {Morris}, {Mulone}, {Muraveva}, {Musella},
  {Nelemans}, {Nicastro}, {Noval}, {O'Mullane}, {Ord{\'e}novic},
  {Ord{\'o}{\~n}ez-Blanco}, {Osborne}, {Pagani}, {Pagano}, {Pailler},
  {Palacin}, {Palaversa}, {Panahi}, {Pawlak}, {Piersimoni}, {Pineau}, {Plachy},
  {Plum}, {Poggio}, {Poujoulet}, {Pr{\v{s}}a}, {Pulone}, {Racero}, {Ragaini},
  {Rambaux}, {Ramos-Lerate}, {Regibo}, {Reyl{\'e}}, {Riclet}, {Ripepi}, {Riva},
  {Rivard}, {Rixon}, {Roegiers}, {Roelens}, {Romero-G{\'o}mez}, {Rowell},
  {Royer}, {Ruiz-Dern}, {Sadowski}, {Sagrist{\`a} Sell{\'e}s}, {Sahlmann},
  {Salgado}, {Salguero}, {Sanna}, {Santana-Ros}, {Sarasso}, {Savietto},
  {Schultheis}, {Sciacca}, {Segol}, {Segovia}, {S{\'e}gransan}, {Shih},
  {Siltala}, {Silva}, {Smart}, {Smith}, {Solano}, {Solitro}, {Sordo}, {Soria
  Nieto}, {Souchay}, {Spagna}, {Spoto}, {Stampa}, {Steele},
  {Steidelm{\"u}ller}, {Stephenson}, {Stoev}, {Suess}, {Surdej}, {Szabados},
  {Szegedi-Elek}, {Tapiador}, {Taris}, {Tauran}, {Taylor}, {Teixeira},
  {Terrett}, {Teyssandier}, {Thuillot}, {Titarenko}, {Torra Clotet}, {Turon},
  {Ulla}, {Utrilla}, {Uzzi}, {Vaillant}, {Valentini}, {Valette}, {van Elteren},
  {Van Hemelryck}, {van Leeuwen}, {Vaschetto}, {Vecchiato}, {Veljanoski},
  {Viala}, {Vicente}, {Vogt}, {von Essen}, {Voss}, {Votruba}, {Voutsinas},
  {Walmsley}, {Weiler}, {Wertz}, {Wevers}, {Wyrzykowski}, {Yoldas},
  {{\v{Z}}erjal}, {Ziaeepour}, {Zorec}, {Zschocke}, {Zucker}, {Zurbach}, and
  {Zwitter}]{2018A&A...616A...1G}
{Gaia Collaboration}.; {Brown}, A.G.A.; {Vallenari}, A.; {Prusti}, T.; {de
  Bruijne}, J.H.J.; {Babusiaux}, C.; {Bailer-Jones}, C.A.L.; {Biermann}, M.;
  {Evans}, D.W.; {Eyer}, L.;  et~al.
\newblock {Gaia Data Release 2. Summary of the contents and survey properties}.
\newblock {\em \aap} {\bf 2018}, {\em 616},~A1,
  \href{http://arxiv.org/abs/1804.09365}{{\normalfont
  [arXiv:astro-ph.GA/1804.09365]}}.
\newblock {\url{https://doi.org/10.1051/0004-6361/201833051}}.

\bibitem[{Makarov} et~al.(2023){Makarov}, {Khoperskov}, {Makarov}, {Makarova},
  {Libeskind}, and {Salomon}]{2023MNRAS.521.3540M}
{Makarov}, D.; {Khoperskov}, S.; {Makarov}, D.; {Makarova}, L.; {Libeskind},
  N.; {Salomon}, J.B.
\newblock {The LMC impact on the kinematics of the Milky Way satellites: clues
  from the running solar apex}.
\newblock {\em \mnras} {\bf 2023}, {\em 521},~3540--3552,
  \href{http://arxiv.org/abs/2303.06175}{{\normalfont
  [arXiv:astro-ph.GA/2303.06175]}}.
\newblock {\url{https://doi.org/10.1093/mnras/stad757}}.

\bibitem[{Akhmetov} et~al.(2024){Akhmetov}, {Bucciarelli}, {Crosta},
  {Lattanzi}, {Spagna}, {Re Fiorentin}, and {Bannikova}]{2024MNRAS.530..710A}
{Akhmetov}, V.S.; {Bucciarelli}, B.; {Crosta}, M.; {Lattanzi}, M.G.; {Spagna},
  A.; {Re Fiorentin}, P.; {Bannikova}, E.Y.
\newblock {A new kinematic model of the Galaxy: analysis of the stellar
  velocity field from Gaia Data Release 3}.
\newblock {\em \mnras} {\bf 2024}, {\em 530},~710--729,
  \href{http://arxiv.org/abs/2307.08527}{{\normalfont
  [arXiv:astro-ph.GA/2307.08527]}}.
\newblock {\url{https://doi.org/10.1093/mnras/stae772}}.

\bibitem[{Reid} and {Brunthaler}(2020)]{2020ApJ...892...39R}
{Reid}, M.J.; {Brunthaler}, A.
\newblock {The Proper Motion of Sagittarius A*. III. The Case for a
  Supermassive Black Hole}.
\newblock {\em \apj} {\bf 2020}, {\em 892},~39,
  \href{http://arxiv.org/abs/2001.04386}{{\normalfont
  [arXiv:astro-ph.GA/2001.04386]}}.
\newblock {\url{https://doi.org/10.3847/1538-4357/ab76cd}}.

\bibitem[{GRAVITY Collaboration} et~al.(2021){GRAVITY Collaboration}, {Abuter},
  {Amorim}, {Baub{\"o}ck}, {Berger}, {Bonnet}, {Brandner}, {Cl{\'e}net},
  {Davies}, {de Zeeuw}, {Dexter}, {Dallilar}, {Drescher}, {Eckart},
  {Eisenhauer}, {F{\"o}rster Schreiber}, {Garcia}, {Gao}, {Gendron}, {Genzel},
  {Gillessen}, {Habibi}, {Haubois}, {Hei{\ss}el}, {Henning}, {Hippler},
  {Horrobin}, {Jim{\'e}nez-Rosales}, {Jochum}, {Jocou}, {Kaufer}, {Kervella},
  {Lacour}, {Lapeyr{\`e}re}, {Le Bouquin}, {L{\'e}na}, {Lutz}, {Nowak}, {Ott},
  {Paumard}, {Perraut}, {Perrin}, {Pfuhl}, {Rabien}, {Rodr{\'\i}guez-Coira},
  {Shangguan}, {Shimizu}, {Scheithauer}, {Stadler}, {Straub}, {Straubmeier},
  {Sturm}, {Tacconi}, {Vincent}, {von Fellenberg}, {Waisberg}, {Widmann},
  {Wieprecht}, {Wiezorrek}, {Woillez}, {Yazici}, {Young}, and
  {Zins}]{2021A&A...647A..59G}
{GRAVITY Collaboration}.; {Abuter}, R.; {Amorim}, A.; {Baub{\"o}ck}, M.;
  {Berger}, J.P.; {Bonnet}, H.; {Brandner}, W.; {Cl{\'e}net}, Y.; {Davies}, R.;
  {de Zeeuw}, P.T.;  et~al.
\newblock {Improved GRAVITY astrometric accuracy from modeling optical
  aberrations}.
\newblock {\em \aap} {\bf 2021}, {\em 647},~A59,
  \href{http://arxiv.org/abs/2101.12098}{{\normalfont
  [arXiv:astro-ph.GA/2101.12098]}}.
\newblock {\url{https://doi.org/10.1051/0004-6361/202040208}}.

\bibitem[{Bajkova} and {Bobylev}(2017)]{2017ARep...61..727B}
{Bajkova}, A.T.; {Bobylev}, V.V.
\newblock {Galactic orbits of selected companions of the Milky Way}.
\newblock {\em Astronomy Reports} {\bf 2017}, {\em 61},~727--738,
  \href{http://arxiv.org/abs/1707.04168}{{\normalfont
  [arXiv:astro-ph.GA/1707.04168]}}.
\newblock {\url{https://doi.org/10.1134/S1063772917080017}}.

\bibitem[{Boylan-Kolchin} et~al.(2013){Boylan-Kolchin}, {Bullock}, {Sohn},
  {Besla}, and {van der Marel}]{2013ApJ...768..140B}
{Boylan-Kolchin}, M.; {Bullock}, J.S.; {Sohn}, S.T.; {Besla}, G.; {van der
  Marel}, R.P.
\newblock {The Space Motion of Leo I: The Mass of the Milky Way's Dark Matter
  Halo}.
\newblock {\em \apj} {\bf 2013}, {\em 768},~140,
  \href{http://arxiv.org/abs/1210.6046}{{\normalfont
  [arXiv:astro-ph.CO/1210.6046]}}.
\newblock {\url{https://doi.org/10.1088/0004-637X/768/2/140}}.

\bibitem[{Pacucci} et~al.(2023){Pacucci}, {Ni}, and
  {Loeb}]{2023ApJ...956L..37P}
{Pacucci}, F.; {Ni}, Y.; {Loeb}, A.
\newblock {Extreme Tidal Stripping May Explain the Overmassive Black Hole in
  Leo I: A Proof of Concept}.
\newblock {\em \apjl} {\bf 2023}, {\em 956},~L37,
  \href{http://arxiv.org/abs/2309.02487}{{\normalfont
  [arXiv:astro-ph.GA/2309.02487]}}.
\newblock {\url{https://doi.org/10.3847/2041-8213/acff5e}}.

\bibitem[{Bland-Hawthorn} and {Gerhard}(2016)]{2016ARA&A..54..529B}
{Bland-Hawthorn}, J.; {Gerhard}, O.
\newblock {The Galaxy in Context: Structural, Kinematic, and Integrated
  Properties}.
\newblock {\em \araa} {\bf 2016}, {\em 54},~529--596,
  \href{http://arxiv.org/abs/1602.07702}{{\normalfont
  [arXiv:astro-ph.GA/1602.07702]}}.
\newblock {\url{https://doi.org/10.1146/annurev-astro-081915-023441}}.

\bibitem[{Hunt} and {Vasiliev}(2025)]{2025NewAR.10001721H}
{Hunt}, J.A.S.; {Vasiliev}, E.
\newblock {Milky Way dynamics in light of Gaia}.
\newblock {\em \nar} {\bf 2025}, {\em 100},~101721,
  \href{http://arxiv.org/abs/2501.04075}{{\normalfont
  [arXiv:astro-ph.GA/2501.04075]}}.
\newblock {\url{https://doi.org/10.1016/j.newar.2024.101721}}.

\bibitem[{van der Marel} et~al.(2019){van der Marel}, {Fardal}, {Sohn},
  {Patel}, {Besla}, {del Pino}, {Sahlmann}, and {Watkins}]{2019ApJ...872...24V}
{van der Marel}, R.P.; {Fardal}, M.A.; {Sohn}, S.T.; {Patel}, E.; {Besla}, G.;
  {del Pino}, A.; {Sahlmann}, J.; {Watkins}, L.L.
\newblock {First Gaia Dynamics of the Andromeda System: DR2 Proper Motions,
  Orbits, and Rotation of M31 and M33}.
\newblock {\em \apj} {\bf 2019}, {\em 872},~24,
  \href{http://arxiv.org/abs/1805.04079}{{\normalfont
  [arXiv:astro-ph.GA/1805.04079]}}.
\newblock {\url{https://doi.org/10.3847/1538-4357/ab001b}}.

\bibitem[{Salomon} et~al.(2021){Salomon}, {Ibata}, {Reyl{\'e}}, {Famaey},
  {Libeskind}, {McConnachie}, and {Hoffman}]{2021MNRAS.507.2592S}
{Salomon}, J.B.; {Ibata}, R.; {Reyl{\'e}}, C.; {Famaey}, B.; {Libeskind}, N.I.;
  {McConnachie}, A.W.; {Hoffman}, Y.
\newblock {The proper motion of Andromeda from Gaia EDR3: confirming a nearly
  radial orbit}.
\newblock {\em \mnras} {\bf 2021}, {\em 507},~2592--2601,
  \href{http://arxiv.org/abs/2012.09204}{{\normalfont
  [arXiv:astro-ph.GA/2012.09204]}}.
\newblock {\url{https://doi.org/10.1093/mnras/stab2253}}.

\bibitem[{Patel} and {Mandel}(2023)]{2023ApJ...948..104P}
{Patel}, E.; {Mandel}, K.S.
\newblock {Evidence for a Massive Andromeda Galaxy Using Satellite Galaxy
  Proper Motions}.
\newblock {\em \apj} {\bf 2023}, {\em 948},~104,
  \href{http://arxiv.org/abs/2211.15928}{{\normalfont
  [arXiv:astro-ph.GA/2211.15928]}}.
\newblock {\url{https://doi.org/10.3847/1538-4357/acc029}}.

\bibitem[{Casetti-Dinescu} et~al.(2024){Casetti-Dinescu}, {Pawlowski},
  {Girard}, {Kanehisa}, {Petroski}, {Martone}, {Kozhurina-Platais}, and
  {Platais}]{2024ApJ...975..138C}
{Casetti-Dinescu}, D.I.; {Pawlowski}, M.S.; {Girard}, T.M.; {Kanehisa}, K.J.;
  {Petroski}, A.; {Martone}, M.; {Kozhurina-Platais}, V.; {Platais}, I.
\newblock {HST Proper Motion of Andromeda III. Another Satellite Coorbiting the
  M31 Satellite Plane}.
\newblock {\em \apj} {\bf 2024}, {\em 975},~138,
  \href{http://arxiv.org/abs/2409.08252}{{\normalfont
  [arXiv:astro-ph.GA/2409.08252]}}.
\newblock {\url{https://doi.org/10.3847/1538-4357/ad7b10}}.

\bibitem[{Collins} et~al.(2024){Collins}, {Karim}, {Martinez-Delgado},
  {Monelli}, {Tollerud}, {Donatiello}, {Navabi}, {Charles}, and
  {Boschin}]{2024MNRAS.528.2614C}
{Collins}, M.L.M.; {Karim}, N.; {Martinez-Delgado}, D.; {Monelli}, M.;
  {Tollerud}, E.J.; {Donatiello}, G.; {Navabi}, M.; {Charles}, E.; {Boschin},
  W.
\newblock {Pisces VII/Triangulum III - M33's second dwarf satellite galaxy}.
\newblock {\em \mnras} {\bf 2024}, {\em 528},~2614--2620,
  \href{http://arxiv.org/abs/2305.13966}{{\normalfont
  [arXiv:astro-ph.GA/2305.13966]}}.
\newblock {\url{https://doi.org/10.1093/mnras/stae199}}.

\bibitem[{Patel} et~al.(2018){Patel}, {Carlin}, {Tollerud}, {Collins}, and
  {Dooley}]{2018MNRAS.480.1883P}
{Patel}, E.; {Carlin}, J.L.; {Tollerud}, E.J.; {Collins}, M.L.M.; {Dooley},
  G.A.
\newblock {{\ensuremath{\Lambda}}CDM predictions for the satellite population
  of M33}.
\newblock {\em \mnras} {\bf 2018}, {\em 480},~1883--1897,
  \href{http://arxiv.org/abs/1807.05318}{{\normalfont
  [arXiv:astro-ph.GA/1807.05318]}}.
\newblock {\url{https://doi.org/10.1093/mnras/sty1946}}.

\bibitem[{Makarova} et~al.(2017){Makarova}, {Makarov}, {Karachentsev}, {Tully},
  and {Rizzi}]{2017MNRAS.464.2281M}
{Makarova}, L.N.; {Makarov}, D.I.; {Karachentsev}, I.D.; {Tully}, R.B.;
  {Rizzi}, L.
\newblock {Star formation history of And XVIII: a dwarf spheroidal galaxy in
  isolation}.
\newblock {\em \mnras} {\bf 2017}, {\em 464},~2281--2289,
  \href{http://arxiv.org/abs/1609.09706}{{\normalfont
  [arXiv:astro-ph.GA/1609.09706]}}.
\newblock {\url{https://doi.org/10.1093/mnras/stw2502}}.

\bibitem[{Drlica-Wagner} et~al.(2015){Drlica-Wagner}, {Bechtol}, {Rykoff},
  {Luque}, {Queiroz}, {Mao}, {Wechsler}, {Simon}, {Santiago}, {Yanny},
  {Balbinot}, {Dodelson}, {Fausti Neto}, {James}, {Li}, {Maia}, {Marshall},
  {Pieres}, {Stringer}, {Walker}, {Abbott}, {Abdalla}, {Allam},
  {Benoit-L{\'e}vy}, {Bernstein}, {Bertin}, {Brooks}, {Buckley-Geer}, {Burke},
  {Carnero Rosell}, {Carrasco Kind}, {Carretero}, {Crocce}, {da Costa},
  {Desai}, {Diehl}, {Dietrich}, {Doel}, {Eifler}, {Evrard}, {Finley},
  {Flaugher}, {Fosalba}, {Frieman}, {Gaztanaga}, {Gerdes}, {Gruen}, {Gruendl},
  {Gutierrez}, {Honscheid}, {Kuehn}, {Kuropatkin}, {Lahav}, {Martini},
  {Miquel}, {Nord}, {Ogando}, {Plazas}, {Reil}, {Roodman}, {Sako}, {Sanchez},
  {Scarpine}, {Schubnell}, {Sevilla-Noarbe}, {Smith}, {Soares-Santos},
  {Sobreira}, {Suchyta}, {Swanson}, {Tarle}, {Tucker}, {Vikram}, {Wester},
  {Zhang}, {Zuntz}, and {DES Collaboration}]{2015ApJ...813..109D}
{Drlica-Wagner}, A.; {Bechtol}, K.; {Rykoff}, E.S.; {Luque}, E.; {Queiroz}, A.;
  {Mao}, Y.Y.; {Wechsler}, R.H.; {Simon}, J.D.; {Santiago}, B.; {Yanny}, B.;
  et~al.
\newblock {Eight Ultra-faint Galaxy Candidates Discovered in Year Two of the
  Dark Energy Survey}.
\newblock {\em \apj} {\bf 2015}, {\em 813},~109,
  \href{http://arxiv.org/abs/1508.03622}{{\normalfont
  [arXiv:astro-ph.GA/1508.03622]}}.
\newblock {\url{https://doi.org/10.1088/0004-637X/813/2/109}}.

\bibitem[{Karachentsev} et~al.(2004){Karachentsev}, {Karachentseva},
  {Huchtmeier}, and {Makarov}]{2004AJ....127.2031K}
{Karachentsev}, I.D.; {Karachentseva}, V.E.; {Huchtmeier}, W.K.; {Makarov},
  D.I.
\newblock {A Catalog of Neighboring Galaxies}.
\newblock {\em \aj} {\bf 2004}, {\em 127},~2031--2068.
\newblock {\url{https://doi.org/10.1086/382905}}.

\bibitem[{Walker} et~al.(2009){Walker}, {Mateo}, {Olszewski}, {Sen}, and
  {Woodroofe}]{2009AJ....137.3109W}
{Walker}, M.G.; {Mateo}, M.; {Olszewski}, E.W.; {Sen}, B.; {Woodroofe}, M.
\newblock {Clean Kinematic Samples in Dwarf Spheroidals: An Algorithm for
  Evaluating Membership and Estimating Distribution Parameters When
  Contamination is Present}.
\newblock {\em \aj} {\bf 2009}, {\em 137},~3109--3138,
  \href{http://arxiv.org/abs/0811.1990}{{\normalfont
  [arXiv:astro-ph/0811.1990]}}.
\newblock {\url{https://doi.org/10.1088/0004-6256/137/2/3109}}.

\bibitem[{Conn} et~al.(2018){Conn}, {Jerjen}, {Kim}, and
  {Schirmer}]{2018ApJ...852...68C}
{Conn}, B.C.; {Jerjen}, H.; {Kim}, D.; {Schirmer}, M.
\newblock {On the Nature of Ultra-faint Dwarf Galaxy Candidates. I. DES1,
  Eridanus III, and Tucana V}.
\newblock {\em \apj} {\bf 2018}, {\em 852},~68,
  \href{http://arxiv.org/abs/1712.01439}{{\normalfont
  [arXiv:astro-ph.GA/1712.01439]}}.
\newblock {\url{https://doi.org/10.3847/1538-4357/aa9eda}}.

\bibitem[{Cerny} et~al.(2021){Cerny}, {Pace}, {Drlica-Wagner}, {Ferguson},
  {Mau}, {Adam{\'o}w}, {Carlin}, {Choi}, {Erkal}, {Johnson}, {Li},
  {Mart{\'\i}nez-V{\'a}zquez}, {Mutlu-Pakdil}, {Nidever}, {Olsen}, {Pieres},
  {Tollerud}, {Simon}, {Vivas}, {James}, {Kuropatkin}, {Majewski},
  {Mart{\'\i}nez-Delgado}, {Massana}, {Miller}, {Neilsen}, {No{\"e}l}, {Riley},
  {Sand}, {Santana-Silva}, {Stringfellow}, {Tucker}, and {Delve
  Collaboration}]{2021ApJ...910...18C}
{Cerny}, W.; {Pace}, A.B.; {Drlica-Wagner}, A.; {Ferguson}, P.S.; {Mau}, S.;
  {Adam{\'o}w}, M.; {Carlin}, J.L.; {Choi}, Y.; {Erkal}, D.; {Johnson}, L.C.;
  et~al.
\newblock {Discovery of an Ultra-faint Stellar System near the Magellanic
  Clouds with the DECam Local Volume Exploration Survey}.
\newblock {\em \apj} {\bf 2021}, {\em 910},~18,
  \href{http://arxiv.org/abs/2009.08550}{{\normalfont
  [arXiv:astro-ph.GA/2009.08550]}}.
\newblock {\url{https://doi.org/10.3847/1538-4357/abe1af}}.

\bibitem[{Carlin} et~al.(2017){Carlin}, {Sand}, {Mu{\~n}oz}, {Spekkens},
  {Willman}, {Crnojevi{\'c}}, {Forbes}, {Hargis}, {Kirby}, {Peter},
  {Romanowsky}, and {Strader}]{2017AJ....154..267C}
{Carlin}, J.L.; {Sand}, D.J.; {Mu{\~n}oz}, R.R.; {Spekkens}, K.; {Willman}, B.;
  {Crnojevi{\'c}}, D.; {Forbes}, D.A.; {Hargis}, J.; {Kirby}, E.; {Peter},
  A.H.G.;  et~al.
\newblock {Deep Subaru Hyper Suprime-Cam Observations of Milky Way Satellites
  Columba I and Triangulum II}.
\newblock {\em \aj} {\bf 2017}, {\em 154},~267,
  \href{http://arxiv.org/abs/1710.06444}{{\normalfont
  [arXiv:astro-ph.GA/1710.06444]}}.
\newblock {\url{https://doi.org/10.3847/1538-3881/aa94d0}}.

\bibitem[{Luque} et~al.(2017){Luque}, {Pieres}, {Santiago}, {Yanny}, {Vivas},
  {Queiroz}, {Drlica-Wagner}, {Morganson}, {Balbinot}, {Marshall}, {Li},
  {Neto}, {da Costa}, {Maia}, {Bechtol}, {Kim}, {Bernstein}, {Dodelson},
  {Whiteway}, {Diehl}, {Finley}, {Abbott}, {Abdalla}, {Allam}, {Annis},
  {Benoit-L{\'e}vy}, {Bertin}, {Brooks}, {Burke}, {Rosell}, {Kind},
  {Carretero}, {Cunha}, {D'Andrea}, {Desai}, {Doel}, {Evrard}, {Flaugher},
  {Fosalba}, {Gerdes}, {Goldstein}, {Gruen}, {Gruendl}, {Gutierrez}, {James},
  {Kuehn}, {Kuropatkin}, {Lahav}, {Martini}, {Miquel}, {Nord}, {Ogando},
  {Plazas}, {Romer}, {Sanchez}, {Scarpine}, {Schubnell}, {Sevilla-Noarbe},
  {Smith}, {Soares-Santos}, {Sobreira}, {Suchyta}, {Swanson}, {Tarle},
  {Thomas}, and {Walker}]{2017MNRAS.468...97L}
{Luque}, E.; {Pieres}, A.; {Santiago}, B.; {Yanny}, B.; {Vivas}, A.K.;
  {Queiroz}, A.; {Drlica-Wagner}, A.; {Morganson}, E.; {Balbinot}, E.;
  {Marshall}, J.L.;  et~al.
\newblock {The Dark Energy Survey view of the Sagittarius stream: discovery of
  two faint stellar system candidates}.
\newblock {\em \mnras} {\bf 2017}, {\em 468},~97--108,
  \href{http://arxiv.org/abs/1608.04033}{{\normalfont
  [arXiv:astro-ph.GA/1608.04033]}}.
\newblock {\url{https://doi.org/10.1093/mnras/stx405}}.

\bibitem[{Ferrarese} et~al.(2000){Ferrarese}, {Mould}, {Kennicutt}, {Huchra},
  {Ford}, {Freedman}, {Stetson}, {Madore}, {Sakai}, {Gibson}, {Graham},
  {Hughes}, {Illingworth}, {Kelson}, {Macri}, {Sebo}, and
  {Silbermann}]{2000ApJ...529..745F}
{Ferrarese}, L.; {Mould}, J.R.; {Kennicutt}, Jr., R.C.; {Huchra}, J.; {Ford},
  H.C.; {Freedman}, W.L.; {Stetson}, P.B.; {Madore}, B.F.; {Sakai}, S.;
  {Gibson}, B.K.;  et~al.
\newblock {The Hubble Space Telescope Key Project on the Extragalactic Distance
  Scale. XXVI. The Calibration of Population II Secondary Distance Indicators
  and the Value of the Hubble Constant}.
\newblock {\em \apj} {\bf 2000}, {\em 529},~745--767,
  \href{http://arxiv.org/abs/astro-ph/9908192}{{\normalfont
  [arXiv:astro-ph/astro-ph/9908192]}}.
\newblock {\url{https://doi.org/10.1086/308309}}.

\bibitem[{Strauss} et~al.(1992){Strauss}, {Huchra}, {Davis}, {Yahil}, {Fisher},
  and {Tonry}]{1992ApJS...83...29S}
{Strauss}, M.A.; {Huchra}, J.P.; {Davis}, M.; {Yahil}, A.; {Fisher}, K.B.;
  {Tonry}, J.
\newblock {A Redshift Survey of IRAS Galaxies. VII. The Infrared and Redshift
  Data for the 1.936 Jansky Sample}.
\newblock {\em \apjs} {\bf 1992}, {\em 83},~29.
\newblock {\url{https://doi.org/10.1086/191730}}.

\bibitem[{Hargis} et~al.(2016){Hargis}, {Kimmig}, {Willman}, {Caldwell},
  {Walker}, {Strader}, {Sand}, {Grillmair}, and {Yoon}]{2016ApJ...818...39H}
{Hargis}, J.R.; {Kimmig}, B.; {Willman}, B.; {Caldwell}, N.; {Walker}, M.G.;
  {Strader}, J.; {Sand}, D.J.; {Grillmair}, C.J.; {Yoon}, J.H.
\newblock {Evidence That Hydra I is a Tidally Disrupting Milky Way Dwarf
  Galaxy}.
\newblock {\em \apj} {\bf 2016}, {\em 818},~39,
  \href{http://arxiv.org/abs/1509.06391}{{\normalfont
  [arXiv:astro-ph.GA/1509.06391]}}.
\newblock {\url{https://doi.org/10.3847/0004-637X/818/1/39}}.

\bibitem[{Torrealba} et~al.(2019){Torrealba}, {Belokurov}, {Koposov}, {Li},
  {Walker}, {Sanders}, {Geringer-Sameth}, {Zucker}, {Kuehn}, {Evans}, and
  {Dehnen}]{2019MNRAS.488.2743T}
{Torrealba}, G.; {Belokurov}, V.; {Koposov}, S.E.; {Li}, T.S.; {Walker}, M.G.;
  {Sanders}, J.L.; {Geringer-Sameth}, A.; {Zucker}, D.B.; {Kuehn}, K.; {Evans},
  N.W.;  et~al.
\newblock {The hidden giant: discovery of an enormous Galactic dwarf satellite
  in Gaia DR2}.
\newblock {\em \mnras} {\bf 2019}, {\em 488},~2743--2766,
  \href{http://arxiv.org/abs/1811.04082}{{\normalfont
  [arXiv:astro-ph.GA/1811.04082]}}.
\newblock {\url{https://doi.org/10.1093/mnras/stz1624}}.

\bibitem[{Ji} et~al.(2021){Ji}, {Koposov}, {Li}, {Erkal}, {Pace}, {Simon},
  {Belokurov}, {Cullinane}, {Da Costa}, {Kuehn}, {Lewis}, {Mackey}, {Shipp},
  {Simpson}, {Zucker}, {Hansen}, {Bland-Hawthorn}, and {S5
  Collaboration}]{2021ApJ...921...32J}
{Ji}, A.P.; {Koposov}, S.E.; {Li}, T.S.; {Erkal}, D.; {Pace}, A.B.; {Simon},
  J.D.; {Belokurov}, V.; {Cullinane}, L.R.; {Da Costa}, G.S.; {Kuehn}, K.;
  et~al.
\newblock {Kinematics of Antlia 2 and Crater 2 from the Southern Stellar Stream
  Spectroscopic Survey (S$^{5}$)}.
\newblock {\em \apj} {\bf 2021}, {\em 921},~32,
  \href{http://arxiv.org/abs/2106.12656}{{\normalfont
  [arXiv:astro-ph.GA/2106.12656]}}.
\newblock {\url{https://doi.org/10.3847/1538-4357/ac1869}}.

\bibitem[{McQuinn} et~al.(2024){McQuinn}, {Mao}, {Tollerud}, {Cohen}, {Shih},
  {Buckley}, and {Dolphin}]{2024ApJ...967..161M}
{McQuinn}, K.B.W.; {Mao}, Y.Y.; {Tollerud}, E.J.; {Cohen}, R.E.; {Shih}, D.;
  {Buckley}, M.R.; {Dolphin}, A.E.
\newblock {Discovery and Characterization of Two Ultrafaint Dwarfs outside the
  Halo of the Milky Way: Leo M and Leo K}.
\newblock {\em \apj} {\bf 2024}, {\em 967},~161,
  \href{http://arxiv.org/abs/2307.08738}{{\normalfont
  [arXiv:astro-ph.GA/2307.08738]}}.
\newblock {\url{https://doi.org/10.3847/1538-4357/ad429b}}.

\bibitem[{Laevens} et~al.(2014){Laevens}, {Martin}, {Sesar}, {Bernard}, {Rix},
  {Slater}, {Bell}, {Ferguson}, {Schlafly}, {Burgett}, {Chambers}, {Denneau},
  {Draper}, {Kaiser}, {Kudritzki}, {Magnier}, {Metcalfe}, {Morgan}, {Price},
  {Sweeney}, {Tonry}, {Wainscoat}, and {Waters}]{2014ApJ...786L...3L}
{Laevens}, B.P.M.; {Martin}, N.F.; {Sesar}, B.; {Bernard}, E.J.; {Rix}, H.W.;
  {Slater}, C.T.; {Bell}, E.F.; {Ferguson}, A.M.N.; {Schlafly}, E.F.;
  {Burgett}, W.S.;  et~al.
\newblock {A New Distant Milky Way Globular Cluster in the Pan-STARRS1
  3{\ensuremath{\pi}} Survey}.
\newblock {\em \apjl} {\bf 2014}, {\em 786},~L3,
  \href{http://arxiv.org/abs/1403.6593}{{\normalfont
  [arXiv:astro-ph.GA/1403.6593]}}.
\newblock {\url{https://doi.org/10.1088/2041-8205/786/1/L3}}.

\bibitem[{Kirby} et~al.(2015){Kirby}, {Simon}, and
  {Cohen}]{2015ApJ...810...56K}
{Kirby}, E.N.; {Simon}, J.D.; {Cohen}, J.G.
\newblock {Spectroscopic Confirmation of the Dwarf Galaxies Hydra II and Pisces
  II and the Globular Cluster Laevens 1}.
\newblock {\em \apj} {\bf 2015}, {\em 810},~56,
  \href{http://arxiv.org/abs/1506.01021}{{\normalfont
  [arXiv:astro-ph.GA/1506.01021]}}.
\newblock {\url{https://doi.org/10.1088/0004-637X/810/1/56}}.

\bibitem[{Mau} et~al.(2020){Mau}, {Cerny}, {Pace}, {Choi}, {Drlica-Wagner},
  {Santana-Silva}, {Riley}, {Erkal}, {Stringfellow}, {Adam{\'o}w}, {Carlin},
  {Gruendl}, {Hernandez-Lang}, {Kuropatkin}, {Li}, {Mart{\'\i}nez-V{\'a}zquez},
  {Morganson}, {Mutlu-Pakdil}, {Neilsen}, {Nidever}, {Olsen}, {Sand},
  {Tollerud}, {Tucker}, {Yanny}, {Zenteno}, {Allam}, {Barkhouse}, {Bechtol},
  {Bell}, {Balaji}, {Crnojevi{\'c}}, {Esteves}, {Ferguson}, {Gallart},
  {Hughes}, {James}, {Jethwa}, {Johnson}, {Kuehn}, {Majewski}, {Mao},
  {Massana}, {McNanna}, {Monachesi}, {Nadler}, {No{\"e}l}, {Palmese},
  {Paz-Chinchon}, {Pieres}, {Sanchez}, {Shipp}, {Simon}, {Soares-Santos},
  {Tavangar}, {van der Marel}, {Vivas}, {Walker}, and
  {Wechsler}]{2020ApJ...890..136M}
{Mau}, S.; {Cerny}, W.; {Pace}, A.B.; {Choi}, Y.; {Drlica-Wagner}, A.;
  {Santana-Silva}, L.; {Riley}, A.H.; {Erkal}, D.; {Stringfellow}, G.S.;
  {Adam{\'o}w}, M.;  et~al.
\newblock {Two Ultra-faint Milky Way Stellar Systems Discovered in Early Data
  from the DECam Local Volume Exploration Survey}.
\newblock {\em \apj} {\bf 2020}, {\em 890},~136,
  \href{http://arxiv.org/abs/1912.03301}{{\normalfont
  [arXiv:astro-ph.GA/1912.03301]}}.
\newblock {\url{https://doi.org/10.3847/1538-4357/ab6c67}}.

\bibitem[{Simon} and {Geha}(2007)]{2007ApJ...670..313S}
{Simon}, J.D.; {Geha}, M.
\newblock {The Kinematics of the Ultra-faint Milky Way Satellites: Solving the
  Missing Satellite Problem}.
\newblock {\em \apj} {\bf 2007}, {\em 670},~313--331,
  \href{http://arxiv.org/abs/0706.0516}{{\normalfont
  [arXiv:astro-ph/0706.0516]}}.
\newblock {\url{https://doi.org/10.1086/521816}}.

\bibitem[{Grillmair}(2009)]{2009ApJ...693.1118G}
{Grillmair}, C.J.
\newblock {Four New Stellar Debris Streams in the Galactic Halo}.
\newblock {\em \apj} {\bf 2009}, {\em 693},~1118--1127,
  \href{http://arxiv.org/abs/0811.3965}{{\normalfont
  [arXiv:astro-ph/0811.3965]}}.
\newblock {\url{https://doi.org/10.1088/0004-637X/693/2/1118}}.

\bibitem[{Gregory} et~al.(2020){Gregory}, {Collins}, {Erkal}, {Tollerud},
  {Delorme}, {Hill}, {Sand}, {Strader}, and {Willman}]{2020MNRAS.496.1092G}
{Gregory}, A.L.; {Collins}, M.L.M.; {Erkal}, D.; {Tollerud}, E.; {Delorme}, M.;
  {Hill}, L.; {Sand}, D.J.; {Strader}, J.; {Willman}, B.
\newblock {Uncovering the orbit of the hercules dwarf galaxy}.
\newblock {\em \mnras} {\bf 2020}, {\em 496},~1092--1104,
  \href{http://arxiv.org/abs/1912.00156}{{\normalfont
  [arXiv:astro-ph.GA/1912.00156]}}.
\newblock {\url{https://doi.org/10.1093/mnras/staa1553}}.

\bibitem[{GRAVITY Collaboration} et~al.(2018){GRAVITY Collaboration}, {Abuter},
  {Amorim}, {Anugu}, {Baub{\"o}ck}, {Benisty}, {Berger}, {Blind}, {Bonnet},
  {Brandner}, {Buron}, {Collin}, {Chapron}, {Cl{\'e}net}, {Coud{\'e} Du
  Foresto}, {de Zeeuw}, {Deen}, {Delplancke-Str{\"o}bele}, {Dembet}, {Dexter},
  {Duvert}, {Eckart}, {Eisenhauer}, {Finger}, {F{\"o}rster Schreiber},
  {F{\'e}dou}, {Garcia}, {Garcia Lopez}, {Gao}, {Gendron}, {Genzel},
  {Gillessen}, {Gordo}, {Habibi}, {Haubois}, {Haug}, {Hau{\ss}mann}, {Henning},
  {Hippler}, {Horrobin}, {Hubert}, {Hubin}, {Jimenez Rosales}, {Jochum},
  {Jocou}, {Kaufer}, {Kellner}, {Kendrew}, {Kervella}, {Kok}, {Kulas},
  {Lacour}, {Lapeyr{\`e}re}, {Lazareff}, {Le Bouquin}, {L{\'e}na}, {Lippa},
  {Lenzen}, {M{\'e}rand}, {M{\"u}ler}, {Neumann}, {Ott}, {Palanca}, {Paumard},
  {Pasquini}, {Perraut}, {Perrin}, {Pfuhl}, {Plewa}, {Rabien}, {Ram{\'\i}rez},
  {Ramos}, {Rau}, {Rodr{\'\i}guez-Coira}, {Rohloff}, {Rousset},
  {Sanchez-Bermudez}, {Scheithauer}, {Sch{\"o}ller}, {Schuler}, {Spyromilio},
  {Straub}, {Straubmeier}, {Sturm}, {Tacconi}, {Tristram}, {Vincent}, {von
  Fellenberg}, {Wank}, {Waisberg}, {Widmann}, {Wieprecht}, {Wiest},
  {Wiezorrek}, {Woillez}, {Yazici}, {Ziegler}, and {Zins}]{2018AA...615L..15G}
{GRAVITY Collaboration}.; {Abuter}, R.; {Amorim}, A.; {Anugu}, N.;
  {Baub{\"o}ck}, M.; {Benisty}, M.; {Berger}, J.P.; {Blind}, N.; {Bonnet}, H.;
  {Brandner}, W.;  et~al.
\newblock {Detection of the gravitational redshift in the orbit of the star S2
  near the Galactic centre massive black hole}.
\newblock {\em \aap} {\bf 2018}, {\em 615},~L15,
  \href{http://arxiv.org/abs/1807.09409}{{\normalfont
  [arXiv:astro-ph.GA/1807.09409]}}.
\newblock {\url{https://doi.org/10.1051/0004-6361/201833718}}.

\bibitem[{Monaco} et~al.(2004){Monaco}, {Bellazzini}, {Ferraro}, and
  {Pancino}]{2004MNRAS.353..874M}
{Monaco}, L.; {Bellazzini}, M.; {Ferraro}, F.R.; {Pancino}, E.
\newblock {The distance to the Sagittarius dwarf spheroidal galaxy from the red
  giant branch tip}.
\newblock {\em \mnras} {\bf 2004}, {\em 353},~874--878,
  \href{http://arxiv.org/abs/astro-ph/0406350}{{\normalfont
  [arXiv:astro-ph/astro-ph/0406350]}}.
\newblock {\url{https://doi.org/10.1111/j.1365-2966.2004.08122.x}}.

\bibitem[{Kirby} et~al.(2014){Kirby}, {Bullock}, {Boylan-Kolchin},
  {Kaplinghat}, and {Cohen}]{2014MNRAS.439.1015K}
{Kirby}, E.N.; {Bullock}, J.S.; {Boylan-Kolchin}, M.; {Kaplinghat}, M.;
  {Cohen}, J.G.
\newblock {The dynamics of isolated Local Group galaxies}.
\newblock {\em \mnras} {\bf 2014}, {\em 439},~1015--1027,
  \href{http://arxiv.org/abs/1401.1208}{{\normalfont
  [arXiv:astro-ph.GA/1401.1208]}}.
\newblock {\url{https://doi.org/10.1093/mnras/stu025}}.

\bibitem[{Longeard} et~al.(2021){Longeard}, {Martin}, {Ibata}, {Starkenburg},
  {Jablonka}, {Aguado}, {Carlberg}, {C{\^o}t{\'e}}, {Gonz{\'a}lez
  Hern{\'a}ndez}, {Lucchesi}, {Malhan}, {Navarro}, {S{\'a}nchez-Janssen},
  {Thomas}, {Venn}, and {McConnachie}]{2021MNRAS.503.2754L}
{Longeard}, N.; {Martin}, N.; {Ibata}, R.A.; {Starkenburg}, E.; {Jablonka}, P.;
  {Aguado}, D.S.; {Carlberg}, R.G.; {C{\^o}t{\'e}}, P.; {Gonz{\'a}lez
  Hern{\'a}ndez}, J.I.; {Lucchesi}, R.;  et~al.
\newblock {The pristine dwarf-galaxy survey - III. Revealing the nature of the
  Milky Way globular cluster Sagittarius II}.
\newblock {\em \mnras} {\bf 2021}, {\em 503},~2754--2762,
  \href{http://arxiv.org/abs/2005.05976}{{\normalfont
  [arXiv:astro-ph.GA/2005.05976]}}.
\newblock {\url{https://doi.org/10.1093/mnras/stab604}}.

\bibitem[{Fadely} et~al.(2011){Fadely}, {Willman}, {Geha}, {Walsh},
  {Mu{\~n}oz}, {Jerjen}, {Vargas}, and {Da Costa}]{2011AJ....142...88F}
{Fadely}, R.; {Willman}, B.; {Geha}, M.; {Walsh}, S.; {Mu{\~n}oz}, R.R.;
  {Jerjen}, H.; {Vargas}, L.C.; {Da Costa}, G.S.
\newblock {Segue 3: An Old, Extremely Low Luminosity Star Cluster in the Milky
  Way's Halo}.
\newblock {\em \aj} {\bf 2011}, {\em 142},~88,
  \href{http://arxiv.org/abs/1107.3151}{{\normalfont
  [arXiv:astro-ph.GA/1107.3151]}}.
\newblock {\url{https://doi.org/10.1088/0004-6256/142/3/88}}.

\bibitem[{Fritz} et~al.(2019){Fritz}, {Carrera}, {Battaglia}, and
  {Taibi}]{2019AA...623A.129F}
{Fritz}, T.K.; {Carrera}, R.; {Battaglia}, G.; {Taibi}, S.
\newblock {Gaia DR 2 and VLT/FLAMES search for new satellites of the LMC}.
\newblock {\em \aap} {\bf 2019}, {\em 623},~A129,
  \href{http://arxiv.org/abs/1805.07350}{{\normalfont
  [arXiv:astro-ph.GA/1805.07350]}}.
\newblock {\url{https://doi.org/10.1051/0004-6361/201833458}}.

\bibitem[{Huxor} et~al.(2014){Huxor}, {Mackey}, {Ferguson}, {Irwin}, {Martin},
  {Tanvir}, {Veljanoski}, {McConnachie}, {Fishlock}, {Ibata}, and
  {Lewis}]{2014MNRAS.442.2165H}
{Huxor}, A.P.; {Mackey}, A.D.; {Ferguson}, A.M.N.; {Irwin}, M.J.; {Martin},
  N.F.; {Tanvir}, N.R.; {Veljanoski}, J.; {McConnachie}, A.; {Fishlock}, C.K.;
  {Ibata}, R.;  et~al.
\newblock {The outer halo globular cluster system of M31 - I. The final PAndAS
  catalogue}.
\newblock {\em \mnras} {\bf 2014}, {\em 442},~2165--2187,
  \href{http://arxiv.org/abs/1404.5807}{{\normalfont
  [arXiv:astro-ph.GA/1404.5807]}}.
\newblock {\url{https://doi.org/10.1093/mnras/stu771}}.

\bibitem[{Weisz} et~al.(2019){Weisz}, {Dolphin}, {Martin}, {Albers}, {Collins},
  {Ferguson}, {Lewis}, {Mackey}, {McConnachie}, {Rich}, and
  {Skillman}]{2019MNRAS.489..763W}
{Weisz}, D.R.; {Dolphin}, A.E.; {Martin}, N.F.; {Albers}, S.M.; {Collins},
  M.L.M.; {Ferguson}, A.M.N.; {Lewis}, G.F.; {Mackey}, A.D.; {McConnachie}, A.;
  {Rich}, R.M.;  et~al.
\newblock {A rogues gallery of Andromeda's dwarf galaxies - II. Precise
  distances to 17 faint satellites}.
\newblock {\em \mnras} {\bf 2019}, {\em 489},~763--770,
  \href{http://arxiv.org/abs/1909.02017}{{\normalfont
  [arXiv:astro-ph.GA/1909.02017]}}.
\newblock {\url{https://doi.org/10.1093/mnras/stz1984}}.

\bibitem[{Watkins} et~al.(2013){Watkins}, {Evans}, and {van de
  Ven}]{2013MNRAS.430..971W}
{Watkins}, L.L.; {Evans}, N.W.; {van de Ven}, G.
\newblock {A census of orbital properties of the M31 satellites}.
\newblock {\em \mnras} {\bf 2013}, {\em 430},~971--985,
  \href{http://arxiv.org/abs/1211.2638}{{\normalfont
  [arXiv:astro-ph.CO/1211.2638]}}.
\newblock {\url{https://doi.org/10.1093/mnras/sts634}}.

\bibitem[{Karachentsev} et~al.(2014){Karachentsev}, {Kaisina}, and
  {Makarov}]{2014AJ....147...13K}
{Karachentsev}, I.D.; {Kaisina}, E.I.; {Makarov}, D.I.
\newblock {Suites of Dwarfs around nearby Giant Galaxies}.
\newblock {\em \aj} {\bf 2014}, {\em 147},~13,
  \href{http://arxiv.org/abs/1310.6838}{{\normalfont
  [arXiv:astro-ph.CO/1310.6838]}}.
\newblock {\url{https://doi.org/10.1088/0004-6256/147/1/13}}.

\bibitem[{Oh} et~al.(2015){Oh}, {Hunter}, {Brinks}, {Elmegreen}, {Schruba},
  {Walter}, {Rupen}, {Young}, {Simpson}, {Johnson}, {Herrmann}, {Ficut-Vicas},
  {Cigan}, {Heesen}, {Ashley}, and {Zhang}]{2015AJ....149..180O}
{Oh}, S.H.; {Hunter}, D.A.; {Brinks}, E.; {Elmegreen}, B.G.; {Schruba}, A.;
  {Walter}, F.; {Rupen}, M.P.; {Young}, L.M.; {Simpson}, C.E.; {Johnson}, M.C.;
   et~al.
\newblock {High-resolution Mass Models of Dwarf Galaxies from LITTLE THINGS}.
\newblock {\em \aj} {\bf 2015}, {\em 149},~180,
  \href{http://arxiv.org/abs/1502.01281}{{\normalfont
  [arXiv:astro-ph.GA/1502.01281]}}.
\newblock {\url{https://doi.org/10.1088/0004-6256/149/6/180}}.

\bibitem[{Martin} et~al.(2014){Martin}, {Chambers}, {Collins}, {Ibata}, {Rich},
  {Bell}, {Bernard}, {Ferguson}, {Flewelling}, {Kaiser}, {Magnier}, {Tonry},
  and {Wainscoat}]{2014ApJ...793L..14M}
{Martin}, N.F.; {Chambers}, K.C.; {Collins}, M.L.M.; {Ibata}, R.A.; {Rich},
  R.M.; {Bell}, E.F.; {Bernard}, E.J.; {Ferguson}, A.M.N.; {Flewelling}, H.;
  {Kaiser}, N.;  et~al.
\newblock {Spectroscopy of the Three Distant Andromedan Satellites Cassiopeia
  III, Lacerta I, and Perseus I}.
\newblock {\em \apjl} {\bf 2014}, {\em 793},~L14,
  \href{http://arxiv.org/abs/1408.5130}{{\normalfont
  [arXiv:astro-ph.GA/1408.5130]}}.
\newblock {\url{https://doi.org/10.1088/2041-8205/793/1/L14}}.

\bibitem[{Richardson} et~al.(2011){Richardson}, {Irwin}, {McConnachie},
  {Martin}, {Dotter}, {Ferguson}, {Ibata}, {Chapman}, {Lewis}, {Tanvir}, and
  {Rich}]{2011ApJ...732...76R}
{Richardson}, J.C.; {Irwin}, M.J.; {McConnachie}, A.W.; {Martin}, N.F.;
  {Dotter}, A.L.; {Ferguson}, A.M.N.; {Ibata}, R.A.; {Chapman}, S.C.; {Lewis},
  G.F.; {Tanvir}, N.R.;  et~al.
\newblock {PAndAS' Progeny: Extending the M31 Dwarf Galaxy Cabal}.
\newblock {\em \apj} {\bf 2011}, {\em 732},~76,
  \href{http://arxiv.org/abs/1102.2902}{{\normalfont
  [arXiv:astro-ph.CO/1102.2902]}}.
\newblock {\url{https://doi.org/10.1088/0004-637X/732/2/76}}.

\bibitem[{Martin} et~al.(2006){Martin}, {Ibata}, {Irwin}, {Chapman}, {Lewis},
  {Ferguson}, {Tanvir}, and {McConnachie}]{2006MNRAS.371.1983M}
{Martin}, N.F.; {Ibata}, R.A.; {Irwin}, M.J.; {Chapman}, S.; {Lewis}, G.F.;
  {Ferguson}, A.M.N.; {Tanvir}, N.; {McConnachie}, A.W.
\newblock {Discovery and analysis of three faint dwarf galaxies and a globular
  cluster in the outer halo of the Andromeda galaxy}.
\newblock {\em \mnras} {\bf 2006}, {\em 371},~1983--1991,
  \href{http://arxiv.org/abs/astro-ph/0607472}{{\normalfont
  [arXiv:astro-ph/astro-ph/0607472]}}.
\newblock {\url{https://doi.org/10.1111/j.1365-2966.2006.10823.x}}.

\bibitem[{Sakari} and {Wallerstein}(2016)]{2016MNRAS.456..831S}
{Sakari}, C.M.; {Wallerstein}, G.
\newblock {The integrated calcium II triplet as a metallicity indicator:
  comparisons with high-resolution [Fe/H] in M31 globular clusters}.
\newblock {\em \mnras} {\bf 2016}, {\em 456},~831--843,
  \href{http://arxiv.org/abs/1511.06766}{{\normalfont
  [arXiv:astro-ph.GA/1511.06766]}}.
\newblock {\url{https://doi.org/10.1093/mnras/stv2711}}.

\bibitem[{Mackey} et~al.(2013){Mackey}, {Huxor}, {Martin}, {Ferguson},
  {Dotter}, {McConnachie}, {Ibata}, {Irwin}, {Lewis}, {Sakari}, {Tanvir}, and
  {Venn}]{2013ApJ...770L..17M}
{Mackey}, A.D.; {Huxor}, A.P.; {Martin}, N.F.; {Ferguson}, A.M.N.; {Dotter},
  A.; {McConnachie}, A.W.; {Ibata}, R.A.; {Irwin}, M.J.; {Lewis}, G.F.;
  {Sakari}, C.M.;  et~al.
\newblock {A Peculiar Faint Satellite in the Remote Outer Halo of M31}.
\newblock {\em \apjl} {\bf 2013}, {\em 770},~L17,
  \href{http://arxiv.org/abs/1304.7826}{{\normalfont
  [arXiv:astro-ph.GA/1304.7826]}}.
\newblock {\url{https://doi.org/10.1088/2041-8205/770/2/L17}}.

\bibitem[{Collins} et~al.(2013){Collins}, {Chapman}, {Rich}, {Ibata}, {Martin},
  {Irwin}, {Bate}, {Lewis}, {Pe{\~n}arrubia}, {Arimoto}, {Casey}, {Ferguson},
  {Koch}, {McConnachie}, and {Tanvir}]{2013ApJ...768..172C}
{Collins}, M.L.M.; {Chapman}, S.C.; {Rich}, R.M.; {Ibata}, R.A.; {Martin},
  N.F.; {Irwin}, M.J.; {Bate}, N.F.; {Lewis}, G.F.; {Pe{\~n}arrubia}, J.;
  {Arimoto}, N.;  et~al.
\newblock {A Kinematic Study of the Andromeda Dwarf Spheroidal System}.
\newblock {\em \apj} {\bf 2013}, {\em 768},~172,
  \href{http://arxiv.org/abs/1302.6590}{{\normalfont
  [arXiv:astro-ph.CO/1302.6590]}}.
\newblock {\url{https://doi.org/10.1088/0004-637X/768/2/172}}.

\bibitem[{Alam} et~al.(2015){Alam}, {Albareti}, {Allende Prieto}, {Anders},
  {Anderson}, {Anderton}, {Andrews}, {Armengaud}, {Aubourg}, {Bailey}, {Basu},
  {Bautista}, {Beaton}, {Beers}, {Bender}, {Berlind}, {Beutler}, {Bhardwaj},
  {Bird}, {Bizyaev}, {Blake}, {Blanton}, {Blomqvist}, {Bochanski}, {Bolton},
  {Bovy}, {Shelden Bradley}, {Brandt}, {Brauer}, {Brinkmann}, {Brown},
  {Brownstein}, {Burden}, {Burtin}, {Busca}, {Cai}, {Capozzi}, {Carnero
  Rosell}, {Carr}, {Carrera}, {Chambers}, {Chaplin}, {Chen}, {Chiappini},
  {Chojnowski}, {Chuang}, {Clerc}, {Comparat}, {Covey}, {Croft}, {Cuesta},
  {Cunha}, {da Costa}, {Da Rio}, {Davenport}, {Dawson}, {De Lee}, {Delubac},
  {Deshpande}, {Dhital}, {Dutra-Ferreira}, {Dwelly}, {Ealet}, {Ebelke},
  {Edmondson}, {Eisenstein}, {Ellsworth}, {Elsworth}, {Epstein}, {Eracleous},
  {Escoffier}, {Esposito}, {Evans}, {Fan}, {Fern{\'a}ndez-Alvar}, {Feuillet},
  {Filiz Ak}, {Finley}, {Finoguenov}, {Flaherty}, {Fleming}, {Font-Ribera},
  {Foster}, {Frinchaboy}, {Galbraith-Frew}, {Garc{\'\i}a},
  {Garc{\'\i}a-Hern{\'a}ndez}, {Garc{\'\i}a P{\'e}rez}, {Gaulme}, {Ge},
  {G{\'e}nova-Santos}, {Georgakakis}, {Ghezzi}, {Gillespie}, {Girardi},
  {Goddard}, {Gontcho}, {Gonz{\'a}lez Hern{\'a}ndez}, {Grebel}, {Green},
  {Grieb}, {Grieves}, {Gunn}, {Guo}, {Harding}, {Hasselquist}, {Hawley},
  {Hayden}, {Hearty}, {Hekker}, {Ho}, {Hogg}, {Holley-Bockelmann}, {Holtzman},
  {Honscheid}, {Huber}, {Huehnerhoff}, {Ivans}, {Jiang}, {Johnson},
  {Kinemuchi}, {Kirkby}, {Kitaura}, {Klaene}, {Knapp}, {Kneib}, {Koenig},
  {Lam}, {Lan}, {Lang}, {Laurent}, {Le Goff}, {Leauthaud}, {Lee}, {Lee},
  {Licquia}, {Liu}, {Long}, {L{\'o}pez-Corredoira}, {Lorenzo-Oliveira},
  {Lucatello}, {Lundgren}, {Lupton}, {Mack}, {Mahadevan}, {Maia}, {Majewski},
  {Malanushenko}, {Malanushenko}, {Manchado}, {Manera}, {Mao}, {Maraston},
  {Marchwinski}, {Margala}, {Martell}, {Martig}, {Masters}, {Mathur},
  {McBride}, {McGehee}, {McGreer}, {McMahon}, {M{\'e}nard}, {Menzel},
  {Merloni}, {M{\'e}sz{\'a}ros}, {Miller}, {Miralda-Escud{\'e}}, {Miyatake},
  {Montero-Dorta}, {More}, {Morganson}, {Morice-Atkinson}, {Morrison},
  {Mosser}, {Muna}, {Myers}, {Nandra}, {Newman}, {Neyrinck}, {Nguyen},
  {Nichol}, {Nidever}, {Noterdaeme}, {Nuza}, {O'Connell}, {O'Connell},
  {O'Connell}, {Ogando}, {Olmstead}, {Oravetz}, {Oravetz}, {Osumi}, {Owen},
  {Padgett}, {Padmanabhan}, {Paegert}, {Palanque-Delabrouille}, and
  {Pan}]{2015ApJS..219...12A}
{Alam}, S.; {Albareti}, F.D.; {Allende Prieto}, C.; {Anders}, F.; {Anderson},
  S.F.; {Anderton}, T.; {Andrews}, B.H.; {Armengaud}, E.; {Aubourg}, {\'E}.;
  {Bailey}, S.;  et~al.
\newblock {The Eleventh and Twelfth Data Releases of the Sloan Digital Sky
  Survey: Final Data from SDSS-III}.
\newblock {\em \apjs} {\bf 2015}, {\em 219},~12,
  \href{http://arxiv.org/abs/1501.00963}{{\normalfont
  [arXiv:astro-ph.IM/1501.00963]}}.
\newblock {\url{https://doi.org/10.1088/0067-0049/219/1/12}}.

\bibitem[{Prudil} et~al.(2022){Prudil}, {Koch-Hansen}, {Lemasle}, {Grebel},
  {Marchetti}, {Hansen}, {Crestani}, {Braga}, {Bono}, {Chaboyer}, {Fabrizio},
  {Dall'Ora}, and {Mart{\'\i}nez-V{\'a}zquez}]{2022AandA...664A.148P}
{Prudil}, Z.; {Koch-Hansen}, A.J.; {Lemasle}, B.; {Grebel}, E.K.; {Marchetti},
  T.; {Hansen}, C.J.; {Crestani}, J.; {Braga}, V.F.; {Bono}, G.; {Chaboyer},
  B.;  et~al.
\newblock {Milky Way archaeology using RR Lyrae and type II Cepheids. II.
  High-velocity RR Lyrae stars and Milky Way mass}.
\newblock {\em \aap} {\bf 2022}, {\em 664},~A148,
  \href{http://arxiv.org/abs/2206.00417}{{\normalfont
  [arXiv:astro-ph.GA/2206.00417]}}.
\newblock {\url{https://doi.org/10.1051/0004-6361/202142251}}.

\bibitem[{Roche} et~al.(2024){Roche}, {Necib}, {Lin}, {Ou}, and
  {Nguyen}]{2024ApJ...972...70R}
{Roche}, C.; {Necib}, L.; {Lin}, T.; {Ou}, X.; {Nguyen}, T.
\newblock {The Escape Velocity Profile of the Milky Way from Gaia DR3}.
\newblock {\em \apj} {\bf 2024}, {\em 972},~70,
  \href{http://arxiv.org/abs/2402.00108}{{\normalfont
  [arXiv:astro-ph.GA/2402.00108]}}.
\newblock {\url{https://doi.org/10.3847/1538-4357/ad58d7}}.

\bibitem[{McMillan}(2011)]{2011MNRAS.414.2446M}
{McMillan}, P.J.
\newblock {Mass models of the Milky Way}.
\newblock {\em \mnras} {\bf 2011}, {\em 414},~2446--2457,
  \href{http://arxiv.org/abs/1102.4340}{{\normalfont
  [arXiv:astro-ph.GA/1102.4340]}}.
\newblock {\url{https://doi.org/10.1111/j.1365-2966.2011.18564.x}}.

\bibitem[{Bovy} et~al.(2012){Bovy}, {Allende Prieto}, {Beers}, {Bizyaev}, {da
  Costa}, {Cunha}, {Ebelke}, {Eisenstein}, {Frinchaboy}, {Garc{\'\i}a
  P{\'e}rez}, {Girardi}, {Hearty}, {Hogg}, {Holtzman}, {Maia}, {Majewski},
  {Malanushenko}, {Malanushenko}, {M{\'e}sz{\'a}ros}, {Nidever}, {O'Connell},
  {O'Donnell}, {Oravetz}, {Pan}, {Rocha-Pinto}, {Schiavon}, {Schneider},
  {Schultheis}, {Skrutskie}, {Smith}, {Weinberg}, {Wilson}, and
  {Zasowski}]{2012ApJ...759..131B}
{Bovy}, J.; {Allende Prieto}, C.; {Beers}, T.C.; {Bizyaev}, D.; {da Costa},
  L.N.; {Cunha}, K.; {Ebelke}, G.L.; {Eisenstein}, D.J.; {Frinchaboy}, P.M.;
  {Garc{\'\i}a P{\'e}rez}, A.E.;  et~al.
\newblock {The Milky Way's Circular-velocity Curve between 4 and 14 kpc from
  APOGEE data}.
\newblock {\em \apj} {\bf 2012}, {\em 759},~131,
  \href{http://arxiv.org/abs/1209.0759}{{\normalfont
  [arXiv:astro-ph.GA/1209.0759]}}.
\newblock {\url{https://doi.org/10.1088/0004-637X/759/2/131}}.

\bibitem[{Huang} et~al.(2016){Huang}, {Liu}, {Yuan}, {Xiang}, {Zhang}, {Chen},
  {Ren}, {Wang}, {Zhang}, {Hou}, {Wang}, and {Cao}]{2016MNRAS.463.2623H}
{Huang}, Y.; {Liu}, X.W.; {Yuan}, H.B.; {Xiang}, M.S.; {Zhang}, H.W.; {Chen},
  B.Q.; {Ren}, J.J.; {Wang}, C.; {Zhang}, Y.; {Hou}, Y.H.;  et~al.
\newblock {The Milky Way's rotation curve out to 100 kpc and its constraint on
  the Galactic mass distribution}.
\newblock {\em \mnras} {\bf 2016}, {\em 463},~2623--2639,
  \href{http://arxiv.org/abs/1604.01216}{{\normalfont
  [arXiv:astro-ph.GA/1604.01216]}}.
\newblock {\url{https://doi.org/10.1093/mnras/stw2096}}.

\bibitem[{Eilers} et~al.(2019){Eilers}, {Hogg}, {Rix}, and
  {Ness}]{2019ApJ...871..120E}
{Eilers}, A.C.; {Hogg}, D.W.; {Rix}, H.W.; {Ness}, M.K.
\newblock {The Circular Velocity Curve of the Milky Way from 5 to 25 kpc}.
\newblock {\em \apj} {\bf 2019}, {\em 871},~120,
  \href{http://arxiv.org/abs/1810.09466}{{\normalfont
  [arXiv:astro-ph.GA/1810.09466]}}.
\newblock {\url{https://doi.org/10.3847/1538-4357/aaf648}}.

\bibitem[{Cautun} et~al.(2020){Cautun}, {Ben{\'\i}tez-Llambay}, {Deason},
  {Frenk}, {Fattahi}, {G{\'o}mez}, {Grand}, {Oman}, {Navarro}, and
  {Simpson}]{2020MNRAS.494.4291C}
{Cautun}, M.; {Ben{\'\i}tez-Llambay}, A.; {Deason}, A.J.; {Frenk}, C.S.;
  {Fattahi}, A.; {G{\'o}mez}, F.A.; {Grand}, R.J.J.; {Oman}, K.A.; {Navarro},
  J.F.; {Simpson}, C.M.
\newblock {The milky way total mass profile as inferred from Gaia DR2}.
\newblock {\em \mnras} {\bf 2020}, {\em 494},~4291--4313,
  \href{http://arxiv.org/abs/1911.04557}{{\normalfont
  [arXiv:astro-ph.GA/1911.04557]}}.
\newblock {\url{https://doi.org/10.1093/mnras/staa1017}}.

\bibitem[{Ablimit} et~al.(2020){Ablimit}, {Zhao}, {Flynn}, and
  {Bird}]{2020ApJ...895L..12A}
{Ablimit}, I.; {Zhao}, G.; {Flynn}, C.; {Bird}, S.A.
\newblock {The Rotation Curve, Mass Distribution, and Dark Matter Content of
  the Milky Way from Classical Cepheids}.
\newblock {\em \apjl} {\bf 2020}, {\em 895},~L12,
  \href{http://arxiv.org/abs/2004.13768}{{\normalfont
  [arXiv:astro-ph.GA/2004.13768]}}.
\newblock {\url{https://doi.org/10.3847/2041-8213/ab8d45}}.

\bibitem[{Sylos Labini} et~al.(2023){Sylos Labini}, {Chrob{\'a}kov{\'a}},
  {Capuzzo-Dolcetta}, and {L{\'o}pez-Corredoira}]{2023ApJ...945....3S}
{Sylos Labini}, F.; {Chrob{\'a}kov{\'a}}, {\v{Z}}.; {Capuzzo-Dolcetta}, R.;
  {L{\'o}pez-Corredoira}, M.
\newblock {Mass Models of the Milky Way and Estimation of Its Mass from the
  Gaia DR3 Data Set}.
\newblock {\em \apj} {\bf 2023}, {\em 945},~3,
  \href{http://arxiv.org/abs/2302.01379}{{\normalfont
  [arXiv:astro-ph.GA/2302.01379]}}.
\newblock {\url{https://doi.org/10.3847/1538-4357/acb92c}}.

\bibitem[{Kla{\v{c}}ka} et~al.(2024){Kla{\v{c}}ka}, {{\v{S}}turc}, and
  {Puha}]{2024arXiv240712551K}
{Kla{\v{c}}ka}, J.; {{\v{S}}turc}, M.; {Puha}, E.
\newblock {Milky Way: New Galactic mass model for orbit computations}.
\newblock {\em arXiv e-prints} {\bf 2024}, p. arXiv:2407.12551,
  \href{http://arxiv.org/abs/2407.12551}{{\normalfont
  [arXiv:astro-ph.GA/2407.12551]}}.
\newblock {\url{https://doi.org/10.48550/arXiv.2407.12551}}.

\bibitem[{Craig} et~al.(2022){Craig}, {Chakrabarti}, {Baum}, and
  {Lewis}]{2022MNRAS.517.1737C}
{Craig}, P.A.; {Chakrabarti}, S.; {Baum}, S.; {Lewis}, B.T.
\newblock {An estimate of the mass of the Milky Way from the Magellanic
  Stream}.
\newblock {\em \mnras} {\bf 2022}, {\em 517},~1737--1749.
\newblock {\url{https://doi.org/10.1093/mnras/stac2308}}.

\bibitem[{Gnedin} et~al.(2010){Gnedin}, {Brown}, {Geller}, and
  {Kenyon}]{2010ApJ...720L.108G}
{Gnedin}, O.Y.; {Brown}, W.R.; {Geller}, M.J.; {Kenyon}, S.J.
\newblock {The Mass Profile of the Galaxy to 80 kpc}.
\newblock {\em \apjl} {\bf 2010}, {\em 720},~L108--L112,
  \href{http://arxiv.org/abs/1005.2619}{{\normalfont
  [arXiv:astro-ph.GA/1005.2619]}}.
\newblock {\url{https://doi.org/10.1088/2041-8205/720/1/L108}}.

\bibitem[{Kafle} et~al.(2012){Kafle}, {Sharma}, {Lewis}, and
  {Bland-Hawthorn}]{2012ApJ...761...98K}
{Kafle}, P.R.; {Sharma}, S.; {Lewis}, G.F.; {Bland-Hawthorn}, J.
\newblock {Kinematics of the Stellar Halo and the Mass Distribution of the
  Milky Way Using Blue Horizontal Branch Stars}.
\newblock {\em \apj} {\bf 2012}, {\em 761},~98,
  \href{http://arxiv.org/abs/1210.7527}{{\normalfont
  [arXiv:astro-ph.GA/1210.7527]}}.
\newblock {\url{https://doi.org/10.1088/0004-637X/761/2/98}}.

\bibitem[{Kafle} et~al.(2014){Kafle}, {Sharma}, {Lewis}, and
  {Bland-Hawthorn}]{2014ApJ...794...59K}
{Kafle}, P.R.; {Sharma}, S.; {Lewis}, G.F.; {Bland-Hawthorn}, J.
\newblock {On the Shoulders of Giants: Properties of the Stellar Halo and the
  Milky Way Mass Distribution}.
\newblock {\em \apj} {\bf 2014}, {\em 794},~59,
  \href{http://arxiv.org/abs/1408.1787}{{\normalfont
  [arXiv:astro-ph.GA/1408.1787]}}.
\newblock {\url{https://doi.org/10.1088/0004-637X/794/1/59}}.

\bibitem[{Zhai} et~al.(2018){Zhai}, {Xue}, {Zhang}, {Li}, {Zhao}, and
  {Yang}]{2018RAA....18..113Z}
{Zhai}, M.; {Xue}, X.X.; {Zhang}, L.; {Li}, C.D.; {Zhao}, G.; {Yang}, C.Q.
\newblock {The mass of the Galactic dark matter halo from
  {\ensuremath{\sim}}9000 LAMOST DR5 K giants}.
\newblock {\em Research in Astronomy and Astrophysics} {\bf 2018}, {\em
  18},~113.
\newblock {\url{https://doi.org/10.1088/1674-4527/18/9/113}}.

\bibitem[{Bird} et~al.(2022){Bird}, {Xue}, {Liu}, {Flynn}, {Shen}, {Wang},
  {Yang}, {Zhai}, {Zhu}, {Zhao}, and {Tian}]{2022MNRAS.516..731B}
{Bird}, S.A.; {Xue}, X.X.; {Liu}, C.; {Flynn}, C.; {Shen}, J.; {Wang}, J.;
  {Yang}, C.; {Zhai}, M.; {Zhu}, L.; {Zhao}, G.;  et~al.
\newblock {Milky Way mass with K giants and BHB stars using LAMOST, SDSS/SEGUE,
  and Gaia: 3D spherical Jeans equation and tracer mass estimator}.
\newblock {\em \mnras} {\bf 2022}, {\em 516},~731--748,
  \href{http://arxiv.org/abs/2207.08839}{{\normalfont
  [arXiv:astro-ph.GA/2207.08839]}}.
\newblock {\url{https://doi.org/10.1093/mnras/stac2036}}.

\bibitem[{Eadie} et~al.(2015){Eadie}, {Harris}, and
  {Widrow}]{2015ApJ...806...54E}
{Eadie}, G.M.; {Harris}, W.E.; {Widrow}, L.M.
\newblock {Estimating the Galactic Mass Profile in the Presence of Incomplete
  Data}.
\newblock {\em \apj} {\bf 2015}, {\em 806},~54,
  \href{http://arxiv.org/abs/1503.07176}{{\normalfont
  [arXiv:astro-ph.GA/1503.07176]}}.
\newblock {\url{https://doi.org/10.1088/0004-637X/806/1/54}}.

\bibitem[{Eadie} and {Harris}(2016)]{2016ApJ...829..108E}
{Eadie}, G.M.; {Harris}, W.E.
\newblock {Bayesian Mass Estimates of the Milky Way: The Dark and Light Sides
  of Parameter Assumptions}.
\newblock {\em \apj} {\bf 2016}, {\em 829},~108,
  \href{http://arxiv.org/abs/1608.04757}{{\normalfont
  [arXiv:astro-ph.GA/1608.04757]}}.
\newblock {\url{https://doi.org/10.3847/0004-637X/829/2/108}}.

\bibitem[{Sohn} et~al.(2018){Sohn}, {Watkins}, {Fardal}, {van der Marel},
  {Deason}, {Besla}, and {Bellini}]{2018ApJ...862...52S}
{Sohn}, S.T.; {Watkins}, L.L.; {Fardal}, M.A.; {van der Marel}, R.P.; {Deason},
  A.J.; {Besla}, G.; {Bellini}, A.
\newblock {Absolute Hubble Space Telescope Proper Motion (HSTPROMO) of Distant
  Milky Way Globular Clusters: Galactocentric Space Velocities and the Milky
  Way Mass}.
\newblock {\em \apj} {\bf 2018}, {\em 862},~52,
  \href{http://arxiv.org/abs/1804.01994}{{\normalfont
  [arXiv:astro-ph.GA/1804.01994]}}.
\newblock {\url{https://doi.org/10.3847/1538-4357/aacd0b}}.

\bibitem[{Watkins} et~al.(2019){Watkins}, {van der Marel}, {Sohn}, and
  {Evans}]{2019ApJ...873..118W}
{Watkins}, L.L.; {van der Marel}, R.P.; {Sohn}, S.T.; {Evans}, N.W.
\newblock {Evidence for an Intermediate-mass Milky Way from Gaia DR2 Halo
  Globular Cluster Motions}.
\newblock {\em \apj} {\bf 2019}, {\em 873},~118,
  \href{http://arxiv.org/abs/1804.11348}{{\normalfont
  [arXiv:astro-ph.GA/1804.11348]}}.
\newblock {\url{https://doi.org/10.3847/1538-4357/ab089f}}.

\bibitem[{Posti} and {Helmi}(2019)]{2019AandA...621A..56P}
{Posti}, L.; {Helmi}, A.
\newblock {Mass and shape of the Milky Way's dark matter halo with globular
  clusters from Gaia and Hubble}.
\newblock {\em \aap} {\bf 2019}, {\em 621},~A56,
  \href{http://arxiv.org/abs/1805.01408}{{\normalfont
  [arXiv:astro-ph.GA/1805.01408]}}.
\newblock {\url{https://doi.org/10.1051/0004-6361/201833355}}.

\bibitem[{Li} et~al.(2020){Li}, {Qian}, {Han}, {Li}, {Wang}, and
  {Jing}]{2020ApJ...894...10L}
{Li}, Z.Z.; {Qian}, Y.Z.; {Han}, J.; {Li}, T.S.; {Wang}, W.; {Jing}, Y.P.
\newblock {Constraining the Milky Way Mass Profile with Phase-space
  Distribution of Satellite Galaxies}.
\newblock {\em \apj} {\bf 2020}, {\em 894},~10,
  \href{http://arxiv.org/abs/1912.02086}{{\normalfont
  [arXiv:astro-ph.GA/1912.02086]}}.
\newblock {\url{https://doi.org/10.3847/1538-4357/ab84f0}}.

\bibitem[{Deason} et~al.(2021){Deason}, {Erkal}, {Belokurov}, {Fattahi},
  {G{\'o}mez}, {Grand}, {Pakmor}, {Xue}, {Liu}, {Yang}, {Zhang}, and
  {Zhao}]{2021MNRAS.501.5964D}
{Deason}, A.J.; {Erkal}, D.; {Belokurov}, V.; {Fattahi}, A.; {G{\'o}mez}, F.A.;
  {Grand}, R.J.J.; {Pakmor}, R.; {Xue}, X.X.; {Liu}, C.; {Yang}, C.;  et~al.
\newblock {The mass of the Milky Way out to 100 kpc using halo stars}.
\newblock {\em \mnras} {\bf 2021}, {\em 501},~5964--5972,
  \href{http://arxiv.org/abs/2010.13801}{{\normalfont
  [arXiv:astro-ph.GA/2010.13801]}}.
\newblock {\url{https://doi.org/10.1093/mnras/staa3984}}.

\bibitem[{Shen} et~al.(2022){Shen}, {Eadie}, {Murray}, {Zaritsky}, {Speagle},
  {Ting}, {Conroy}, {Cargile}, {Johnson}, {Naidu}, and
  {Han}]{2022ApJ...925....1S}
{Shen}, J.; {Eadie}, G.M.; {Murray}, N.; {Zaritsky}, D.; {Speagle}, J.S.;
  {Ting}, Y.S.; {Conroy}, C.; {Cargile}, P.A.; {Johnson}, B.D.; {Naidu}, R.P.;
  et~al.
\newblock {The Mass of the Milky Way from the H3 Survey}.
\newblock {\em \apj} {\bf 2022}, {\em 925},~1,
  \href{http://arxiv.org/abs/2111.09327}{{\normalfont
  [arXiv:astro-ph.GA/2111.09327]}}.
\newblock {\url{https://doi.org/10.3847/1538-4357/ac3a7a}}.

\bibitem[{Busha} et~al.(2011){Busha}, {Marshall}, {Wechsler}, {Klypin}, and
  {Primack}]{2011ApJ...743...40B}
{Busha}, M.T.; {Marshall}, P.J.; {Wechsler}, R.H.; {Klypin}, A.; {Primack}, J.
\newblock {The Mass Distribution and Assembly of the Milky Way from the
  Properties of the Magellanic Clouds}.
\newblock {\em \apj} {\bf 2011}, {\em 743},~40,
  \href{http://arxiv.org/abs/1011.2203}{{\normalfont
  [arXiv:astro-ph.GA/1011.2203]}}.
\newblock {\url{https://doi.org/10.1088/0004-637X/743/1/40}}.

\bibitem[{Gonz{\'a}lez} et~al.(2013){Gonz{\'a}lez}, {Kravtsov}, and
  {Gnedin}]{2013ApJ...770...96G}
{Gonz{\'a}lez}, R.E.; {Kravtsov}, A.V.; {Gnedin}, N.Y.
\newblock {Satellites in Milky-Way-like Hosts: Environment Dependence and Close
  Pairs}.
\newblock {\em \apj} {\bf 2013}, {\em 770},~96,
  \href{http://arxiv.org/abs/1301.2605}{{\normalfont
  [arXiv:astro-ph.CO/1301.2605]}}.
\newblock {\url{https://doi.org/10.1088/0004-637X/770/2/96}}.

\bibitem[{Cautun} et~al.(2014){Cautun}, {Frenk}, {van de Weygaert}, {Hellwing},
  and {Jones}]{2014MNRAS.445.2049C}
{Cautun}, M.; {Frenk}, C.S.; {van de Weygaert}, R.; {Hellwing}, W.A.; {Jones},
  B.J.T.
\newblock {Milky Way mass constraints from the Galactic satellite gap}.
\newblock {\em \mnras} {\bf 2014}, {\em 445},~2049--2060,
  \href{http://arxiv.org/abs/1405.7697}{{\normalfont
  [arXiv:astro-ph.CO/1405.7697]}}.
\newblock {\url{https://doi.org/10.1093/mnras/stu1849}}.

\bibitem[{Barber} et~al.(2014){Barber}, {Starkenburg}, {Navarro},
  {McConnachie}, and {Fattahi}]{2014MNRAS.437..959B}
{Barber}, C.; {Starkenburg}, E.; {Navarro}, J.F.; {McConnachie}, A.W.;
  {Fattahi}, A.
\newblock {The orbital ellipticity of satellite galaxies and the mass of the
  Milky Way}.
\newblock {\em \mnras} {\bf 2014}, {\em 437},~959--967,
  \href{http://arxiv.org/abs/1310.0466}{{\normalfont
  [arXiv:astro-ph.GA/1310.0466]}}.
\newblock {\url{https://doi.org/10.1093/mnras/stt1959}}.

\bibitem[{Patel} et~al.(2017){Patel}, {Besla}, and {Sohn}]{2017MNRAS.464.3825P}
{Patel}, E.; {Besla}, G.; {Sohn}, S.T.
\newblock {Orbits of massive satellite galaxies - I. A close look at the Large
  Magellanic Cloud and a new orbital history for M33}.
\newblock {\em \mnras} {\bf 2017}, {\em 464},~3825--3849,
  \href{http://arxiv.org/abs/1609.04823}{{\normalfont
  [arXiv:astro-ph.GA/1609.04823]}}.
\newblock {\url{https://doi.org/10.1093/mnras/stw2616}}.

\bibitem[{Patel} et~al.(2018){Patel}, {Besla}, {Mandel}, and
  {Sohn}]{2018ApJ...857...78P}
{Patel}, E.; {Besla}, G.; {Mandel}, K.; {Sohn}, S.T.
\newblock {Estimating the Mass of the Milky Way Using the Ensemble of Classical
  Satellite Galaxies}.
\newblock {\em \apj} {\bf 2018}, {\em 857},~78,
  \href{http://arxiv.org/abs/1803.01878}{{\normalfont
  [arXiv:astro-ph.GA/1803.01878]}}.
\newblock {\url{https://doi.org/10.3847/1538-4357/aab78f}}.

\bibitem[{Fritz} et~al.(2020){Fritz}, {Di Cintio}, {Battaglia}, {Brook}, and
  {Taibi}]{2020MNRAS.494.5178F}
{Fritz}, T.K.; {Di Cintio}, A.; {Battaglia}, G.; {Brook}, C.; {Taibi}, S.
\newblock {The mass of our Galaxy from satellite proper motions in the Gaia
  era}.
\newblock {\em \mnras} {\bf 2020}, {\em 494},~5178--5193,
  \href{http://arxiv.org/abs/2001.02651}{{\normalfont
  [arXiv:astro-ph.GA/2001.02651]}}.
\newblock {\url{https://doi.org/10.1093/mnras/staa1040}}.

\bibitem[{Rodriguez Wimberly} et~al.(2022){Rodriguez Wimberly}, {Cooper},
  {Baxter}, {Boylan-Kolchin}, {Bullock}, {Fillingham}, {Ji}, {Sales}, and
  {Simon}]{2022MNRAS.513.4968R}
{Rodriguez Wimberly}, M.K.; {Cooper}, M.C.; {Baxter}, D.C.; {Boylan-Kolchin},
  M.; {Bullock}, J.S.; {Fillingham}, S.P.; {Ji}, A.P.; {Sales}, L.V.; {Simon},
  J.D.
\newblock {Sizing from the smallest scales: the mass of the Milky Way}.
\newblock {\em \mnras} {\bf 2022}, {\em 513},~4968--4982,
  \href{http://arxiv.org/abs/2109.00633}{{\normalfont
  [arXiv:astro-ph.GA/2109.00633]}}.
\newblock {\url{https://doi.org/10.1093/mnras/stac1265}}.

\bibitem[{Tamm} et~al.(2012){Tamm}, {Tempel}, {Tenjes}, {Tihhonova}, and
  {Tuvikene}]{2012AandA...546A...4T}
{Tamm}, A.; {Tempel}, E.; {Tenjes}, P.; {Tihhonova}, O.; {Tuvikene}, T.
\newblock {Stellar mass map and dark matter distribution in M 31}.
\newblock {\em \aap} {\bf 2012}, {\em 546},~A4,
  \href{http://arxiv.org/abs/1208.5712}{{\normalfont
  [arXiv:astro-ph.CO/1208.5712]}}.
\newblock {\url{https://doi.org/10.1051/0004-6361/201220065}}.

\bibitem[{Hayashi} and {Chiba}(2014)]{2014ApJ...789...62H}
{Hayashi}, K.; {Chiba}, M.
\newblock {The Prolate Dark Matter Halo of the Andromeda Galaxy}.
\newblock {\em \apj} {\bf 2014}, {\em 789},~62,
  \href{http://arxiv.org/abs/1405.4606}{{\normalfont
  [arXiv:astro-ph.GA/1405.4606]}}.
\newblock {\url{https://doi.org/10.1088/0004-637X/789/1/62}}.

\bibitem[{Sofue}(2015)]{2015PASJ...67...75S}
{Sofue}, Y.
\newblock {Dark halos of M 31 and the Milky Way}.
\newblock {\em \pasj} {\bf 2015}, {\em 67},~75,
  \href{http://arxiv.org/abs/1504.05368}{{\normalfont
  [arXiv:astro-ph.GA/1504.05368]}}.
\newblock {\url{https://doi.org/10.1093/pasj/psv042}}.

\bibitem[{Zhang} et~al.(2024){Zhang}, {Chen}, {Chen}, {Sun}, and
  {Tian}]{2024MNRAS.528.2653Z}
{Zhang}, X.; {Chen}, B.; {Chen}, P.; {Sun}, J.; {Tian}, Z.
\newblock {The rotation curve and mass distribution of M31}.
\newblock {\em \mnras} {\bf 2024}, {\em 528},~2653--2666,
  \href{http://arxiv.org/abs/2401.01517}{{\normalfont
  [arXiv:astro-ph.GA/2401.01517]}}.
\newblock {\url{https://doi.org/10.1093/mnras/stae025}}.

\bibitem[{Fardal} et~al.(2013){Fardal}, {Weinberg}, {Babul}, {Irwin},
  {Guhathakurta}, {Gilbert}, {Ferguson}, {Ibata}, {Lewis}, {Tanvir}, and
  {Huxor}]{2013MNRAS.434.2779F}
{Fardal}, M.A.; {Weinberg}, M.D.; {Babul}, A.; {Irwin}, M.J.; {Guhathakurta},
  P.; {Gilbert}, K.M.; {Ferguson}, A.M.N.; {Ibata}, R.A.; {Lewis}, G.F.;
  {Tanvir}, N.R.;  et~al.
\newblock {Inferring the Andromeda Galaxy's mass from its giant southern stream
  with Bayesian simulation sampling}.
\newblock {\em \mnras} {\bf 2013}, {\em 434},~2779--2802,
  \href{http://arxiv.org/abs/1307.3219}{{\normalfont
  [arXiv:astro-ph.CO/1307.3219]}}.
\newblock {\url{https://doi.org/10.1093/mnras/stt1121}}.

\bibitem[{Veljanoski} et~al.(2013){Veljanoski}, {Ferguson}, {Mackey}, {Huxor},
  {Irwin}, {C{\^o}t{\'e}}, {Tanvir}, {Bernard}, {Chapman}, {Ibata}, {Fardal},
  {Lewis}, {Martin}, {McConnachie}, and {Pe{\~n}arrubia}]{2013ApJ...768L..33V}
{Veljanoski}, J.; {Ferguson}, A.M.N.; {Mackey}, A.D.; {Huxor}, A.P.; {Irwin},
  M.J.; {C{\^o}t{\'e}}, P.; {Tanvir}, N.R.; {Bernard}, E.J.; {Chapman}, S.C.;
  {Ibata}, R.A.;  et~al.
\newblock {Kinematics of Outer Halo Globular Clusters in M31}.
\newblock {\em \apjl} {\bf 2013}, {\em 768},~L33,
  \href{http://arxiv.org/abs/1303.7368}{{\normalfont
  [arXiv:astro-ph.GA/1303.7368]}}.
\newblock {\url{https://doi.org/10.1088/2041-8205/768/2/L33}}.

\bibitem[{Patel} et~al.(2017){Patel}, {Besla}, and
  {Mandel}]{2017MNRAS.468.3428P}
{Patel}, E.; {Besla}, G.; {Mandel}, K.
\newblock {Orbits of massive satellite galaxies - II. Bayesian estimates of the
  Milky Way and Andromeda masses using high-precision astrometry and
  cosmological simulations}.
\newblock {\em \mnras} {\bf 2017}, {\em 468},~3428--3449,
  \href{http://arxiv.org/abs/1703.05767}{{\normalfont
  [arXiv:astro-ph.GA/1703.05767]}}.
\newblock {\url{https://doi.org/10.1093/mnras/stx698}}.

\bibitem[{van der Marel} et~al.(2012){van der Marel}, {Fardal}, {Besla},
  {Beaton}, {Sohn}, {Anderson}, {Brown}, and
  {Guhathakurta}]{2012ApJ...753....8V}
{van der Marel}, R.P.; {Fardal}, M.; {Besla}, G.; {Beaton}, R.L.; {Sohn}, S.T.;
  {Anderson}, J.; {Brown}, T.; {Guhathakurta}, P.
\newblock {The M31 Velocity Vector. II. Radial Orbit toward the Milky Way and
  Implied Local Group Mass}.
\newblock {\em \apj} {\bf 2012}, {\em 753},~8,
  \href{http://arxiv.org/abs/1205.6864}{{\normalfont
  [arXiv:astro-ph.GA/1205.6864]}}.
\newblock {\url{https://doi.org/10.1088/0004-637X/753/1/8}}.

\bibitem[{Diaz} et~al.(2014){Diaz}, {Koposov}, {Irwin}, {Belokurov}, and
  {Evans}]{2014MNRAS.443.1688D}
{Diaz}, J.D.; {Koposov}, S.E.; {Irwin}, M.; {Belokurov}, V.; {Evans}, N.W.
\newblock {Balancing mass and momentum in the Local Group}.
\newblock {\em \mnras} {\bf 2014}, {\em 443},~1688--1703,
  \href{http://arxiv.org/abs/1405.3662}{{\normalfont
  [arXiv:astro-ph.GA/1405.3662]}}.
\newblock {\url{https://doi.org/10.1093/mnras/stu1210}}.

\bibitem[{Pe{\~n}arrubia} et~al.(2016){Pe{\~n}arrubia}, {G{\'o}mez}, {Besla},
  {Erkal}, and {Ma}]{2016MNRAS.456L..54P}
{Pe{\~n}arrubia}, J.; {G{\'o}mez}, F.A.; {Besla}, G.; {Erkal}, D.; {Ma}, Y.Z.
\newblock {A timing constraint on the (total) mass of the Large Magellanic
  Cloud}.
\newblock {\em \mnras} {\bf 2016}, {\em 456},~L54--L58,
  \href{http://arxiv.org/abs/1507.03594}{{\normalfont
  [arXiv:astro-ph.GA/1507.03594]}}.
\newblock {\url{https://doi.org/10.1093/mnrasl/slv160}}.

\end{thebibliography}
%\end{adjustwidth}

%---------------------------------------------------
\clearpage
\appendixstart
\appendix

%\renewcommand{\tabcolsep}{6.3pt}
%\renewcommand\arraystretch{1.5}

%\onecolumn

\section{List of satellites of the Milky Way}
\label{sec:MW_datatable}

{
\begin{ThreePartTable}
\footnotesize

\begin{longtable}{lcl@{~}l@{~}llS@{}lc@{}S@{}l@{~}l}

\caption{Known satellites of the Milky Way galaxy.}
\label{tab:MW_datatable}\\
\hline\hline
Name               & J2000                 &\multicolumn{3}{c}{$(m-M)_0$} & \multicolumn{1}{c}{Method} & \multicolumn{2}{c}{$D$}    && \multicolumn{3}{c}{$\Vh$}   \\
\cline{3-5}\cline{7-8}\cline{10-12}  
                   &                       & \multicolumn{3}{c}{mag}      &                            & \multicolumn{2}{c}{kpc}    && \multicolumn{3}{c}{\kms{}} \\
\hline
\endfirsthead

\hline
Name               & J2000                 &\multicolumn{3}{c}{$(m-M)_0$} & \multicolumn{1}{c}{Method} & \multicolumn{2}{c}{$D$}    && \multicolumn{3}{c}{$\Vh$} \\
\cline{3-5}\cline{7-8}\cline{10-12}  
                   &                       & \multicolumn{3}{c}{mag}      &                            & \multicolumn{2}{c}{kpc}    && \multicolumn{3}{c}{\kms{}} \\
\hline
\endhead

\hline
\endfoot

\hline\hline
\endlastfoot

Tucana IV          & 000255.2$-$605100 & 18.41 & $\pm 0.19 $        & \cite{2015ApJ...813..109D} & TRGB   &  48.1   & $\pm 4.4 $         &&   15.9 & $^{+1.8}_{-1.7}$   & \cite{2020AJ....160..124M} \\
SMC                & 005238.0$-$724801 & 18.99 & $\pm 0.05 $        & \cite{2020AJ....160..124M} & Cep    &  62.8   & $\pm 1.5 $         &&  158   & $\pm 4  $          & \cite{2004AJ....127.2031K} \\
Sculptor           & 010009.4$-$334233 & 19.67 & $\pm 0.14 $        & \cite{2020AJ....160..124M} & TRGB   &  85.9 & $\pm 5.7 $         &&  111.4 & $\pm 0.1$          & \cite{2009AJ....137.3109W} \\
Cetus II           & 011752.8$-$172512 & 17.10 & $\pm 0.10 $        & \cite{2018ApJ...852...68C} & CMD    &  26.3 & $\pm 1.2 $                                                    \\
DELVE 2            & 015505.3$-$681511 & 19.26 & $\pm 0.10 $        & \cite{2021ApJ...910...18C} & HB     &  71.1 & $\pm 3.4 $                                                    \\
Cetus III          & 020519.4$-$041612 & 22.00 & $^{+0.20}_{-0.10}$ & \cite{2020AJ....160..124M} & HB     & 251 & $^{+24}_{-12}$                                              \\
Triangulum II      & 021317.4$+$361042 & 17.27 & $\pm 0.11 $        & \cite{2017AJ....154..267C} & CMD    &  28.4 & $\pm 1.5 $         && -381.70 & $\pm 1.10$          & \cite{2020AJ....160..124M} \\
Segue 2            & 021916.0$+$201031 & 17.70 & $\pm 0.10 $        & \cite{2020AJ....160..124M} & TRGB   &  34.7 & $\pm 1.6 $         &&  -40.20 & $\pm 0.90$          & \cite{2022ApJ...940..136P} \\
Eridanus III       & 022245.5$-$521701 & 19.70 & $\pm 0.15 $        & \cite{2020AJ....160..124M} & HB     &  87.1 & $\pm 6.2 $                                                    \\
DES J0225+0304     & 022542.4$+$030410 & 16.88 & $^{+0.06}_{-0.05}$ & \cite{2017MNRAS.468...97L} & CMD    &  23.8 & $^{+0.7}_{-0.6}$                                                    \\
Hydrus I           & 022933.4$-$784128 & 17.20 & $\pm 0.04 $        & \cite{2020AJ....160..124M} & TRGB   &  27.5 & $\pm 0.5 $         &&   80.4 & $\pm 0.6$          & \cite{2020AJ....160..124M} \\
Fornax             & 023954.7$-$343133 & 20.84 & $\pm 0.18 $        & \cite{2020AJ....160..124M} & TRGB   & 147 & $\pm 13$         &&   55.2 & $\pm 0.1$          & \cite{2009AJ....137.3109W} \\
Horologium I       & 025531.7$-$540708 & 19.50 & $\pm 0.20 $        & \cite{2020AJ....160..124M} & HB     &  79.4 & $\pm 7.7 $         &&  112.8 & $^{+2.5}_{-2.6}$   & \cite{2020AJ....160..124M} \\
Horologium II      & 031632.1$-$500105 & 19.46 & $\pm 0.20 $        & \cite{2020AJ....160..124M} & HB     &  78.0 & $\pm 7.5 $         &&  168.7 & $^{+12.9}_{-12.6}$ & \cite{2020AJ....160..124M} \\
Reticulum II       & 033542.1$-$540257 & 17.40 & $\pm 0.15 $        & \cite{2020AJ....160..124M} & HB     &  30.2 & $\pm 2.2 $         &&   64.7 & $^{+1.3}_{-0.8}$   & \cite{2020AJ....160..124M} \\
Reticulum III      & 034526.4$-$602700 & 19.81 & $\pm 0.31 $        & \cite{2020AJ....160..124M} & CMD    &  92 & $\pm 14$         &&  274.2 & $^{+7.5}_{-7.4}$   & \cite{2020AJ....160..124M} \\
Pictor I           & 044347.4$-$501659 & 20.30 & $\pm 0.15 $        & \cite{2020AJ....160..124M} & HB     & 115 & $\pm 8 $                                                    \\
LMC                & 052334.6$-$694522 & 18.50 & $\pm 0.13 $        & \cite{2000ApJ...529..745F} & Cep    &  50.1 & $\pm 3.1 $         &&  278   & $\pm 2 $           & \cite{1992ApJS...83...29S}\\
Columba I          & 053125.7$-$275727 & 21.31 & $\pm 0.11 $        & \cite{2017AJ....154..267C} & BHB    & 183 & $\pm 10 $         &&  153.7 & $^{+5}_{-4.8}$     & \cite{2020AJ....160..124M} \\
Carina             & 064136.7$-$505758 & 20.11 & $\pm 0.13 $        & \cite{2020AJ....160..124M} & TRGB   & 105 & $\pm 7 $         &&  222.9 & $\pm 0.1$          & \cite{2020AJ....160..124M} \\
Pictor II          & 064443.2$-$595360 & 18.30 & $^{+0.12}_{-0.15}$ & \cite{2020AJ....160..124M} & HB     &  45.7 & $^{+2.6}_{-3.3}$                                                   \\
Carina II          & 073625.6$-$560003 & 17.79 & $\pm 0.05 $        & \cite{2020AJ....160..124M} & HB     &  36.1 & $\pm 0.8 $         &&  477.2 & $\pm 1.2$          & \cite{2020AJ....160..124M} \\
Carina III         & 073831.2$-$560601 & 17.22 & $\pm 0.10 $        & \cite{2020AJ....160..124M} & HB     &  27.8 & $\pm 1.3 $         &&  284.6 & $^{+3.4}_{-3.1}$   & \cite{2020AJ....160..124M} \\
Ursa Major II      & 085130.0$+$630748 & 17.50 & $\pm 0.30 $        & \cite{2020AJ....160..124M} & TRGB   &  31.6 & $\pm 4.7 $         && -116.5 & $\pm 1.9$          & \cite{2020AJ....160..124M} \\
HYDRA 1            & 085536.0$+$033600 & 15.52 & $\pm 0.05 $        & \cite{2016ApJ...818...39H} & MS     &  12.7 & $\pm 0.3 $         &&   89   & $\pm 1.4$          & \cite{2016ApJ...818...39H} \\
Antlia II          & 093532.8$-$364602 & 20.6  & $\pm 0.11 $        & \cite{2019MNRAS.488.2743T} & BHB    & 132 & $\pm 7 $         &&  288.8 & $\pm 0.4$          & \cite{2021ApJ...921...32J} \\
Segue 1            & 100703.2$+$160425 & 16.8  & $\pm 0.20 $        & \cite{2020AJ....160..124M} & CMD    &  22.9 & $\pm 2.2 $         &&  208.5 & $\pm 0.9$          & \cite{2020AJ....160..124M} \\
Leo I              & 100826.9$+$121829 & 22.02 & $\pm 0.13 $        & \cite{2020AJ....160..124M} & TRGB   & 254 & $\pm 16$         &&  282.5 & $\pm 0.1$          & \cite{2020AJ....160..124M} \\
Sextans dSph       & 101303.0$-$013652 & 19.67 & $\pm 0.10 $        & \cite{2020AJ....160..124M} & TRGB   &  85.9 & $\pm 4.0 $         &&  224.2 & $\pm 0.1$          & \cite{2020AJ....160..124M} \\
Sextans II         & 102544.9$-$003752 & 20.50 & $\pm 0.20$ & \cite{2024ApJ...967..161M} & HB     & 126 & $\pm 12 $                                              \\
Ursa Major I       & 103448.8$+$515606 & 19.93 & $\pm 0.10 $        & \cite{2020AJ....160..124M} & TRGB   &  97 & $\pm 5 $         &&  -55.3 & $\pm 1.4$          & \cite{2020AJ....160..124M} \\
Willman 1          & 104921.0$+$510260 & 17.90 & $\pm 0.40 $        & \cite{2020AJ....160..124M} & CMD    &  38.0 & $\pm 7.7 $         &&  -12.8 & $\pm 1  $          & \cite{2022ApJ...940..136P} \\
Leo II             & 111329.2$+$220917 & 21.84 & $\pm 0.13 $        & \cite{2020AJ....160..124M} & TRGB   & 233 & $\pm 14$         &&   78   & $\pm 0.1$          & \cite{2020AJ....160..124M} \\
Leo V              & 113109.6$+$021312 & 21.46 & $\pm 0.16 $        & \cite{2020AJ....160..124M} & TRGB   & 196 & $\pm 15$         &&  170.9 & $^{+2.1}_{-1.9}$   & \cite{2020AJ....160..124M} \\
Leo IV             & 113257.0$+$003160 & 20.94 & $\pm 0.09 $        & \cite{2020AJ....160..124M} & HB     & 154 & $\pm 7 $         &&  132.3 & $\pm 1.4$          & \cite{2020AJ....160..124M} \\
Crater             & 113615.8$-$105240 & 20.81 & $\pm 0.12 $        & \cite{2014ApJ...786L...3L} & HB     & 145 & $\pm 8 $         &&  149.3 & $\pm 1.2$          & \cite{2015ApJ...810...56K} \\
Crater II          & 114914.4$-$182447 & 20.35 & $\pm 0.02 $        & \cite{2020AJ....160..124M} & TRGB   & 118 & $\pm 1 $         &&   89.3 & $\pm 0.3$          & \cite{2021ApJ...921...32J} \\
Virgo I            & 120009.6$+$004048 & 19.80 & $\pm 0.20 $        & \cite{2020AJ....160..124M} & HB     &  91 & $\pm 9 $                                                    \\
Hydra II           & 122142.1$-$315907 & 20.64 & $\pm 0.16 $        & \cite{2020AJ....160..124M} & TRGB   & 134 & $\pm 10$         &&  303.1 & $\pm 1.4$          & \cite{2020AJ....160..124M} \\
Coma Berenices     & 122658.4$+$235442 & 18.13 & $\pm 0.08 $        & \cite{2020AJ....160..124M} & HB     &  42.3 & $\pm 1.6  $        &&  98.1  & $\pm 0.9$          & \cite{2020AJ....160..124M} \\
Centaurus I        & 123820.4$-$405407 & 20.33 & $\pm 0.10 $        & \cite{2020ApJ...890..136M} & HB     & 116 & $\pm 6$                                                   \\
Canes Venatici II  & 125710.0$+$341915 & 21.02 & $\pm 0.06 $        & \cite{2020AJ....160..124M} & HB     & 160 & $\pm 5  $        && -129   & $\pm 1.2$          & \cite{2020AJ....160..124M} \\
Canes Venatici I   & 132803.5$+$333321 & 21.69 & $\pm 0.10 $        & \cite{2020AJ....160..124M} & TRGB   & 218 & $\pm 10 $        &&   30.9 & $\pm 0.6$          & \cite{2007ApJ...670..313S} \\
Bootes III         & 135712.0$+$264800 & 18.35 & $\pm 0.10 $        & \cite{2009ApJ...693.1118G} & HB     &  46.8 & $\pm 2.2  $        &&  197.5 & $\pm 3.8$          & \cite{2012AstBu..67..115K} \\
Bootes II          & 135808.0$+$125054 & 18.10 & $\pm 0.06 $        & \cite{2020AJ....160..124M} & TRGB   &  41.7 & $\pm 1.2  $        && -117   & $\pm 5.2$          & \cite{2020AJ....160..124M} \\
Bootes I           & 140005.0$+$143015 & 19.11 & $\pm 0.08 $        & \cite{2020AJ....160..124M} & HB     &  66.4 & $\pm 2.5  $        &&  101.8 & $\pm 0.7$          & \cite{2022ApJ...940..136P} \\
Ursa Minor         & 150911.3$+$671252 & 19.40 & $\pm 0.10 $        & \cite{2020AJ....160..124M} & TRGB   &  75.9 & $\pm 3.6  $        && -246.9 & $\pm 0.1$          & \cite{2020AJ....160..124M} \\
Bootes IV          & 153445.4$+$434334 & 21.60 & $\pm 0.20 $        & \cite{2020AJ....160..124M} & HB     & 209 & $\pm 20 $                                                   \\
Draco II           & 155247.6$+$643355 & 16.67 & $\pm 0.05 $        & \cite{2020AJ....160..124M} & TRGB   &  21.6 & $\pm 0.5 $        &&  342.5 & $^{+1.1}_{-1.2}$   & \cite{2020AJ....160..124M} \\
DELVE 1            & 163054.0$+$005819 & 16.39 & $\pm 0.10 $        & \cite{2020ApJ...890..136M} & HB     &  19.0 & $\pm 0.9 $                                            \\
Hercules           & 163103.6$+$124724 & 20.84 & $\pm 0.20 $        & \cite{2020AJ....160..124M} & TRGB   & 147 & $\pm 14 $        &&   46.4 & $\pm 1.3$          & \cite{2020MNRAS.496.1092G} \\
Draco              & 172001.4$+$575434 & 19.40 & $\pm 0.17 $        & \cite{2020AJ....160..124M} & TRGB   &  76 & $\pm 6  $        && -291   & $\pm 0.1$          & \cite{2020AJ....160..124M} \\
\textbf{Milky Way} & 174540.0$-$290028 & 14.55 & $\pm 0.01 $        & \cite{2018AA...615L..15G}  & Direct &   8.1 & $\pm 0.0$        &&   -9.5 & $\pm 0.0$          & \cite{2024MNRAS.530..710A} \\
Sagittarius dSph   & 185503.1$-$302842 & 17.10 & $\pm 0.15 $        & \cite{2004MNRAS.353..874M} & TRGB   &  26.3 & $\pm 1.9  $        &&  140   & $\pm 2  $          & \cite{2014MNRAS.439.1015K} \\
Sagittarius II     & 195240.5$-$220405 & 19.32 & $^{+0.03}_{-0.02}$ & \cite{2020AJ....160..124M} & CMD    &  73.1 & $^{+1.0}_{-0.7}$       && -177.2 & $^{+0.5}_{-0.6}$   & \cite{2021MNRAS.503.2754L} \\
Indus II           & 203852.8$-$460936 & 21.65 & $\pm 0.16 $        & \cite{2020AJ....160..124M} & CMD    & 214 & $\pm 16 $                                                   \\
Indus I            & 210850.0$-$510949 & 20.00 & $\pm 0.20 $        & \cite{2020AJ....160..124M} & HB     & 100   & $\pm 10 $                                                   \\
Segue 3            & 212131.0$+$190702 & 16.16 & $\pm 0.09 $        & \cite{2011AJ....142...88F} & CMD    &  17.1 & $\pm 0.7  $        && -167.1 & $\pm 1.5$          & \cite{2011AJ....142...88F}       \\
Grus II            & 220404.8$-$462624 & 18.62 & $\pm 0.21 $        & \cite{2020AJ....160..124M} & CMD    &  53   & $\pm 5  $        && -110   & $\pm 0.5$          & \cite{2020AJ....160..124M} \\
Pegasus III        & 222422.6$+$052512 & 21.56 & $\pm 0.20 $        & \cite{2020AJ....160..124M} & CMD    & 205 & $\pm 20 $        && -222.9 & $\pm 2.6$          & \cite{2020AJ....160..124M} \\
Aquarius II        & 223355.5$-$091939 & 20.16 & $\pm 0.07 $        & \cite{2020AJ....160..124M} & TRGB   & 108 & $\pm 4  $        &&  -71.1 & $\pm 2.5$          & \cite{2020AJ....160..124M} \\
Tucana II          & 225155.1$-$583408 & 18.80 & $\pm 0.20 $        & \cite{2020AJ....160..124M} & HB     &  58 & $\pm 6  $        && -129.1 & $\pm 3.5$          & \cite{2020AJ....160..124M} \\
Grus I             & 225642.4$-$500948 & 20.40 & $\pm 0.20 $        & \cite{2020AJ....160..124M} & TRGB   & 120 & $\pm 12 $        && -140.5 & $^{+2.4}_{-1.6}$   & \cite{2020AJ....160..124M} \\
Pisces II          & 225831.0$+$055709 & 21.31 & $\pm 0.17 $        & \cite{2020AJ....160..124M} & TRGB   & 183 & $\pm 15 $        && -226.5 & $\pm 2.7$          & \cite{2020AJ....160..124M} \\
Tucana V           & 233724.0$-$631612 & 18.71 & $\pm 0.34 $        & \cite{2020AJ....160..124M} & CMD    &  55 & $\pm 9  $        &&  -36.2 & $^{+2.5}_{-2.2}$   & \cite{2020AJ....160..124M} \\
Phoenix II         & 233959.4$-$542422 & 19.60 & $\pm 0.15 $        & \cite{2020AJ....160..124M} & HB     &  83 & $\pm 6  $        &&   32.4 & $^{+3.7}_{-3.8}$   & \cite{2019AA...623A.129F} \\
Tucana III         & 235636.0$-$593600 & 17.01 & $\pm 0.16 $        & \cite{2020AJ....160..124M} & TRGB   &  25.2 & $\pm 1.9  $        && -102.3 & $\pm 0.4$          & \cite{2022ApJ...940..136P} \\

\end{longtable}

\begin{tablenotes}
\item \normalsize\textbf{Notes.} 
The columns contain: 
(1) galaxy name; 
(2) J2000 equatorial coordinates; 
(3--5) distance modulus, $(m-M)_0$, and its source; 
(6) distance determination method:
Blue Horizontal Branch (BHB),
Cepheids (Cep),
Color-Magnitude Diagram (CMD),
the S2 orbit around the supermassive black hole in our Galaxy (Direct),
Horizontal Branch (HB),
Main Sequence (MS),
Tip of the Red Giant Branch (TRGB); 
(7--8) heliocentric distance, $D$, in kpc; 
(9--11) heliocentric line-of-sight velocity, $\Vh$, in \kms{} and its source.
\end{tablenotes}

\end{ThreePartTable}
}

\section{List of satellites of the Andromeda Galaxy}
\label{sec:M31_datatable}

{
\begin{ThreePartTable}
\footnotesize

\begin{longtable}{llcr@{~}l@{~}ll@{~}r@{~}l@{}c@{~~}r@{~}l@{~}l}

\caption{Known satellites of the Andromeda Galaxy.}
\label{tab:M31_datatable}\\

\hline\hline
Name          & Alt~Name  & J2000             & \multicolumn{3}{c}{$(m-M)_0$}  & Method & \multicolumn{2}{c}{$D$} && \multicolumn{3}{c}{$\Vh$} \\
\cline{4-6}\cline{8-9}\cline{11-13}
              &           &                   & \multicolumn{3}{c}{mag}        &        & \multicolumn{2}{c}{kpc} && \multicolumn{3}{c}{\kms{}} \\
\hline
\endfirsthead

\hline
Name          & Alt~Name  & J2000             & \multicolumn{3}{c}{$(m-M)_0$}  & Method & \multicolumn{2}{c}{$D$} && \multicolumn{3}{c}{$\Vh$} \\
\cline{4-6}\cline{8-9}\cline{11-13}
              &           &                   & \multicolumn{3}{c}{mag}        &        & \multicolumn{2}{c}{kpc} && \multicolumn{3}{c}{\kms{}} \\
\hline
\endhead

\hline
\endfoot

\hline\hline
\endlastfoot

PAndAS-05     &           & 000024.1$+$435535 &       &                    &   &       &     &                && $-183.0$  & $\pm 7.0 $       & \cite{2014MNRAS.442.2165H} \\ 
And~XVIII     &           & 000214.5$+$450520 & 25.43 & $^{+0.05}_{-0.03}$ & \cite{2019MNRAS.489..763W} & HB    &1219 & $^{+28}_{-17}$ && $-332.1$  & $\pm 2.7$        & \cite{2013MNRAS.430..971W} \\ 
PAndAS-04     &           & 000442.9$+$472142 &       &                    &   &       &     &                && $-397.0$  & $\pm 7.0$        & \cite{2014MNRAS.442.2165H} \\ 
And~XX        &           & 000730.7$+$350756 & 24.35 & $\pm 0.08$         & \cite{2022ApJ...938..101S} & RR Lyr& 741 & $\pm 28$       && $-456.2$  & $^{+3.0}_{-3.4}$ & \cite{2013MNRAS.430..971W} \\ 
IC~10         &           & 002024.5$+$591730 & 24.50 & $\pm 0.12$         & \cite{2014AJ....147...13K} & TRGB  & 794 & $\pm 45$       && $-348.0$  & $\pm 2.9$        & \cite{2015AJ....149..180O} \\ 
And~XXVI      &           & 002345.6$+$475558 & 24.48 & $^{+0.06}_{-0.07}$ & \cite{2022ApJ...938..101S} & RR Lyr& 787 & $^{+22}_{-26}$ && $-260.6$  & $^{+4.0}_{-3.7}$ & \cite{2013MNRAS.430..971W} \\ 
And~XXV       &           & 003008.9$+$465107 & 24.38 & $^{+0.07}_{-0.06}$ & \cite{2022ApJ...938..101S} & RR Lyr& 752 & $^{+25}_{-21}$ && $-107.8$  & $^{+1.0}_{-0.9}$ & \cite{2013MNRAS.430..971W} \\ 
NGC~147       &           & 003350.8$+$483028 & 24.33 & $\pm 0.06$         & \cite{2022ApJ...938..101S} & RR Lyr& 735 & $\pm 21$       && $-193.0$  & $\pm 3.0$        & \cite{2013MNRAS.430..971W} \\ 
And~III       &           & 003533.8$+$362952 & 24.29 & $\pm 0.05$         & \cite{2022ApJ...938..101S} & RR Lyr& 721 & $\pm 17$       && $-344.3$  & $\pm 1.7$        & \cite{2013MNRAS.430..971W} \\ 
Cas~III       & And~XXXII & 003559.4$+$513335 & 24.52 & $\pm 0.06$         & \cite{2022ApJ...938..101S} & RR Lyr& 802 & $\pm 22$       && $-371.6$  & $\pm 0.7$        & \cite{2014ApJ...793L..14M} \\ 
And~XXX       &           & 003634.9$+$493848 & 23.74 & $\pm 0.06$         & \cite{2022ApJ...938..101S} & RR Lyr& 560 & $\pm 16$       && $-141.4$  & $^{+5.8}_{-6.7}$ & \cite{2013MNRAS.430..971W} \\ 
And~XVII      &           & 003707.0$+$441920 & 24.40 & $\pm 0.07$         & \cite{2022ApJ...938..101S} & RR Lyr& 759 & $\pm 25$       && $-251.1$  & $^{+1.5}_{-1.6}$ & \cite{2013MNRAS.430..971W} \\ 
And~XXVII     &           & 003727.1$+$452313 & 24.59 & $\pm 0.12$         & \cite{2011ApJ...732...76R} & HB    & 828 & $\pm 47$       && $-534.8$  & $^{+5.4}_{-4.9}$ & \cite{2013MNRAS.430..971W} \\ 
NGC~185       &           & 003858.0$+$482010 & 24.06 & $\pm 0.06$         & \cite{2022ApJ...938..101S} & RR Lyr& 649 & $\pm 18$       && $-202.0$  & $\pm 3.0$        & \cite{2013MNRAS.430..971W} \\ 
NGC~205       &           & 004022.5$+$414111 & 24.61 & $\pm 0.06$         & \cite{2022ApJ...938..101S} & RR Lyr& 836 & $\pm 23$       && $-241.0$  & $\pm 3.0$        & \cite{2013MNRAS.430..971W} \\ 
M~32          &           & 004242.1$+$405259 & 24.44 & $\pm 0.06$         & \cite{2022ApJ...938..101S} & RR Lyr& 773 & $\pm 22$       && $-200.0$  & $\pm 6.0$        & \cite{2013MNRAS.430..971W} \\ 
\textbf{M~31} &           & 004244.5$+$411609 & 24.45 & $\pm 0.06$         & \cite{2022ApJ...938..101S} & RR Lyr& 776 & $\pm 22$       && $-301.0$  & $\pm 1.0$        & \cite{2013MNRAS.430..971W} \\ 
And~I         &           & 004540.0$+$380214 & 24.45 & $\pm 0.05$         & \cite{2022ApJ...938..101S} & RR Lyr& 776 & $\pm 18$       && $-376.3$  & $\pm 2.2$        & \cite{2013MNRAS.430..971W} \\ 
And~XI        &           & 004620.0$+$334805 & 24.38 & $\pm 0.07$         & \cite{2022ApJ...938..101S} & RR Lyr& 752 & $\pm 25$       && $-427.0$  & $^{+2.9}_{-2.8}$ & \cite{2013MNRAS.430..971W} \\ 
And~XII       &           & 004727.0$+$342229 & 24.28 & $^{+0.08}_{-0.07}$ & \cite{2022ApJ...938..101S} & RR Lyr& 718 & $^{+27}_{-24}$ && $-557.1$  & $\pm 1.7$        & \cite{2013MNRAS.430..971W} \\ 
Bol~520       &           & 005042.4$+$325559 & 24.00 & $\pm 0.20$         & \cite{2006MNRAS.371.1983M} & TRGB  & 631 & $\pm 61$       && $-370.0$  & $\pm 5.0$        & \cite{2016MNRAS.456..831S} \\ 
And~XIV       &           & 005135.0$+$294149 & 24.44 & $\pm 0.06$         & \cite{2022ApJ...938..101S} & RR Lyr& 773 & $\pm 22$       && $-480.6$  & $\pm 1.2$        & \cite{2013MNRAS.430..971W} \\ 
And~XIII      &           & 005151.0$+$330016 & 24.57 & $\pm 0.07$         & \cite{2022ApJ...938..101S} & RR Lyr& 820 & $\pm 27$       && $-185.4$  & $\pm 2.4$        & \cite{2013MNRAS.430..971W} \\ 
And~IX        &           & 005252.8$+$431200 & 24.60 & $\pm 0.06$         & \cite{2022ApJ...938..101S} & RR Lyr& 832 & $\pm 23$       && $-209.4$  & $\pm 2.5$        & \cite{2013MNRAS.430..971W} \\ 
PAndAS-48     &           & 005928.2$+$312910 & 24.57 & $\pm 0.11$         & \cite{2013ApJ...770L..17M} & HB    & 820 & $\pm 43$       && $-250.0$  & $\pm 5.0$        & \cite{2014MNRAS.442.2165H} \\ 
And~XVI       &           & 005929.8$+$322236 & 23.57 & $\pm 0.08$         & \cite{2022ApJ...938..101S} & RR Lyr& 518 & $\pm 19$       && $-367.3$  & $\pm 2.8$        & \cite{2013MNRAS.430..971W} \\ 
PAndAS-50     &           & 010150.6$+$481819 &       &                    &   &       &     &                && $-323.0$  & $\pm 7.0$        & \cite{2014MNRAS.442.2165H} \\ 
LGS~3         & Pisces~I  & 010355.0$+$215306 & 23.91 & $\pm 0.05$         & \cite{2022ApJ...938..101S} & RR Lyr& 605 & $\pm 14$       && $-286.5$  & $\pm 0.3$        & \cite{2013MNRAS.430..971W} \\ 
And~X         &           & 010633.7$+$444816 & 24.00 & $\pm 0.06$         & \cite{2022ApJ...938..101S} & RR Lyr& 631 & $\pm 18$       && $-164.1$  & $\pm 1.7$        & \cite{2013MNRAS.430..971W} \\ 
And~V         &           & 011017.1$+$473741 & 24.58 & $\pm 0.06$         & \cite{2022ApJ...938..101S} & RR Lyr& 824 & $\pm 23$       && $-397.4$  & $\pm 1.5$        & \cite{2013MNRAS.430..971W} \\ 
And~XV        &           & 011418.7$+$380703 & 24.37 & $\pm 0.05$         & \cite{2022ApJ...938..101S} & RR Lyr& 748 & $\pm 17$       && $-323.0$  & $\pm 1.4$        & \cite{2013MNRAS.430..971W} \\ 
And~II        &           & 011629.8$+$332509 & 24.12 & $\pm 0.05$         & \cite{2022ApJ...938..101S} & RR Lyr& 667 & $^{+16}_{-16}$ && $-193.6$  & $\pm 1.0$        & \cite{2013MNRAS.430..971W} \\ 
And~XXIV      &           & 011830.0$+$462258 & 23.92 & $\pm 0.07$         & \cite{2022ApJ...938..101S} & RR Lyr& 608 & $\pm 20$       && $-127.8$  & $^{+5.3}_{-5.4}$ & \cite{2013MNRAS.430..971W} \\ 
And~XXIX      &           & 011830.0$+$304520 & 24.26 & $\pm 0.06$         & \cite{2022ApJ...938..101S} & RR Lyr& 711 & $\pm 20$       && $-194.4$  & $\pm 1.5$        & \cite{2013ApJ...768..172C} \\ 
Tri~III       & Pisces~VII& 012141.3$+$262332 & 24.81 & $^{+0.15}_{-0.13}$ & \cite{2013ApJ...770L..17M} & TRGB  & 916 & $^{+65}_{-53}$ && $-138.6$  & $\pm 0.5$        & \cite{2015ApJS..219...12A} \\ 
PAndAS-56     &           & 012303.5$+$415511 &       &                    &   &       &     &                && $-239.0$  & $\pm 8.0$        & \cite{2014MNRAS.442.2165H} \\ 
And~XXII      &           & 012740.0$+$280525 & 24.39 & $\pm 0.07$         & \cite{2022ApJ...938..101S} & RR Lyr& 755 & $\pm 25$       && $-129.0$  & $^{+2.1}_{-2.2}$ & \cite{2013MNRAS.430..971W} \\ 
PAndAS-57     &           & 012747.5$+$404047 &       &                    &   &       &     &                && $-186.0$  & $\pm 6.0$        & \cite{2014MNRAS.442.2165H} \\ 
PAndAS-58     &           & 012902.1$+$404708 &       &                    &   &       &     &                && $-167.0$  & $\pm 10.0$       & \cite{2014MNRAS.442.2165H} \\ 
And~XXIII     &           & 012921.8$+$384308 & 24.36 & $\pm 0.07$         & \cite{2022ApJ...938..101S} & RR Lyr& 745 & $\pm 24$       && $-242.7$  & $\pm 1.0$        & \cite{2013MNRAS.430..971W} \\ 
M~33          &           & 013350.8$+$303937 & 24.67 & $\pm 0.06$         & \cite{2022ApJ...938..101S} & RR Lyr& 859 & $\pm 24$       && $-180.0$  & $\pm 1.0$        & \cite{2013MNRAS.430..971W} \\ 
Per~I      	  & And~XXXIII& 030123.6$+$405918 & 24.24 & $\pm 0.06$         & \cite{2022ApJ...938..101S} & RR Lyr& 705 & $\pm 20$       && $-325.9$  & $\pm 3.0$        & \cite{2014ApJ...793L..14M} \\ 
And~XXVIII 	  &           & 223241.2$+$311358 & 24.36 & $\pm 0.05$         & \cite{2022ApJ...938..101S} & RR Lyr& 745 & $\pm 17$       && $-326.2$  & $\pm 2.7$        & \cite{2013ApJ...768..172C} \\ 
Lac~I         & And~XXXI  & 225816.3$+$411728 & 24.36 & $\pm 0.05$         & \cite{2022ApJ...938..101S} & RR Lyr& 745 & $\pm 17$       && $-198.4$  & $\pm 1.1$        & \cite{2014ApJ...793L..14M} \\ 
Cas~dSph      & And~VII   & 232631.8$+$504032 & 24.40 & $\pm 0.06$         & \cite{2022ApJ...938..101S} & RR Lyr& 759 & $\pm 21$       && $-307.2$  & $\pm 1.3$        & \cite{2013MNRAS.430..971W} \\ 
Pegasus       & Peg~DIG   & 232827.6$+$144434 & 24.74 & $\pm 0.05$         & \cite{2022ApJ...938..101S} & RR Lyr& 887 & $\pm 21$       && $-184.5$  & $\pm 0.3$        & \cite{2013MNRAS.430..971W} \\ 
Peg~dSph      & And~VI    & 235146.9$+$243557 & 24.23 & $\pm 0.06$         & \cite{2022ApJ...938..101S} & RR Lyr& 702 & $\pm 20$       && $-340.8$  & $\pm 1.9$        & \cite{2013MNRAS.430..971W} \\ 
And~XXI       &           & 235447.7$+$422815 & 24.44 & $^{+0.06}_{-0.07}$ & \cite{2022ApJ...938..101S} & RR Lyr& 773 & $^{+22}_{-25}$ && $-362.7$  & $\pm 0.8$        & \cite{2013MNRAS.430..971W} \\ 
PAndAS-01     &           & 235712.0$+$433308 &       &                    &   &       &     &                && $-333.0$  & $\pm 21.0$       & \cite{2014MNRAS.442.2165H} \\ 
PAndAS-02     &           & 235755.6$+$414649 &       &                    &   &       &     &                && $-226.0$  & $\pm 4.0$        & \cite{2014MNRAS.442.2165H} \\ 
And~XIX       &           & 235855.6$+$350237 & 24.55 & $^{+0.09}_{-0.08}$ & \cite{2022ApJ...938..101S} & RR Lyr& 813 & $^{+34}_{-31}$ && $-111.2$  & $^{+1.2}_{-1.3}$ & \cite{2013MNRAS.430..971W} \\
\end{longtable}

\begin{tablenotes}
\item \normalsize\textbf{Notes.} 
The columns contain: 
(1) galaxy name; 
(2) alternative name; 
(3) J2000 equatorial coordinates; 
(4--6) distance modulus, $(m-M)_0$, and its source; 
(7) distance measurement method:
Horizontal Branch (HB),
RR Lyrae variables (RR Lyr),
Tip of the Red Giant Branch (TRGB); 
(8--9) heliocentric distance, $D$, in kpc; 
(10--12) heliocentric line-of-sight velocity, $\Vh$, in km/s and its source.
\end{tablenotes}

\end{ThreePartTable}
}

%\newpage

\section{List of the Milky Way mass estimates}

\begin{table}[h!]
\centering
\caption{Milky Way virial mass obtained by different methods since 2010.}
\label{tab:mw_mass_compilation}
\begin{tabular}{lrlrr}
\hline\hline
Authors & \multicolumn{2}{c}{$\Mvir$}                  & \multicolumn{1}{c}{$R_{\max}$} & Ref. \\
        & \multicolumn{2}{c}{$\times 10^{11}$~\Msun{}} & \multicolumn{1}{c}{kpc}        &  \\
\midrule

\textbf{Escape Velocities} & & & \\
\citeauthor{2022AandA...664A.148P} (\citeyear{2022AandA...664A.148P}) & 8.3\phantom{0} & $^{+2.9}_{-1.6}$   & 20  & \cite{2022AandA...664A.148P} \\
\citeauthor{2024ApJ...972...70R} (\citeyear{2024ApJ...972...70R})     & 6.4\phantom{0} & $^{+1.5}_{-1.4}$   & 11  & \cite{2024ApJ...972...70R}   \\
\midrule

\textbf{Rotation Curve} & & & \\
\citeauthor{2011MNRAS.414.2446M} (\citeyear{2011MNRAS.414.2446M})     & 12.6\phantom{0} & $\pm 2.4$           & 8   & \cite{2011MNRAS.414.2446M} \\
\citeauthor{2012ApJ...759..131B} (\citeyear{2012ApJ...759..131B})     & $\sim8.0$\phantom{0} &                & 14  & \cite{2012ApJ...759..131B} \\
\citeauthor{2016MNRAS.463.2623H} (\citeyear{2016MNRAS.463.2623H})     &  8.5\phantom{0} & $^{+0.7}_{-0.8}$    & 25  & \cite{2016MNRAS.463.2623H} \\
\citeauthor{2019ApJ...871..120E} (\citeyear{2019ApJ...871..120E})     &  7.25           & $\pm 0.26$          & 25  & \cite{2019ApJ...871..120E} \\
\citeauthor{2020MNRAS.494.4291C} (\citeyear{2020MNRAS.494.4291C})     & 10.8\phantom{0} & $^{+2.0}_{-1.4}$    & 20  & \cite{2020MNRAS.494.4291C} \\
\citeauthor{2020ApJ...895L..12A} (\citeyear{2020ApJ...895L..12A})     &  8.22           & $\pm 0.52$          & 19  & \cite{2020ApJ...895L..12A} \\
\citeauthor{2023ApJ...946...73Z} (\citeyear{2023ApJ...946...73Z})     &  8.05 & $\pm 1.15$           & 30  & \cite{2023ApJ...946...73Z} \\
\citeauthor{2023ApJ...945....3S} (\citeyear{2023ApJ...945....3S})     &  6.5\phantom{0} & $\pm 0.3$           & 28  & \cite{2023ApJ...945....3S} \\
\citeauthor{2024arXiv240712551K} (\citeyear{2024arXiv240712551K})     & 13.4\phantom{0} & $\pm 0.1$           & 25  & \cite{2024arXiv240712551K} \\
\midrule

\textbf{Streams} & & & \\
\citeauthor{2022MNRAS.517.1737C} (\citeyear{2022MNRAS.517.1737C})     & 15.0\phantom{0} & $\pm 3.2$                        &     & \cite{2022MNRAS.517.1737C} \\
\midrule

\textbf{Spherical Jeans Equation} & & & \\
\citeauthor{2010ApJ...720L.108G} (\citeyear{2010ApJ...720L.108G})     & 16.0\phantom{0} & $\pm 3.0$                        & 80  & \cite{2010ApJ...720L.108G} \\
\citeauthor{2012ApJ...761...98K} (\citeyear{2012ApJ...761...98K})     &  9.0\phantom{0} & $^{+4.0}_{-3.0}$                 & 60  & \cite{2012ApJ...761...98K} \\
\citeauthor{2014ApJ...794...59K} (\citeyear{2014ApJ...794...59K})     &  8.0\phantom{0} & $^{+3.1}_{-1.6}$                 & 160 & \cite{2014ApJ...794...59K} \\
\citeauthor{2018RAA....18..113Z} (\citeyear{2018RAA....18..113Z})     & 10.8\phantom{0} & $^{+1.7}_{-1.4}$                 & 120 & \cite{2018RAA....18..113Z} \\
\citeauthor{2022MNRAS.516..731B} (\citeyear{2022MNRAS.516..731B})     & \multicolumn{2}{c}{$[5.5^{+1.5}_{-1.1}$--$10.0^{+6.7}_{-3.3}]$}  & 70  & \cite{2022MNRAS.516..731B} \\
\midrule

\textbf{Distribution Function} & & & \\
\citeauthor{2015ApJ...806...54E} (\citeyear{2015ApJ...806...54E})     & 13.7\phantom{0} & $^{+1.4}_{-1.0}$    & 261 & \cite{2015ApJ...806...54E}  \\
\citeauthor{2016ApJ...829..108E} (\citeyear{2016ApJ...829..108E})     &  9.02           & $^{+1.7}_{-3.3}$    & 200 & \cite{2016ApJ...829..108E}  \\
\citeauthor{2018ApJ...862...52S} (\citeyear{2018ApJ...862...52S})     & 20.5\phantom{0} & $^{+9.7}_{-7.9}$    & 100 & \cite{2018ApJ...862...52S}  \\
\citeauthor{2019ApJ...873..118W} (\citeyear{2019ApJ...873..118W})     & 15.4\phantom{0} & $^{+7.5}_{-4.4}$    & 40  & \cite{2019ApJ...873..118W}  \\
\citeauthor{2019MNRAS.484.2832V} (\citeyear{2019MNRAS.484.2832V})     & 12\phantom{.00} & $^{+15}_{-5}$       & 50  & \cite{2019MNRAS.484.2832V}  \\
\citeauthor{2019AandA...621A..56P} (\citeyear{2019AandA...621A..56P}) & 13\phantom{.00} & $\pm 3$             & 20  & \cite{2019AandA...621A..56P}\\
\citeauthor{2020ApJ...894...10L} (\citeyear{2020ApJ...894...10L})     & 12.3\phantom{0} & $^{+2.1}_{-1.8}$    & 200 & \cite{2020ApJ...894...10L}  \\
\citeauthor{2021MNRAS.501.5964D} (\citeyear{2021MNRAS.501.5964D})     & 10.1\phantom{0} & $\pm 2.4$           & 100 & \cite{2021MNRAS.501.5964D}  \\
\citeauthor{2022ApJ...925....1S} (\citeyear{2022ApJ...925....1S})     & 10.8\phantom{0} & $^{+1.2}_{-1.1}$    & 145 & \cite{2022ApJ...925....1S}  \\
\midrule

\textbf{Kinematics of Satellites} & & & \\
\citeauthor{2010MNRAS.406..264W} (\citeyear{2010MNRAS.406..264W})     & 14.0\phantom{0} & $\pm 3.0$          & 300 & \cite{2010MNRAS.406..264W} \\
\citeauthor{2011ApJ...743...40B} (\citeyear{2011ApJ...743...40B})     & 12.0\phantom{0} & $^{+7.0}_{-4.0}$   & 300 & \cite{2011ApJ...743...40B} \\
\citeauthor{2013ApJ...770...96G} (\citeyear{2013ApJ...770...96G})     & 11.5\phantom{0} & $^{+4.8}_{-3.4}$   & 200 & \cite{2013ApJ...770...96G} \\
\citeauthor{2013ApJ...768..140B} (\citeyear{2013ApJ...768..140B})     & 16\phantom{.00} & $^{+8}_{-6}$       & 261 & \cite{2013ApJ...768..140B} \\ % 90% confidence 
\citeauthor{2014MNRAS.445.2049C} (\citeyear{2014MNRAS.445.2049C})     &  7.8\phantom{0} & $^{+5.7}_{-3.3}$   & 200 & \cite{2014MNRAS.445.2049C} \\
\citeauthor{2014MNRAS.437..959B} (\citeyear{2014MNRAS.437..959B})     & 11.0\phantom{0} & $^{+4.5}_{-2.9}$   & 200 & \cite{2014MNRAS.437..959B} \\
\citeauthor{2017MNRAS.464.3825P} (\citeyear{2017MNRAS.464.3825P})     &  8.3\phantom{0} & $^{+7.7}_{-5.5}$   & 200 & \cite{2017MNRAS.464.3825P} \\
\citeauthor{2018ApJ...857...78P} (\citeyear{2018ApJ...857...78P})     &  6.8\phantom{0} & $^{+2.3}_{-2.6}$   & 200 & \cite{2018ApJ...857...78P} \\
\citeauthor{2020MNRAS.494.5178F} (\citeyear{2020MNRAS.494.5178F})     & 15.1\phantom{0} & $^{+4.5}_{-4.0}$   & 300 & \cite{2020MNRAS.494.5178F} \\
\citeauthor{2022MNRAS.513.4968R} (\citeyear{2022MNRAS.513.4968R})     & $\sim10$--12    &          & 300 & \cite{2022MNRAS.513.4968R} \\
\citeauthor{2024OJAp....7E..50K} (\citeyear{2024OJAp....7E..50K})     &  9.96           & $\pm 1.45$         & 200 & \cite{2024OJAp....7E..50K} \\
\textbf{this work}                                                    &  7.9\phantom{0} & $\pm 2.3$          & 236 &                            \\
\hline\hline

\end{tabular}
\end{table}

Table~\ref{tab:mw_mass_compilation} presents a compilation of estimates of the virial mass of our Galaxy within its virial radius of 200~kpc, based on publications over the past 15 years, starting from 2010.
The columns provide the following information:
1) a reference to the authors, as presented in Fig.~\ref{fig:MWmassLiterature};
2) the mass of the Galaxy, $\Mvir$, within the virial radius of 200~kpc; 
3) a maximum radius, $R_{\max}$, probed by the mass tracers;
4) a citation to the corresponding article.
Only studies that analyze satellite motion can directly measure the mass within the virial radius. 
In all other cases, the virial mass within $\sim200$~kpc is estimated using a model of the gravitational potential of our Galaxy.
Below is a brief description of the methods presented in the table.

\begin{description}

\item[Escape Velocities.]
This method analyzes the tail of the velocity distribution of the halo stars (so-called high-velocity stars) to estimate the escape speed $v_\mathrm{esc}(r) = \sqrt{2|\Phi(r)|} $ from the Galaxy. 
%This method analyzes high-velocity stars near the escape speed $v_\mathrm{esc}(r) = \sqrt{2|\Phi(r)|} $. 
%A power-law fit \( f(v) \propto (v_{\text{esc}} - v)^k \) (calibrated from simulations) estimates the potential. 
%Contamination by unbound stars and model-dependent assumptions limit accuracy, though refinements account for stellar anisotropy (e.g., the 'Gaia Sausage').

\item[Rotation Curve]
method bases on the measurement of the circular velocities of stars and gas. 
Inner regions use HI/CO gas kinematics, while outer regions rely on tracers like red giants and masers, combined with Gaia proper motions.
%Usually it is possible to trace the rotation curve up to 20--25 kpc from the center of the Galaxy. 
%The relation \( v_c(r) = \sqrt{GM(<r)/r} \) links velocity to enclosed mass, but accuracy depends on correcting for non-circular motions and the Galactic bar’s influence. Modern techniques (e.g., Jeans equation with Gaia data) extend this to the halo, though model degeneracies remain.

\item[Stellar Streams.]
Tidal streams, such as GD-1 and Sagittarius, are sensitive probes of the Galactic potential. 
Orbit-fitting and N-body simulations model their dynamics constraining mass at intermediate radii $~20$-–100~kpc. 
Challenges include separating the host potential’s effects from subhalo perturbations. 
%Future surveys (e.g., LSST) will improve constraints on halo shape and mass distribution.

\item[Spherical Jeans Equation]
connects the radial velocity dispersion, $\sigma_r$, the radial density density profile, $\rho_*$, and the velocity anisotropy, $\beta$, of the mass tracers with the underlying gravitational potential.
It assumes that a system is a spherically symmetric and in a steady-state.
%mass profile to tracer velocity dispersion \( \sigma_r \) and density \( \rho_* \), but suffers from the mass-anisotropy degeneracy. Combining Gaia proper motions with spectroscopy helps break this degeneracy, though deviations from sphericity (e.g., LMC, triaxial halos) bias results. Simplified mass estimators trade accuracy for computational efficiency.

\item[Distribution Function]
approach models the gravitational potential by reconstructing the phase-space distribution of tracers (halo stars, globular clusters) using integrals of motion or action.
%The distribution function $ f(\mathbf{x}, \mathbf{v}) $ connects to integrals of motion (energy, actions). 
%Spherical systems use Eddington inversion, while axisymmetric/triaxial potentials require action-based methods. 
%Gaia data improves fits, but uncertainties arise from disequilibrium (e.g., streams) and the mass-anisotropy degeneracy.

\item[Kinematics of Satellites.]
The motions of satellite galaxies and globular clusters probe the outer halo mass. 
Methods include timing arguments (for bound systems like Leo~I) and tracer mass estimators, often compared with cosmological simulations. %Gaia proper motions enhance precision, but assumptions of equilibrium and the LMC’s gravitational influence introduce uncertainties.

\end{description}

%\newpage

\section{List of the M~31 mass estimates}

\begin{table}[h!]
\centering
\caption{M~31 virial mass estimates obtained by different methods since 2010.\label{tab:m31_mass_compilation}}
\begin{tabular}{lrlrr}
\hline\hline
Authors & \multicolumn{2}{c}{$\Mvir$}                  & \multicolumn{1}{c}{$R_{\max}$} & Ref. \\
        & \multicolumn{2}{c}{$\times 10^{11}$~\Msun{}} & \multicolumn{1}{c}{kpc}        &  \\
\midrule

\textbf{Rotation Curve} & & & \\
\citeauthor{2012AandA...546A...4T} (\citeyear{2012AandA...546A...4T}) &   9.5           & $\pm 1.5$         &  25 & \cite{2012AandA...546A...4T} \\
\citeauthor{2014ApJ...789...62H} (\citeyear{2014ApJ...789...62H})     &  18.2           & $^{+4.9}_{-3.9}$  &  30 & \cite{2014ApJ...789...62H} \\
\citeauthor{2015PASJ...67...75S} (\citeyear{2015PASJ...67...75S})     &  13.7           & $\pm 5.2$         &  31 & \cite{2015PASJ...67...75S} \\
\citeauthor{2024MNRAS.528.2653Z} (\citeyear{2024MNRAS.528.2653Z})     &  11.4           & $^{+5.1}_{-3.5}$  & 125 & \cite{2024MNRAS.528.2653Z} \\
\midrule

\textbf{Substructure} & & & \\
\citeauthor{2013MNRAS.434.2779F} (\citeyear{2013MNRAS.434.2779F})     &  19.9           & $^{+5.2}_{-4.1}$  &     & \cite{2013MNRAS.434.2779F} \\
\midrule

\textbf{Globular Clusters} & & & \\
\citeauthor{2013ApJ...768L..33V} (\citeyear{2013ApJ...768L..33V})     &  13.5           & $\pm 3.5$         & 130 & \cite{2013ApJ...768L..33V} \\
\midrule

\textbf{Kinematics of Satellites} & & & \\
\citeauthor{2010MNRAS.406..264W} (\citeyear{2010MNRAS.406..264W})     &  14\phantom{.0} & $\pm 4$           & 300 & \cite{2010MNRAS.406..264W} \\
\citeauthor{2017MNRAS.468.3428P} (\citeyear{2017MNRAS.468.3428P})     &  13.7           & $^{+13.9}_{-7.5}$ & 300 & \cite{2017MNRAS.468.3428P} \\
\citeauthor{2023ApJ...948..104P} (\citeyear{2023ApJ...948..104P})     &  30.2           & $^{+13.0}_{-6.9}$ & 300 & \cite{2023ApJ...948..104P} \\
\textbf{this work}                                                    & 15.5            & $\pm 3.4$ & 292 & \\
\midrule

\textbf{Local Group Kinematics} & & & \\
\citeauthor{2012ApJ...753....8V} (\citeyear{2012ApJ...753....8V})     &  15.4           & $\pm 3.9$         &     & \cite{2012ApJ...753....8V} \\
\citeauthor{2014MNRAS.443.1688D} (\citeyear{2014MNRAS.443.1688D})     &  17\phantom{.0} & $\pm 3$           &     & \cite{2014MNRAS.443.1688D} \\
\citeauthor{2014MNRAS.443.2204P} (\citeyear{2014MNRAS.443.2204P})     &  15\phantom{.0} & $\pm 3$           &     & \cite{2014MNRAS.443.2204P} \\
\citeauthor{2016MNRAS.456L..54P} (\citeyear{2016MNRAS.456L..54P})     &  13.3           & $^{+3.9}_{-3.3}$  &     & \cite{2016MNRAS.456L..54P} \\
\hline\hline

\end{tabular}
\end{table}

Table~\ref{tab:m31_mass_compilation} contains a compilation from the literature of recent measurements (from the 2010) of the virial mass of the Andromeda Galaxy.
It contains the following information:
1) a reference to the authors, as presented in Fig.~\ref{fig:M31massLiterature};
2) the M~31 mass within the virial radius of 200--300~kpc; 
3) a maximal radius, $R_{\max}$, probed by the mass tracers;
4) a citation to the corresponding article.
A brief description of the methods listed in the table is provided below.

\begin{description}

\item[Rotation Curve]
measures circular velocities of tracers (stars, gas) to determine the mass distribution. 
In the case of M~31, HI observations extend to $~40$ kpc, revealing flat rotation curves that indicate dark matter dominance. 
%Key uncertainties include inclination corrections and stellar mass-to-light ratios.
The virial mass within $\sim200$~kpc is based on a model of the M~31 gravitational potential.

\item[Subsctructure]
Tidal streams (e.g., the Giant Stream) constrain halo mass and shape using N-body simulations. 
Progenitor properties and merger history must be taken into accounted.

\item[Globular Clusters]
(GC) trace mass at intermediate radii of 20--200~kpc. 
%Their larger numbers improve statistical precision, but distinguishing accreted vs. in-situ clusters and tidal effects introduces uncertainty.
The advantage is the large number of known GC.
However, a major challenge lies in distinguishing between \textit{in-situ} and recently accreted clusters, as well as accounting for the tidal effects.

\item[Kinematics of Satellites]
uses the velocities and distances of the dwarf satellites to infer gravitational mass at 100--300~kpc. 
Challenges include small sample sizes ($<30$ satellites) and potential disequilibrium due to ongoing mergers.

\item[Local Group Kinematics]
The M~31–MW motion provides a total mass estimate using the timing argument. 
Modern approaches include proper motions and perturbations from the LMC, but depend on assumed mass ratios.

\end{description}

\end{document}